\documentclass{article}
\usepackage{emulateapj,timesfonts}
\usepackage{float,epsfig,psfig}
\usepackage{pstricks}
   \def\apj#1#2#3#4{\par #4 19#3, {ApJ,\/} {#1}, #2 }

   \def\apjs#1#2#3#4{\par #4 19#3, {ApJ (Supplement Series),\/} { #1}, #2 }
   \def\aj#1#2#3#4{\par #4 19#3, {AJ,\/} { #1}, #2}
   \def\pasj#1#2#3#4{\par #4 19#3, {PASJ,\/} { #1},#2 }

   \def\aa#1#2#3#4{\par #4 19#3, {A\&A,\/} { #1}, #2 }
   
   \def\aas#1#2#3#4{\par #4 19#3, {A\&A (Supplement Series),\/} { #1}, #2 }
   
   \def\mnras#1#2#3#4{\par #4 19#3, {MNRAS,\/} { #1}, #2 }

   \def\apjl#1#2#3#4{\par #4 19#3, {ApJ (Letters),\/} { #1}, #2 }

   \def\BIB {\par}

\def\einstein{{\it Einstein}}

\def\asca{{\it ASCA}}

\def\hea4{{\it HEAO~A4}}
\def\heaoa2{{\it HEAO~A2}}
\def\heao1{{\it HEAO~1}}

\def\arcmin{$^\prime$}

\def\amin{$^\prime$}
\def\asec{$^{\prime\prime}$}

\def\eg{{\it e.g.}~}

\def\deg{$^{\circ}$}

\def\h0{$H_{\rm o}=50$~km~s$^{-1}$~Mpc$^{-1}$}
\def\q0{$q_{\rm o}$}

\def\doublespace{\baselineskip=24pt}

\def\msun     {$M_{\odot}$}
\def\lsun     {$L_{\odot}$}

\def\etal    {{ et~al.}~}

\def\ergscms   {~ergs~cm$^{-2}$~s$^{-1}$}

\def\phscmkev{~{ph~s$^{-1}$~cm$^{-2}$~keV$^{-1}$}}

\def\cms3  {~{cm$^{-3}$}}
\def\nhl{~{$N_{\rm H}$}}

\def\ssd{{\mbox{0.1--0.2~keV}}}

\def\ad{{\mbox{0.2--2.0~keV}}}
\def\nh{N$_{\rm H}$}
\def\nhl{N$_{\rm H}$}
\def\kte{{\it k}T$_{\rm e}$}

\def\tild{^{^{\hspace*{-5pt}\sim}}}

\begin{document}

\submitted{submitted to ApJ in April 1999, revised December 1999, accepted
  March 2000}
\title{{SN Ia Enrichment in Virgo Early-Type Galaxies from ROSAT and ASCA
    Observations}} 
\author{A.~Finoguenov$^{1}$ and C.~Jones$^2$}
\affil{
{$^1$ Space Research Institute, Profsoyuznaya 84/32 117810 Moscow, Russia}\\
{$^2$ Smithsonian Astrophysical Observatory, 60 Garden st., MS 3, Cambridge,
  MA 02138, USA}}
\authoremail{alexis@hea.iki.rssi.ru}

%\doublespace

\begin{abstract}
%\doublespace
  
  We analyzed nine X-ray bright Virgo early-type galaxies observed by both
  ASCA and ROSAT. Through spatially resolved spectroscopy, we determined the
  radial temperature profile and abundances of Mg, Si and Fe for six
  galaxies. The temperature profiles are consistent with isothermal
  temperatures outside of a cooler region at the galaxy center. We present
  new evidence for iron abundance gradients in NGC4472 and NGC4649 and
  confirm the previous results on NGC4636. Mg and Si abundance gradients on
  average are flatter compared to those of iron and correspond to an
  underabundance of alpha-process elements at high Fe values, while at low
  iron, the element ratios favor enrichment by type II SNe. We explain the
  observed trend by the metallicity dependence of SN~Ia metal production and
  present constraints on the available theoretical modeling for
  low-metallicity inhibition of SNe~Ia (Kobayashi \etal 1998). Our results
  imply a cut-off metallicity in the range 0.07--0.3 solar and require a
  lower limit of 0.3 solar on the Fe contribution of SN~Ia. We estimate an
  SN~Ia rate at the centers of the brightest galaxies in our sample of
  $\sim0.08h_{75}^{3}$ SNU (supernova units). The rates inferred from
  optical searches should be corrected for the presence of ``faint'' SN Ia
  events, since these release limited metals and therefore do not contribute
  significantly to the measured metallicity in the X-ray gas.  With this
  correction the present-epoch SN~Ia rate in early-type galaxies is
  $0.10\pm0.06$ $h_{75}^{2}$ SNU (Cappellaro \etal 1997) and is therefore
  comparable with the X-ray estimates. A simple comparison shows the X-ray
  abundances we derive are still discrepant from optically determined
  values.  We attribute this difference to the low spatial resolution of our
  X-ray measurements, radial gradients in the abundances and the importance
  of hydrodynamical effects, particularly the inflow of cooling gas, on the
  measured X-ray abundances.

\end{abstract}


\section{ Introduction}

X-ray observations of the hot interstellar gas in early-type galaxies
provide a unique tool for the study of stellar metallicities and SN~Ia rates
within distances of a few effective radii from the galaxy's center.
Detailed comparison of the metallicities determined from X-ray and optical
observations, as was done for NGC5846 (Finoguenov \etal 1999, hereinafter
F99), demonstrates good agreement.

The abundance pattern detected through X-ray spectroscopy in low-mass
systems like groups and early-type galaxies favors a dominance of SNe~Ia in
the enrichment of the hot gas in these systems (F99, Finoguenov \& Ponman
1999, hereinafter FP99). During the early stages of elliptical galaxy
formation, SN~II products are likely to escape the galaxy, but be contained
by the potentials of clusters of galaxies (see Fukazawa \etal 1998 for a
recent compilation of X-ray data, hereinafter Fu98). The hot gaseous halos
of ellipticals contain elements released via stellar mass loss, plus
elements synthesized in SN~Ia events, {\it after} the cessation of early
star formation (\eg\ Renzini \etal 1993), which is used to constrain the
present-day SN~Ia rate. In previous X-ray studies, though, limits on SN~Ia
activity generally were derived without regard to the metallicity of the
progenitor stars.  While optical searches find only a handful of SNe~Ia in
early-type galaxies (Cappellaro \etal 1997), through X-ray observations we
can study the rate of SN~Ia events as a function of the metallicity of the
host galaxy.

This is the {\it first Paper} in our project to study the X-ray properties
of Virgo galaxies, observed by ASCA. The main goal is to present the results
of gas temperature and element abundance measurements, and to compare these
with current observations and theoretical models. Our approach provides a full
treatment of the instrumental effects, including PSF, as well as gas
projection effects, on the derivation of the spatially resolved temperature
and abundance structure. We consider separately the behavior of iron and
some of the $\alpha$-process elements, that enables us to separate the
contribution from different types of SN.  Discussion of a number of
important quantities, including the luminosity of the hard galactic
component is postponed for the next paper, while we await the completion of
our observational program on low-luminosity Virgo early-types, that should
allow us to present the results for a complete optically selected sample.

This {\it Paper} is organized as follows: in section \ref{sec:dr} we
describe the X-ray analysis, where in subsections we discuss our
measurements and comment on individual galaxies; in section \ref{sec:disc-n}
we attempt a detailed study of the abundance pattern, measured for our
sample. Through a comparison of our results with abundance measurements for
the stars in our Galaxy, as well as for the stellar population of cluster
early-type galaxies, we propose a progenitor star metallicity dependence for
SN~Ia metal yields. In this context, we discuss the model for
low-metallicity inhibition of SN~Ia's (Kobayashi \etal 1998).

\section{Analysis of ROSAT and ASCA Observations} \label{sec:dr}

Since our observations are from the ROSAT and ASCA archives, most of the
observational information is given elsewhere (\eg\  Matsushita
\etal 1997, Forman \etal 1993, Mushotzky \etal 1994, Awaki \etal 1994). A
detailed description of the ASCA observatory, as well as the SIS detectors,
can be found in Tanaka, Inoue \& Holt (1994) and Burke \etal (1991). For the
initial ASCA data reduction, we use the FTOOLS 4.1 package, in which changes
in the instrumental parameters, such as gain and energy resolution are taken
into account. To extract imaging information from ROSAT PSPC observations,
we use the extended source software described in Snowden \etal (1994) with
further references therein.

The combination of ROSAT and ASCA observations, utilized in the present
research, makes use of ROSAT's superior spatial resolution to overcome the
difficulties caused by ASCA's broad, energy dependent point spread function
(PSF) in obtaining high quality spectral information. Moreover, in
early-type galaxies, since the quantities we measure, in particular the gas
temperature and elemental abundance, change rather rapidly with radius, the
modeling must be with high spatial resolution. This imposes potential
problems on the maximum-likelihood solution. In our regularization method,
based on well-known general prescriptions (Press \etal 1992), we retain the
spatially fine binning, but require a smoothness for the obtained solution.
Details of this procedure are described elsewhere (F99, FP99).  We adapted
the XSPEC analysis package to perform the actual spectral fitting. Since the
effects of the broad ASCA PSF introduces a correlation in the error
estimation between adjacent spatial bins, instead of conventional error bars,
we present the determination of the parameter uncertainty as a shaded area,
while the estimation itself is similar to other studies and is based on the
``one parameter of interest'' method (Avni 1976) for every spatial point.

In all our spectral modeling, we use the MEKAL model (Mewe \etal 1985,
Mewe and Kaastra 1995, Liedahl \etal 1995).  For each galaxy, we
determined abundances for Mg, Si and Fe, since only these spectral
features are clearly seen in the data.  In deriving the abundance
data, we use the solar abundance table from Anders \& Grevesse (1989),
where elemental number abundances for O, Ne, Mg, Si, S, Ar, Ca, Fe, Ni
relative to hydrogen are (85.1, 12.3, 3.8, 3.55, 1.62, 0.36, 0.229,
4.68, 0.179)$\times10^{-5}$.  We assume solar abundances for He and
C. For our galaxy sample, we combined elements into five independent
groups: Ne; Mg; Si; S and Ar; and Ca, Fe and Ni. O results are not
presented, due to ASCA calibration uncertainties at low energies. We
chose the energy interval 0.7--3.5 keV for the ASCA spectral analysis
for all the galaxies, except M87. Since the galaxies have a
characteristic temperature of 1.5 keV or less, the high-energy cut-off
is used to reduce the effects of background subtraction.  Modeling M87
is complicated by the presence of a cool (1.3~keV) component within
a radius of $\sim25$ kpc, in addition to the hotter gas at
$\sim2.5$~keV (a typical value within the 200 kpc radius, analyzed
here).  For the range of temperatures of the hotter component, the Fe
L-shell lines are intense at the energies of Mg lines, while the Ne
line becomes more separated from the Fe lines (\eg\  Mushotzky \etal
1996). As a consequence, for M87 we determine the Ne abundance, but
not the Mg. The energy interval, chosen for the spectral analysis of
M87 is 0.7--7 keV.  The S (K-shell) lines are intense in M87, so we
also present results for S.

Due to the choice of low-energy cut-off in our analysis, the results 
are not sensitive to the presence of possible extra absorption and we freeze 
the $N_H$ value at the Galactic value (Stark \etal 1992).

All the ROSAT results, quoted in this paper were obtained using the
Raymond-Smith plasma code (Raymond \& Smith 1977) for spectral modeling.
Temperature determinations obtained from this and the MEKAL code, used in our
ASCA analysis, agree within 10--20 percent (see also Matsushita
\etal 1997).

In our spectral analysis, we took special care to understand and subtract
the Virgo cluster diffuse emission. When possible (except for NGC4472 and
NGC4636), we estimated the background from the same observation, choosing
background regions at a similar distance from the Virgo X-ray center (M87)
as the galaxy. While galaxies are characterized by {\kte} <1.5~keV, the
Virgo cluster emission is significantly hotter, $\sim3$~keV, which makes
this procedure robust.

A hard component at the centers of elliptical galaxies has been found in
many ASCA observations (\eg Matsumoto \etal 1997). Therefore in our spectral
fits, we include an additional spectral component in the center with a fixed
bremsstrahlung temperature of 6.5 keV and allow the normalization to
vary. We will address the analysis of hard components in a follow-up paper.

The apparent problems with spectral analysis of the deep (100 ksec) ASCA
observation of NGC4636, discussed by Matsushita \etal (1997) in terms of
inadequacies in the available plasma codes, and by Buote (1999) in terms of
systematic uncertainties in the calibration of the degrading SIS detectors,
require care, particularly at the low energies. To address these problems,
we introduce a 10 percent systematic error on the $1\sigma$ confidence level
in the spectral band 0.7-1.5 keV.  We also include a 5\% rms flux
uncertainty for wide energy bands that represents the uncertainty in ASCA's
effective area (e.g., Gendreau \& Yaqoob 1997; Markevitch \etal 1998). We
found that after introducing these systematic errors, we are able to reduce
the $\chi^2$ to acceptable values. The systematic effects due to the
uncertainty in ASCA XRT PSF data were estimated by performing the fits with
different spatial binnings.

\subsection{ Temperature and abundance data}

\begin{figure*}

\vspace*{-6cm}

\includegraphics[width=2.2in]{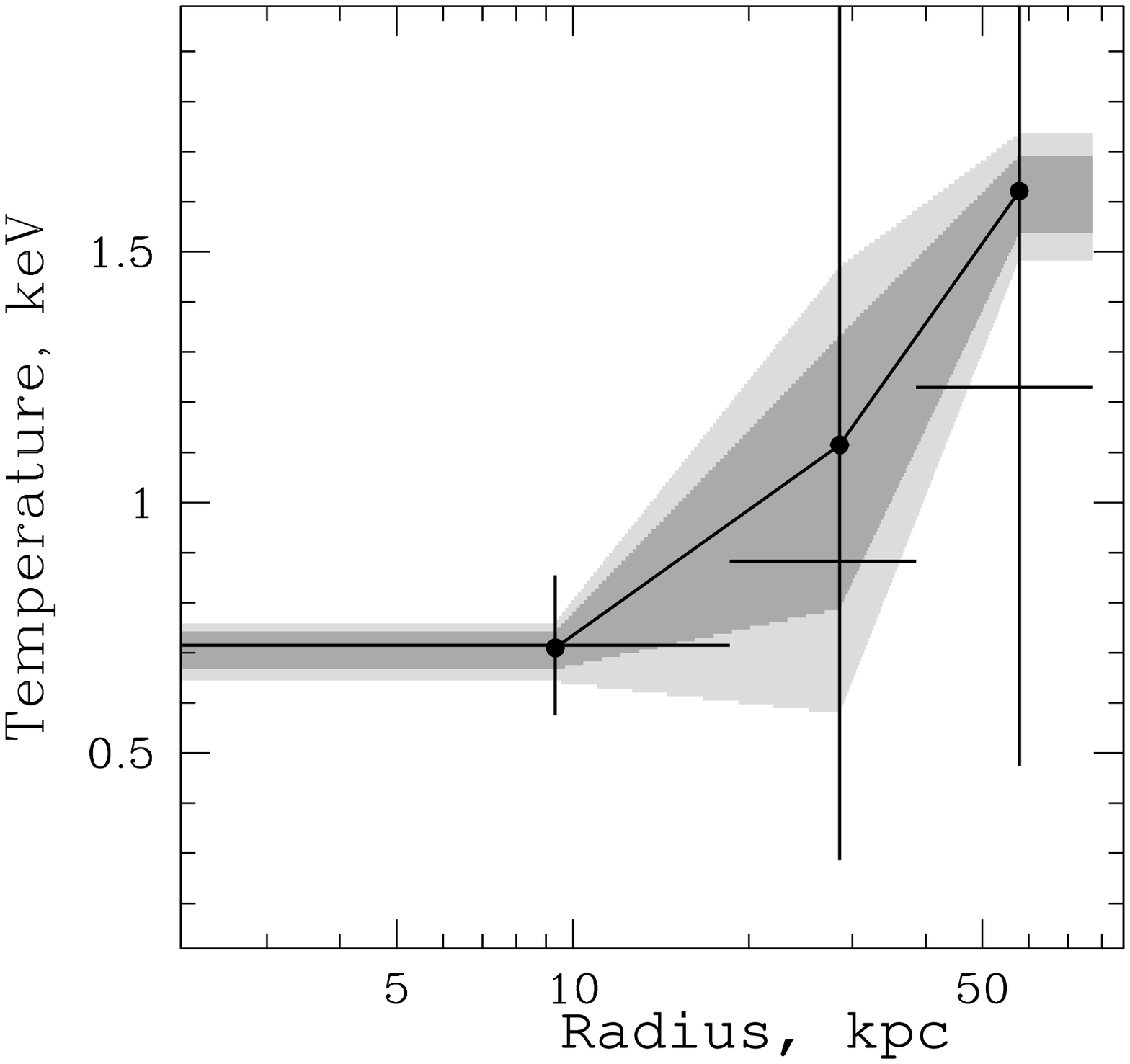} \hfill \includegraphics[width=2.2in]{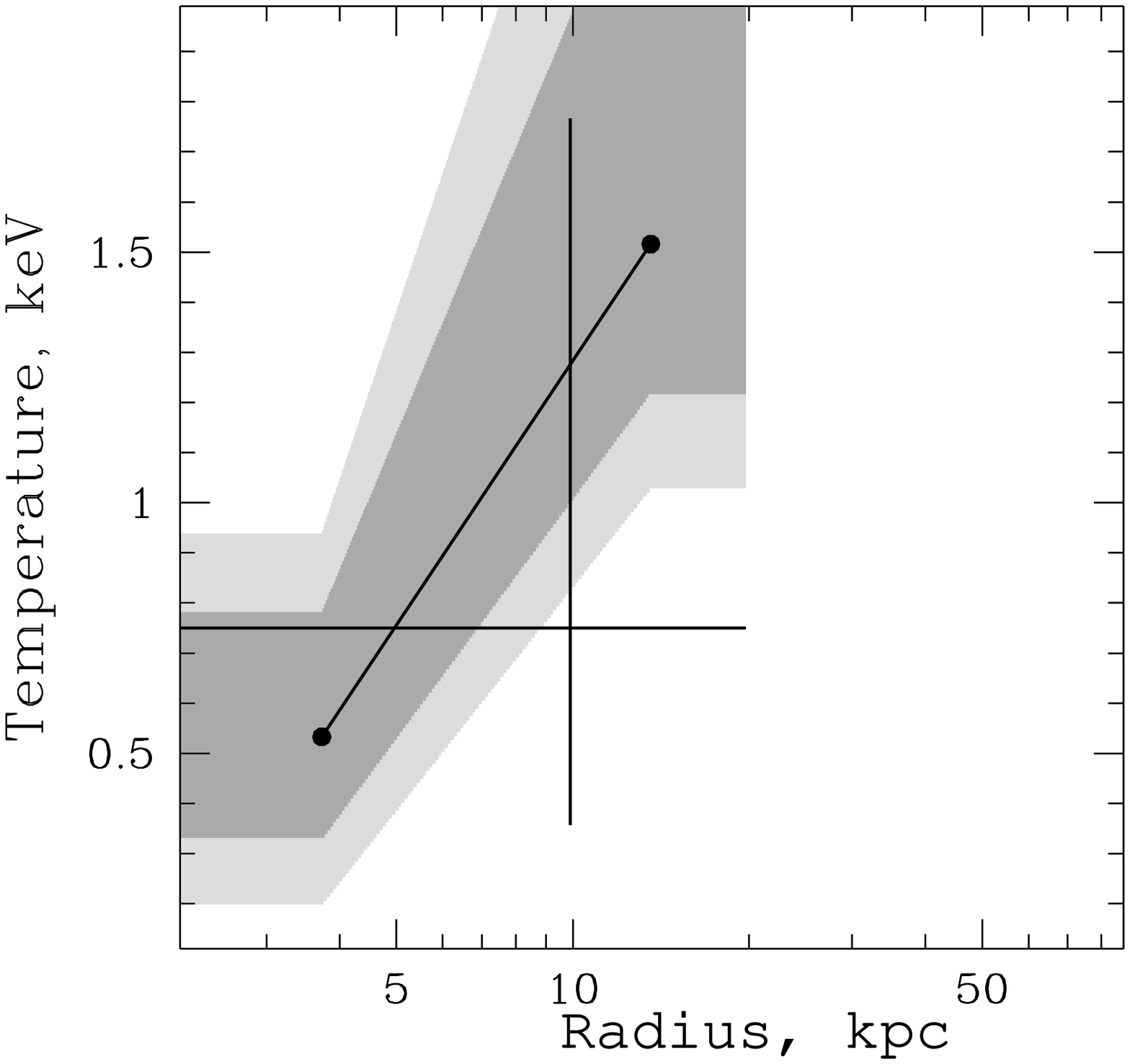} \hfill \includegraphics[width=2.2in]{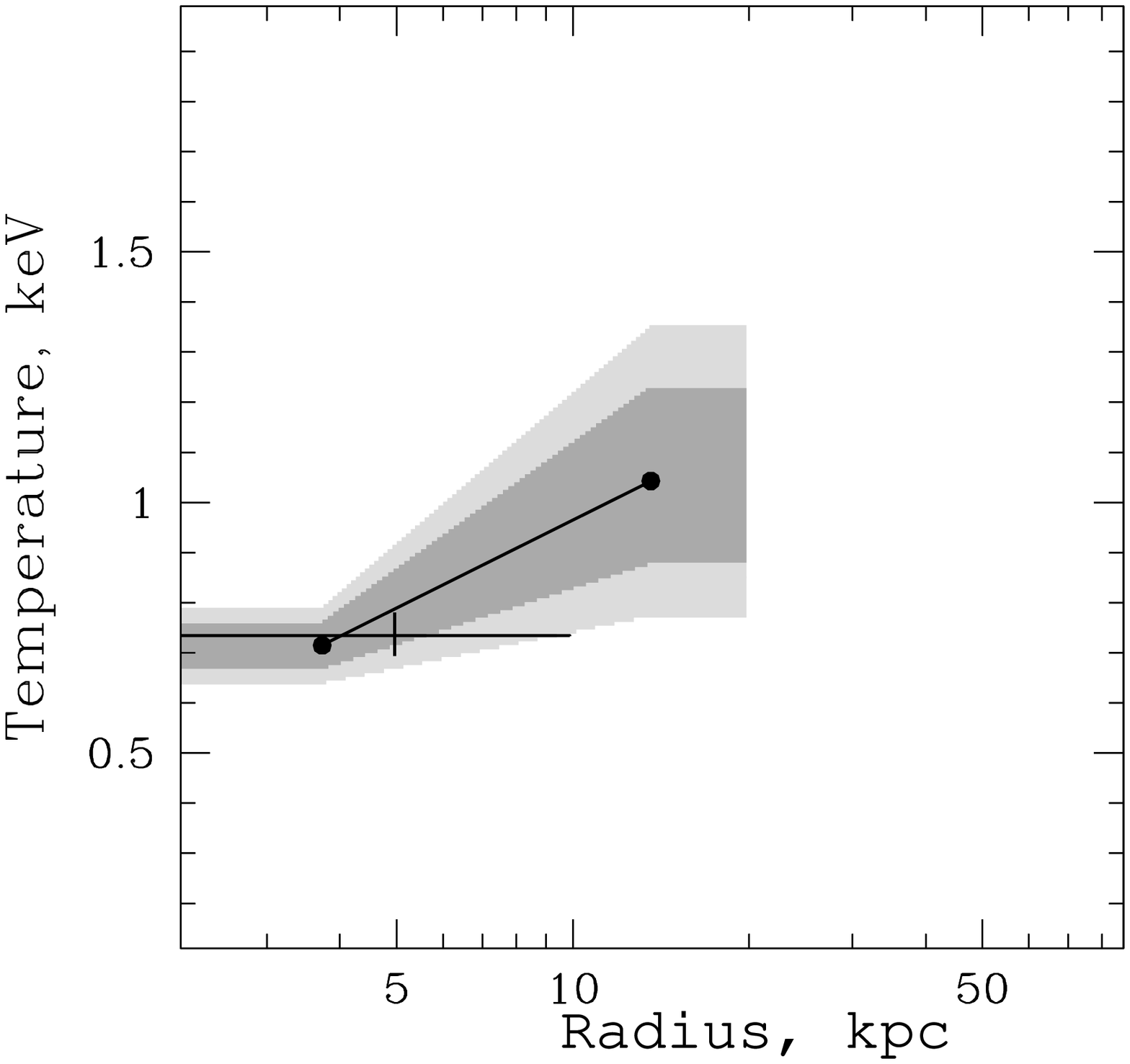}

\includegraphics[width=2.2in]{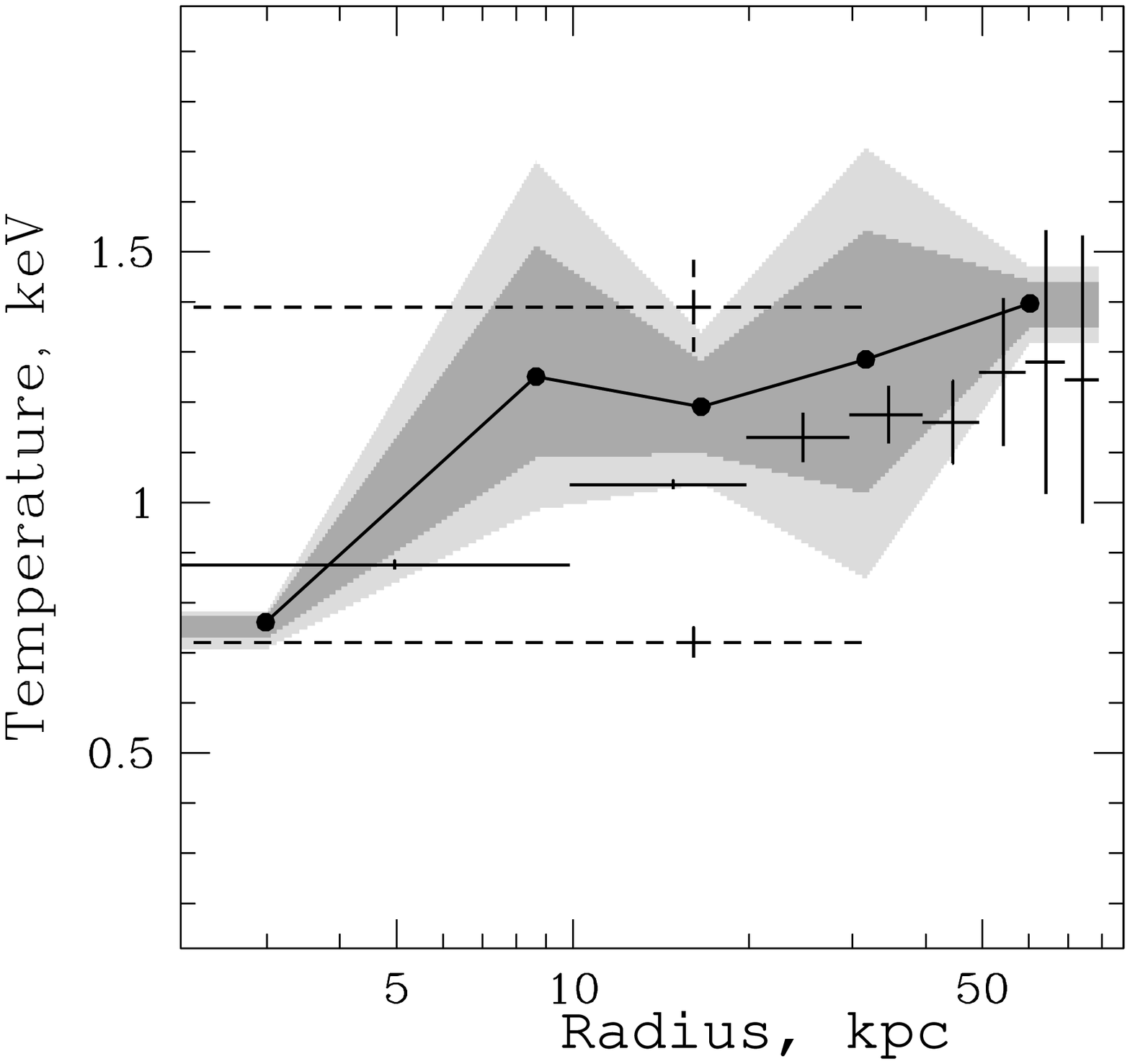} \hfill \includegraphics[width=2.2in]{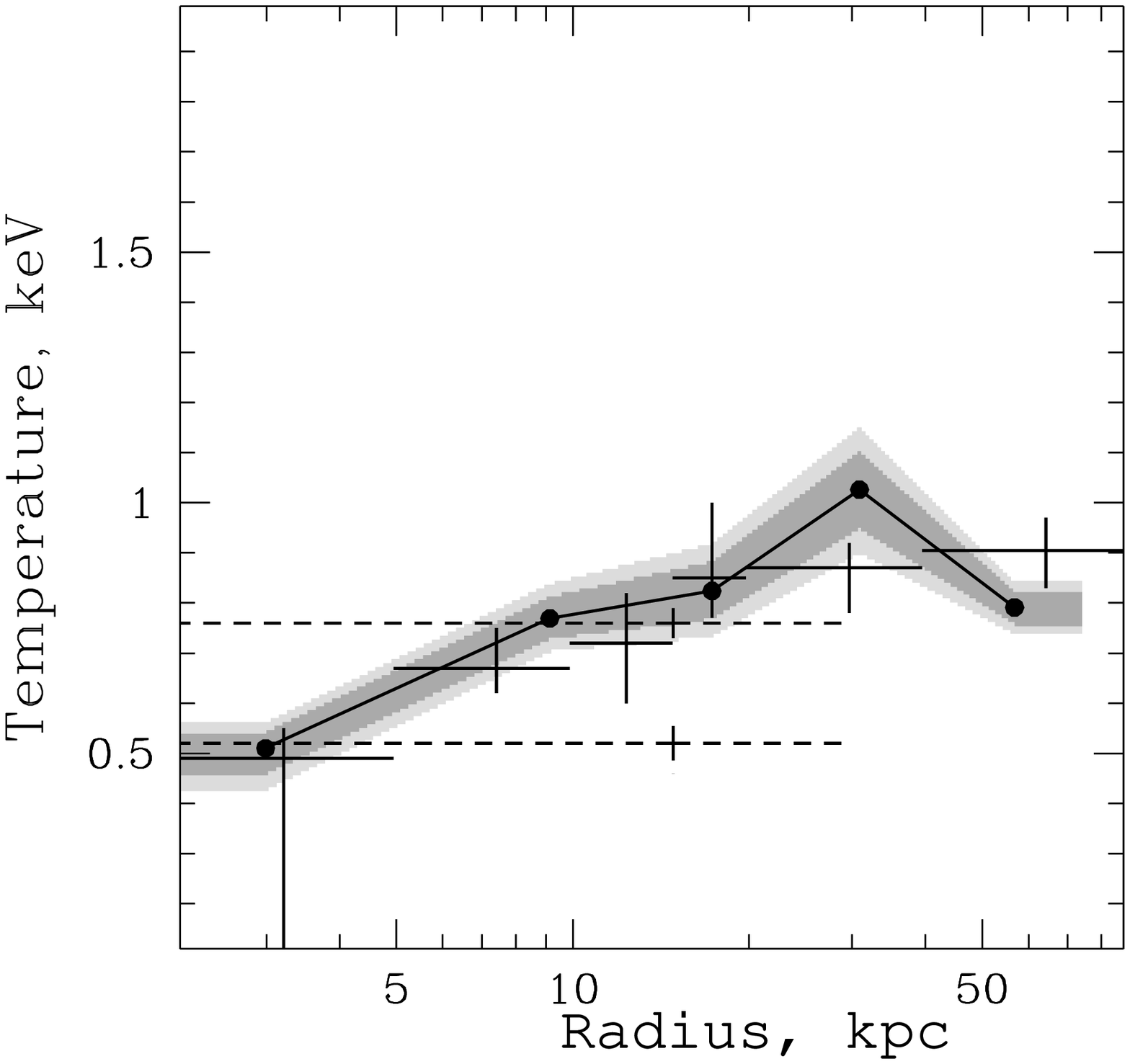} \hfill \includegraphics[width=2.2in]{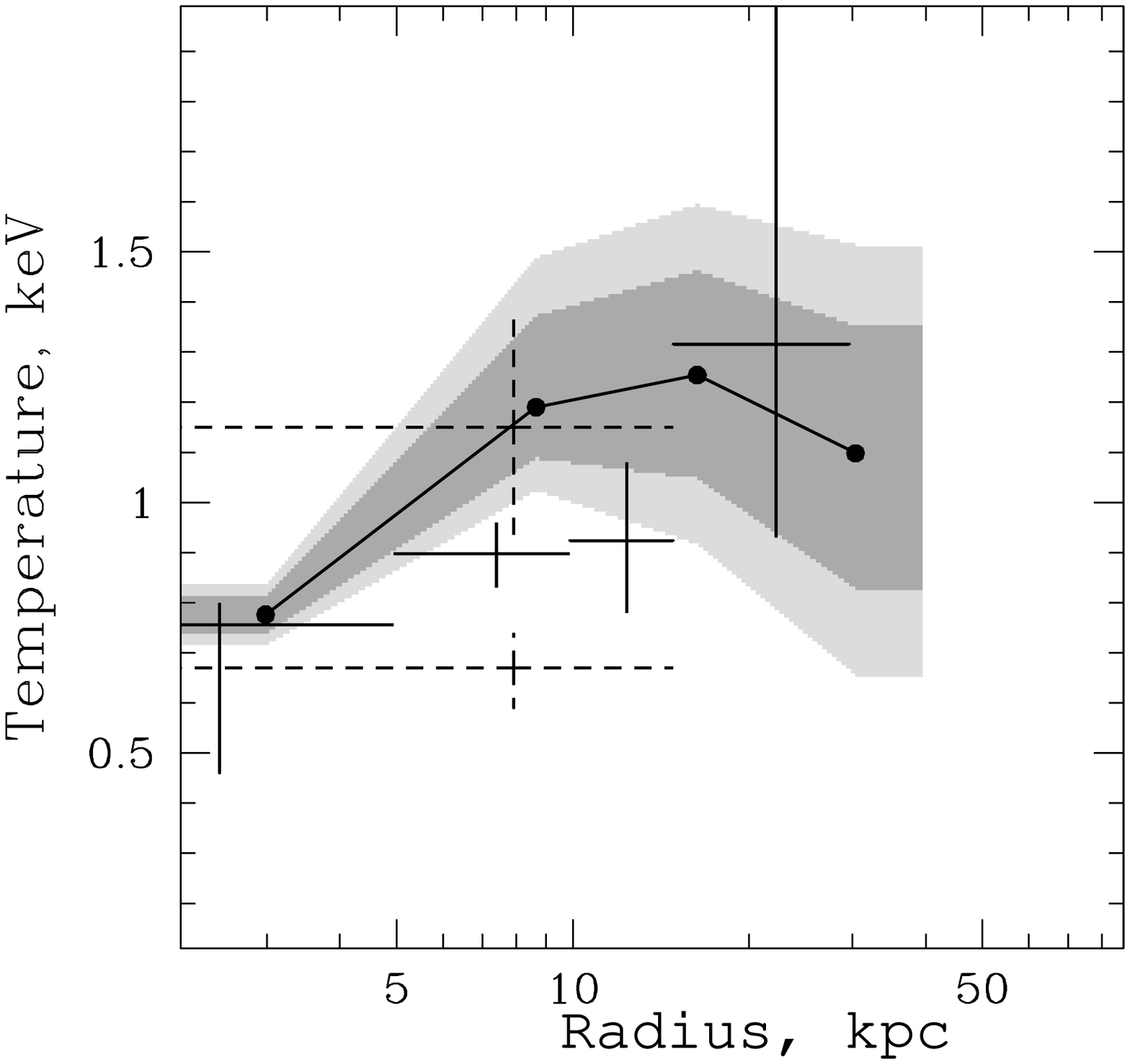}

\figcaption{Temperature profiles of Virgo early-type galaxies. Solid line
  represents the best-fit curve describing ASCA results with filled circles
  indicating the spatial binning used. Dark and light shaded zones around
  the best fit curve denote the 68 and 90 per cent confidence regions,
  respectively. Solid crosses indicate ROSAT data. Dashed crosses indicate
  ASCA ``multi-phase'' temperatures (Buote 1999, Buote \& Fabian 1998).
  Vertical error bars are shown at the 90 \% confidence level.
\label{virgo-te}}

\vspace*{-7.9cm}

{\it \hfill NGC4261\hspace*{1.1cm} \hfill NGC4365\hspace*{1.1cm} \hfill NGC4374\hspace*{0.4cm}}

\vspace*{5cm}

{\it \hfill NGC4472\hspace*{1.1cm} \hfill NGC4636\hspace*{1.1cm} \hfill NGC4649\hspace*{0.4cm}}

\vspace*{2.3cm}

%\end{figure*}

%\begin{figure*}

\includegraphics[width=2.2in]{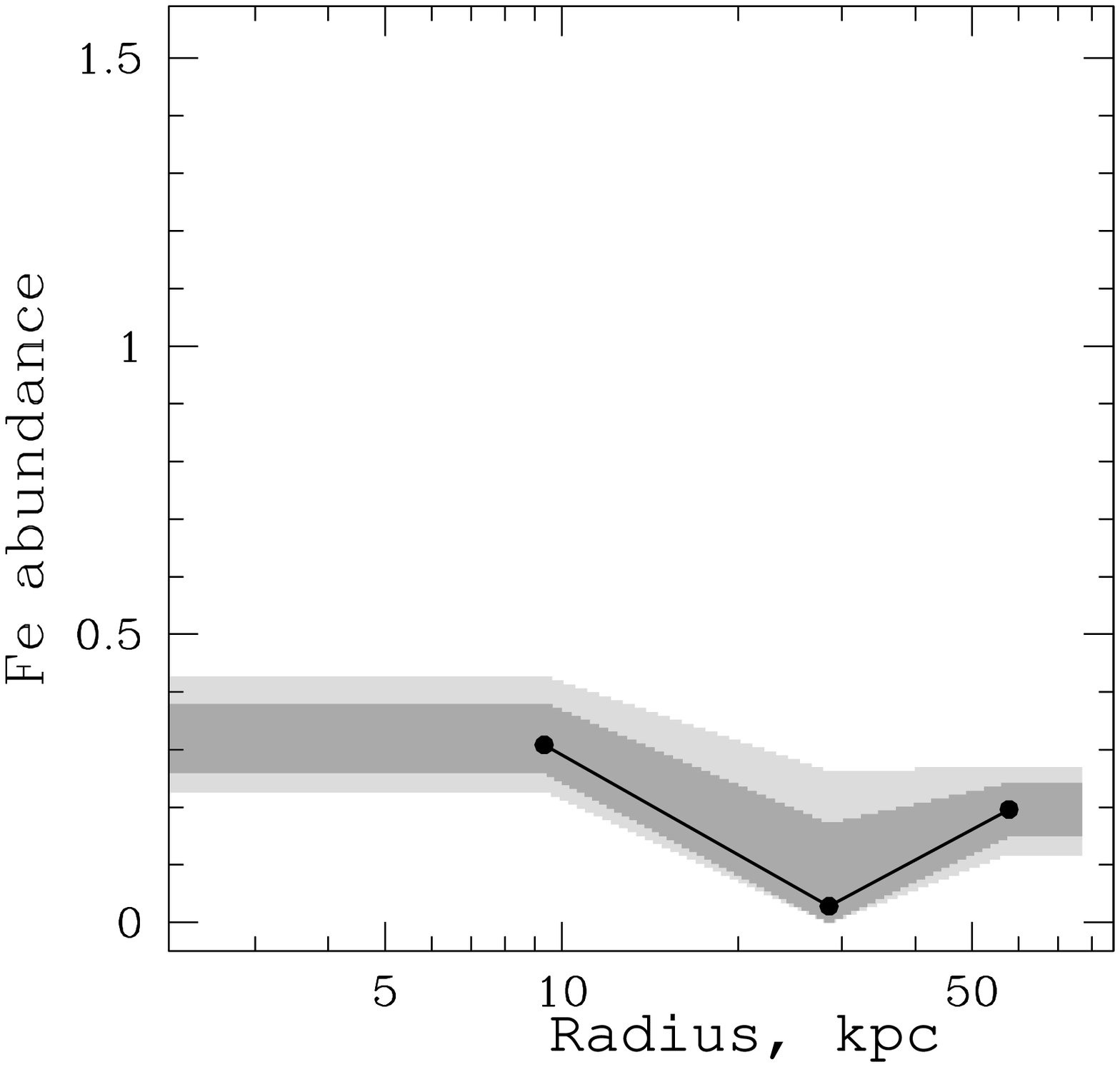} \hfill \includegraphics[width=2.2in]{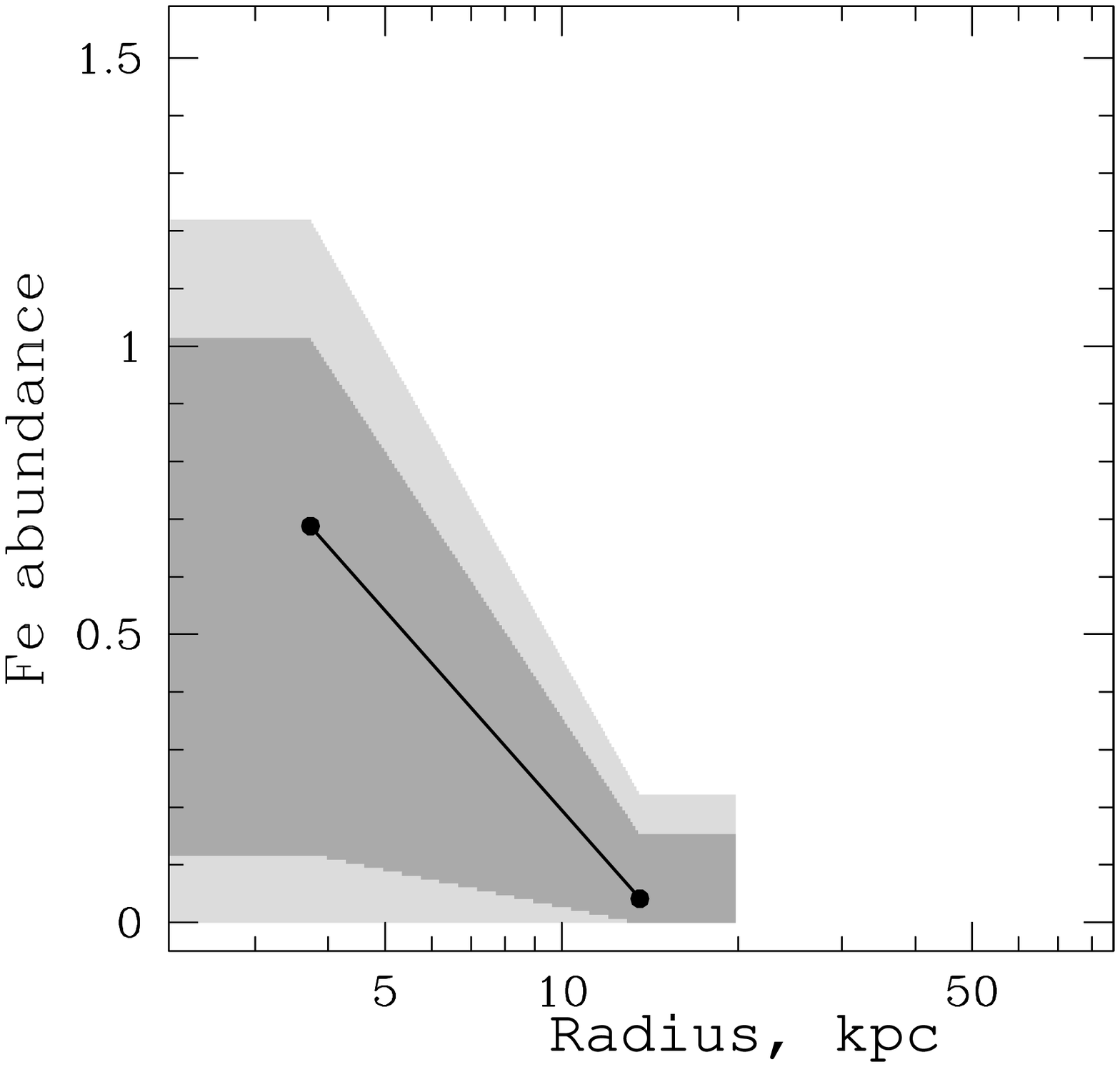} \hfill \includegraphics[width=2.2in]{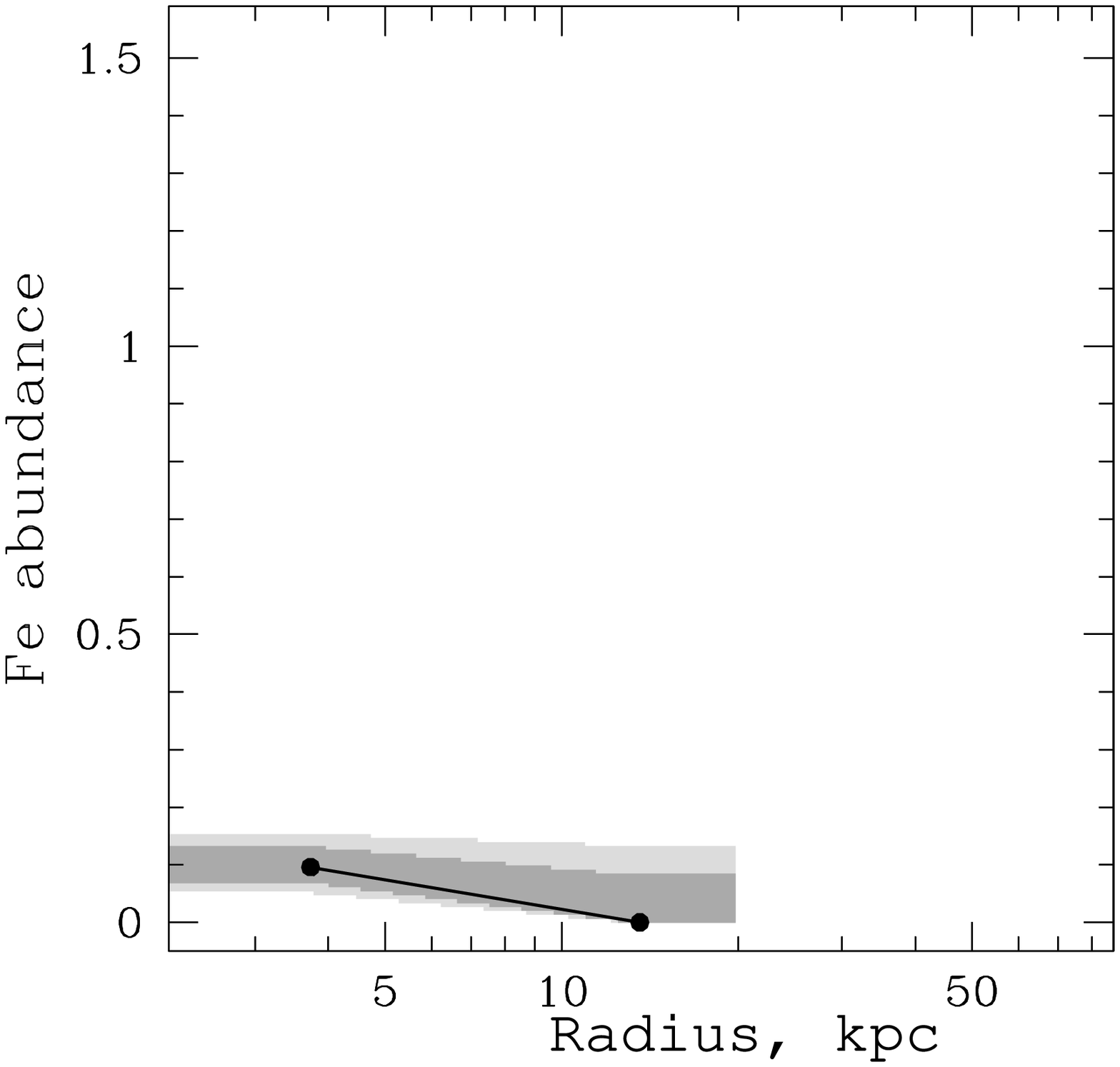}

\includegraphics[width=2.2in]{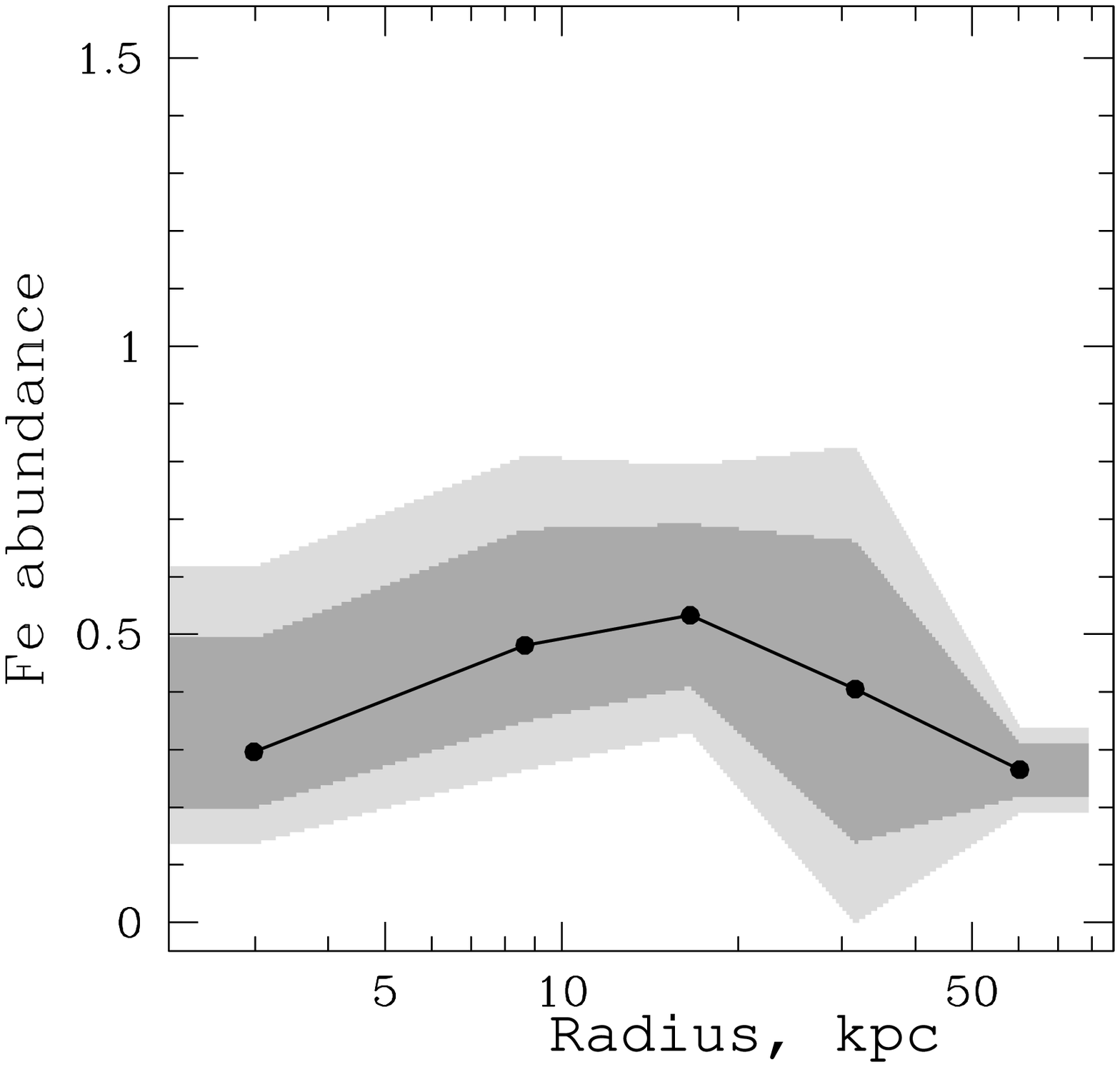} \hfill \includegraphics[width=2.2in]{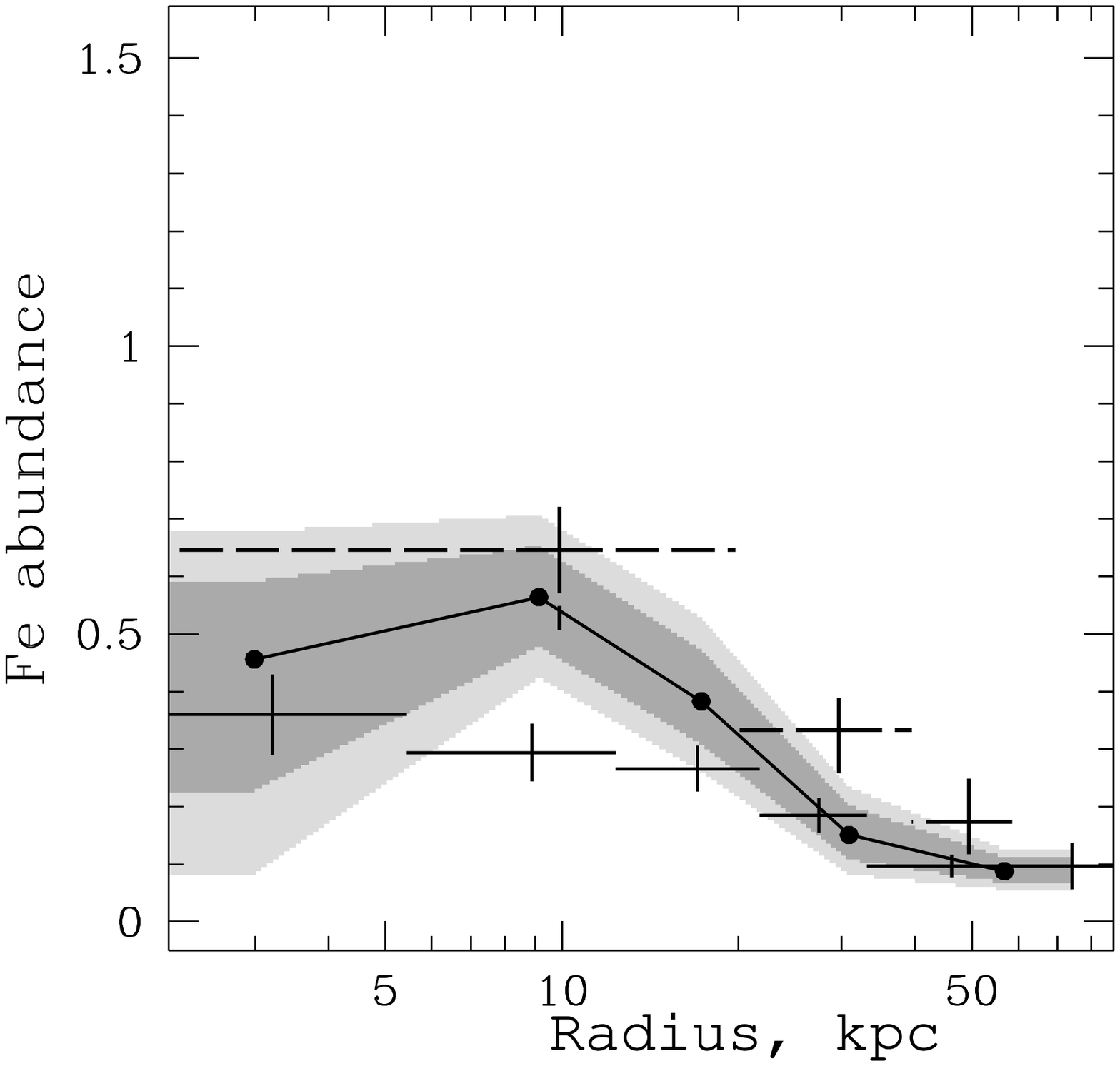} \hfill \includegraphics[width=2.2in]{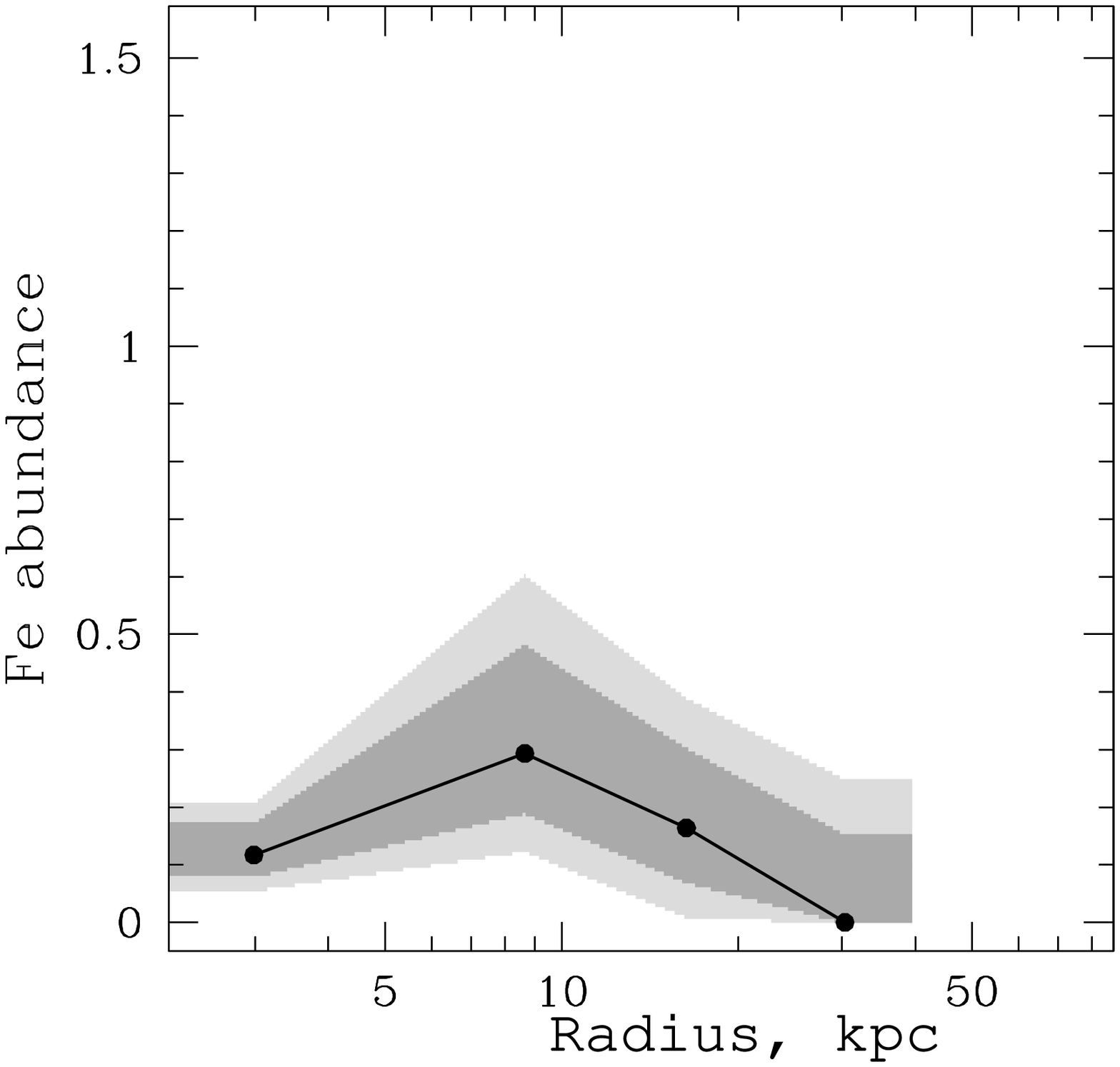}

\figcaption{Fe abundance profiles of Virgo early-type galaxies. Solid line
  represents the best-fit curve describing ASCA results with filled circles
indicating the spatial binning used. Dark and light shaded zones around the
best fit curve denote the 68 and 90 per cent confidence regions, respectively.
On the NGC4636 panel: solid crosses show ASCA SIS measurements by Mushotzky
\etal (1994); dashed crosses present ASCA SIS measurements by Matsushita
\etal (1997). The differences between these results is due to differences
in the exposure time of the observations and how each analysis accounts for
the systematics. Neither the ASCA PSF effects or gas projection was taken
into account in either of these analysis. Vertical error bars are shown at
the 90 \% confidence level.
\label{virgo-fe}}

\vspace*{-11.6cm}

{\it \hfill NGC4261\hspace*{1.1cm} \hfill NGC4365\hspace*{1.1cm} \hfill NGC4374\hspace*{0.4cm}}

\vspace*{5cm}

{\it \hfill NGC4472\hspace*{1.1cm} \hfill NGC4636\hspace*{1.1cm} \hfill NGC4649\hspace*{0.4cm}}

%\vspace*{2cm}
\clearpage

\end{figure*}
\begin{figure*}

\vspace*{-4cm}

\includegraphics[width=2.2in]{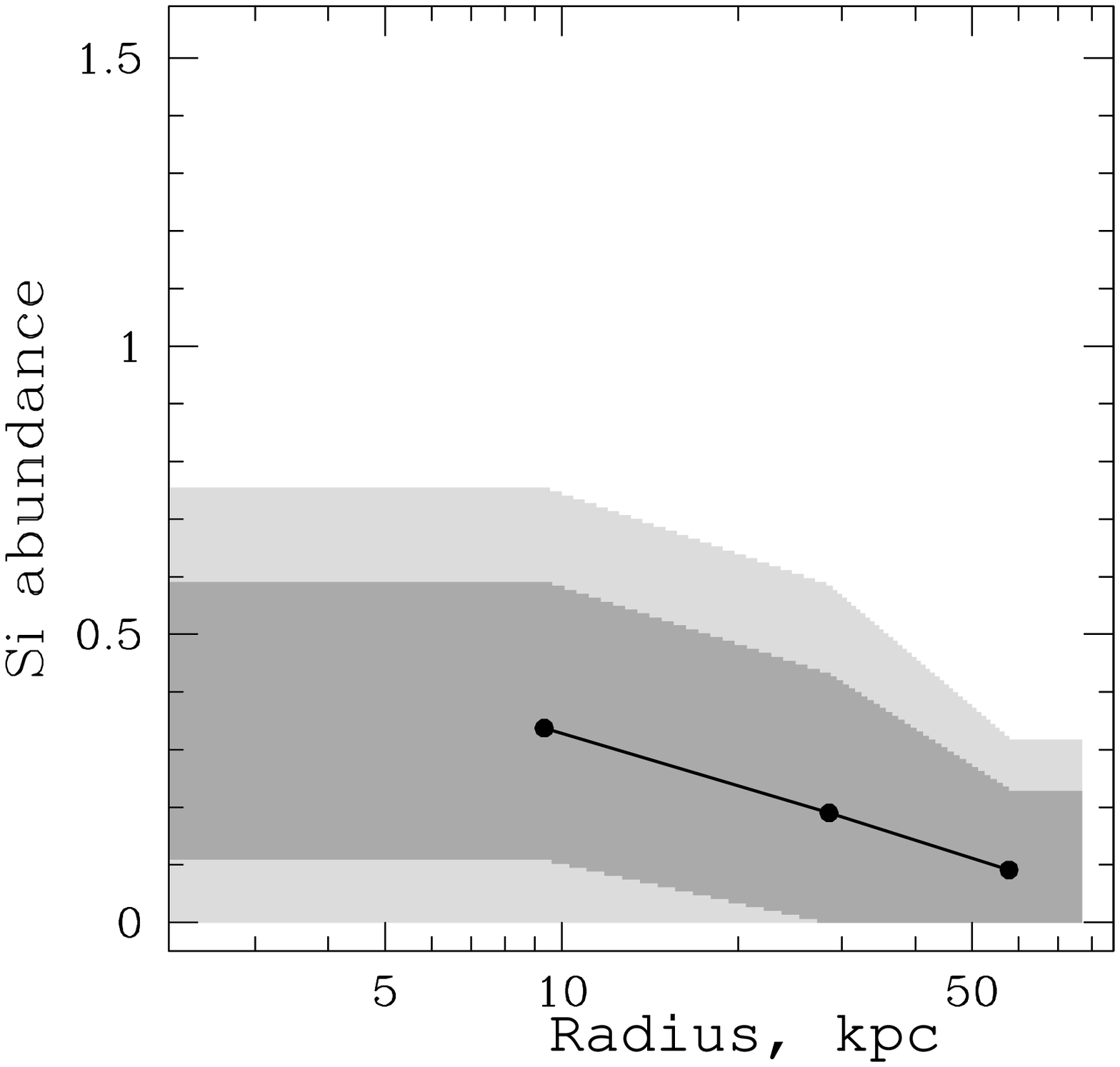} \hfill \includegraphics[width=2.2in]{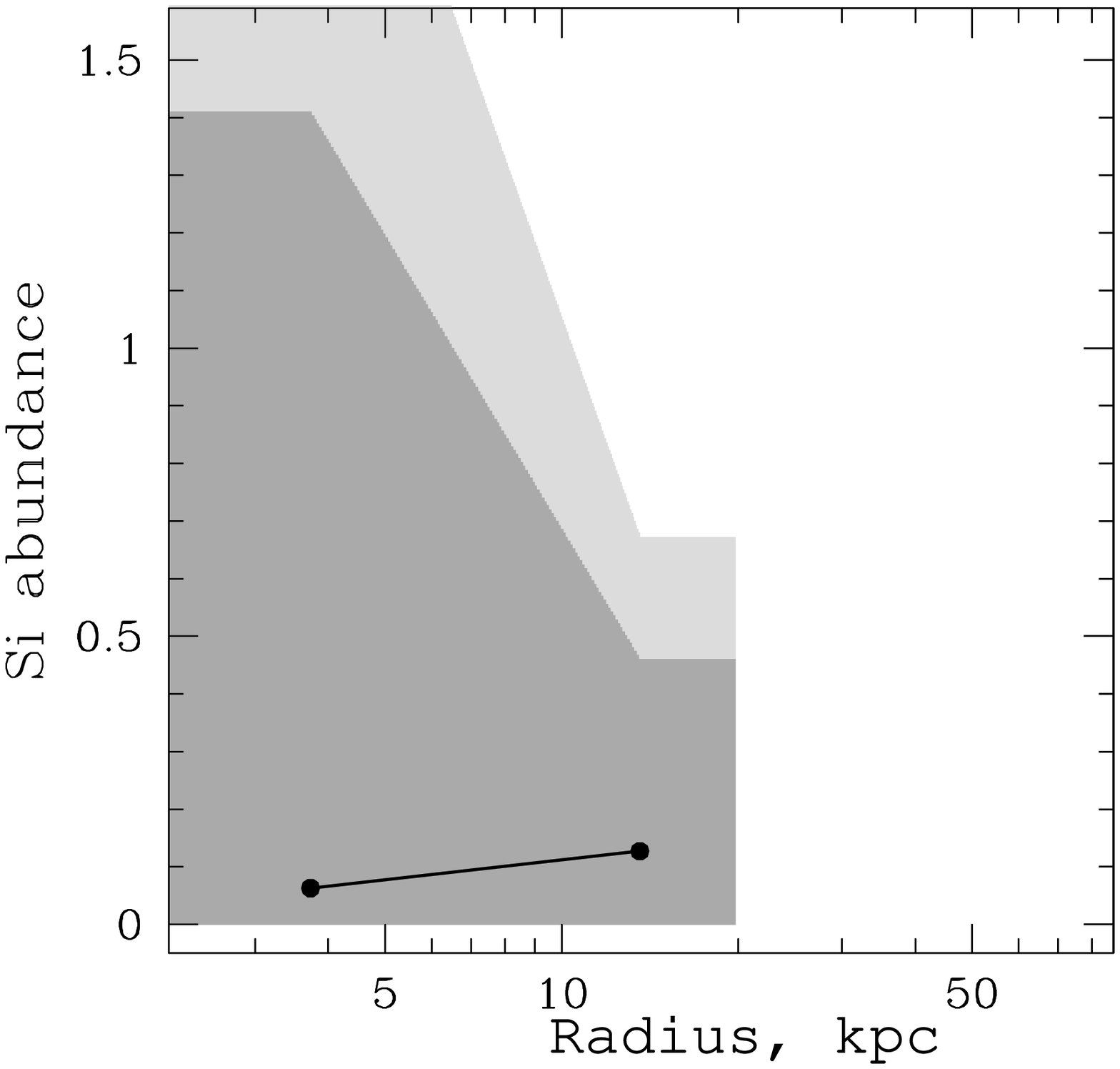} \hfill \includegraphics[width=2.2in]{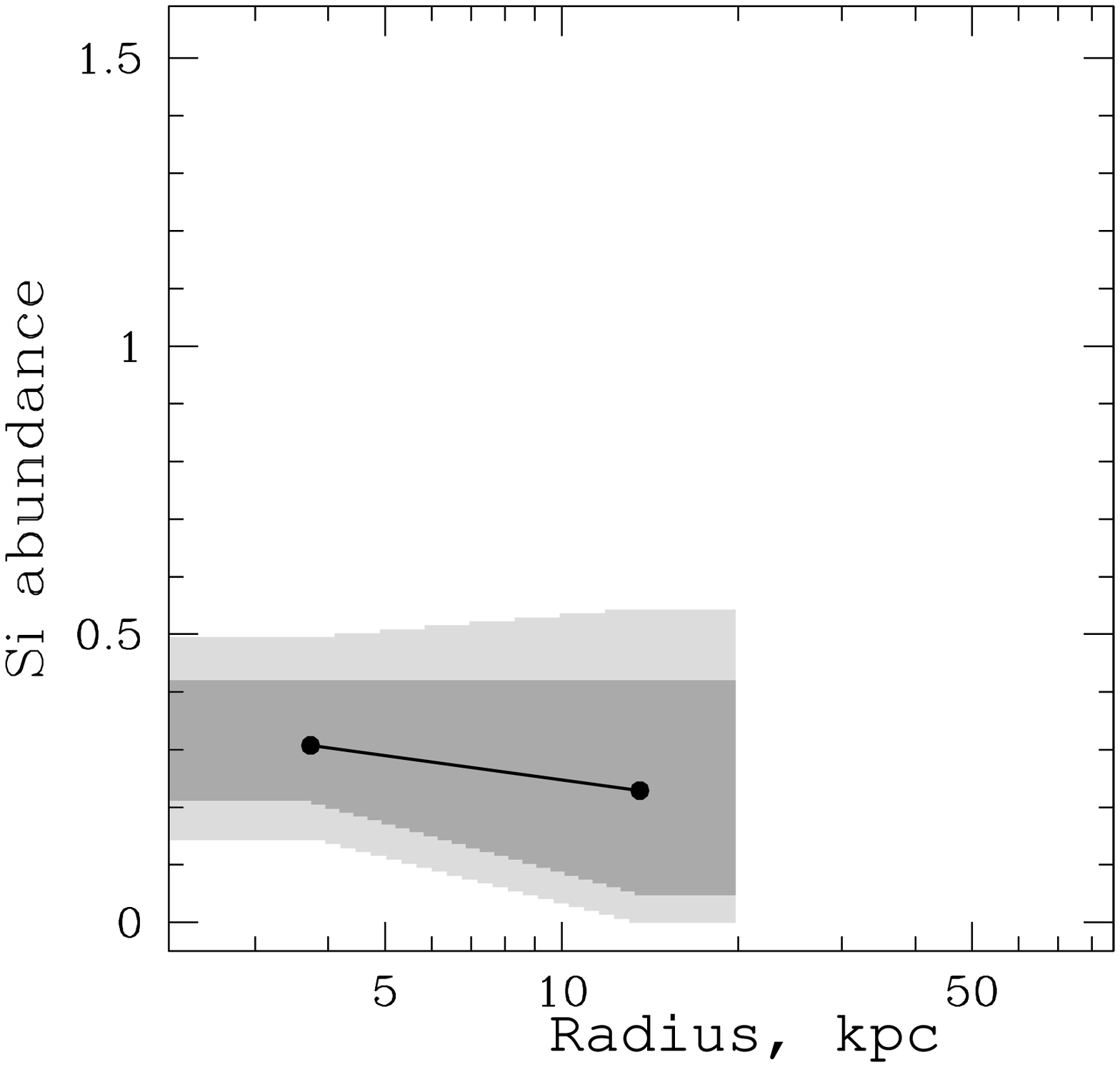}

\includegraphics[width=2.2in]{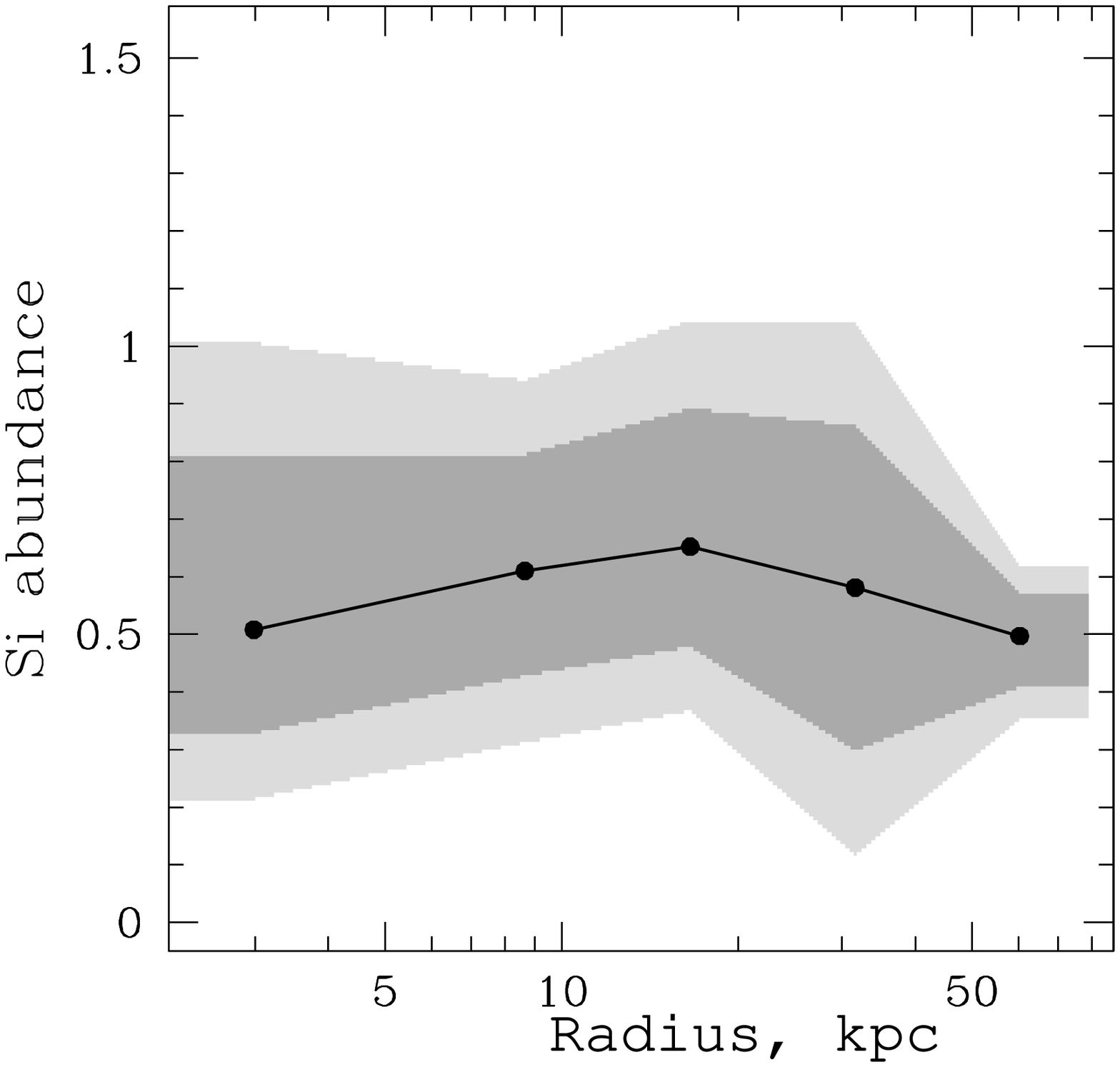} \hfill \includegraphics[width=2.2in]{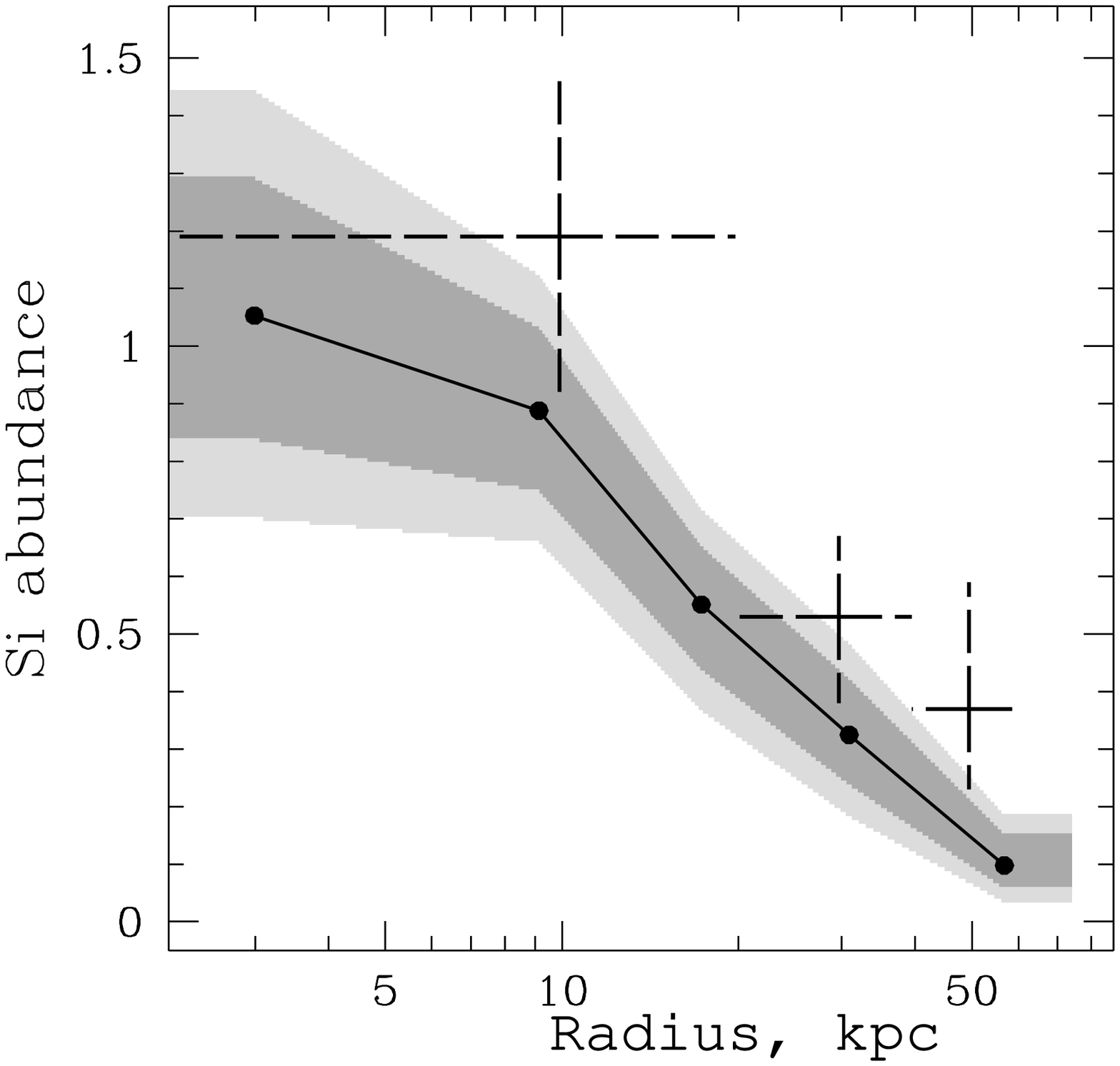} \hfill \includegraphics[width=2.2in]{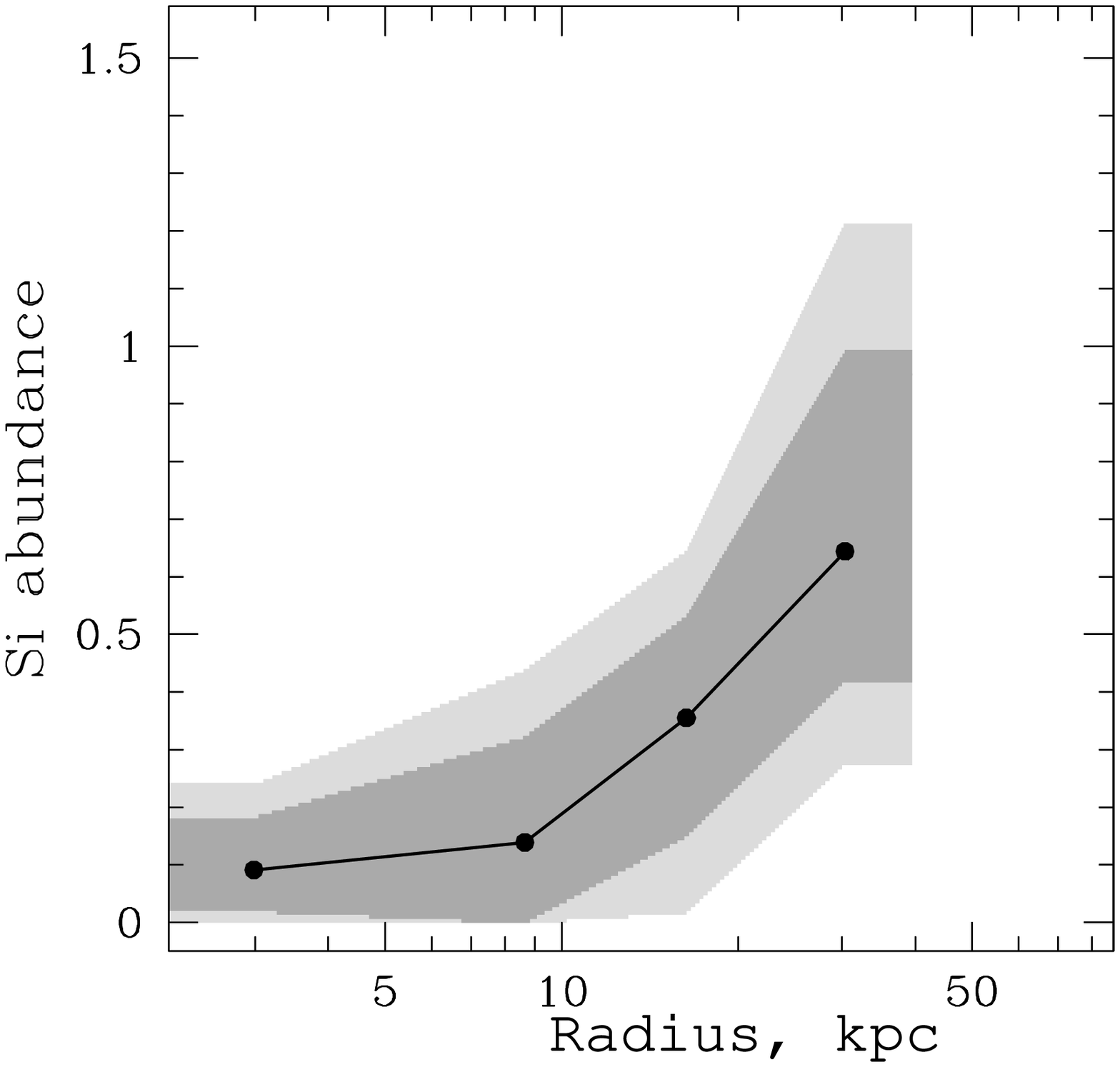}

\figcaption{Si abundance profiles for Virgo early-type galaxies. Solid line
represents the best-fit curve describing ASCA results with filled circles
indicating the spatial binning used. Dark and light shaded zones around the
best fit curve denote the 68 and 90 per cent confidence area, respectively.
On the NGC4636 panel: dashed crosses present ASCA SIS measurements of
alpha-process elements by Matsushita \etal (1997). Effects of the ASCA PSF
and gas projection were not included in their analysis. Vertical error bars
are shown at the 90 \% confidence level.
\label{virgo-si}}

\vspace*{-11.3cm}

{\it \hfill NGC4261\hspace*{1.1cm} \hfill NGC4365\hspace*{1.1cm} \hfill NGC4374\hspace*{0.4cm}}

\vspace*{5cm}

{\it \hfill NGC4472\hspace*{1.1cm} \hfill NGC4636\hspace*{1.1cm} \hfill NGC4649\hspace*{0.4cm}}

\vspace*{6cm}

%\end{figure*}

%\begin{figure*}

\includegraphics[width=2.2in]{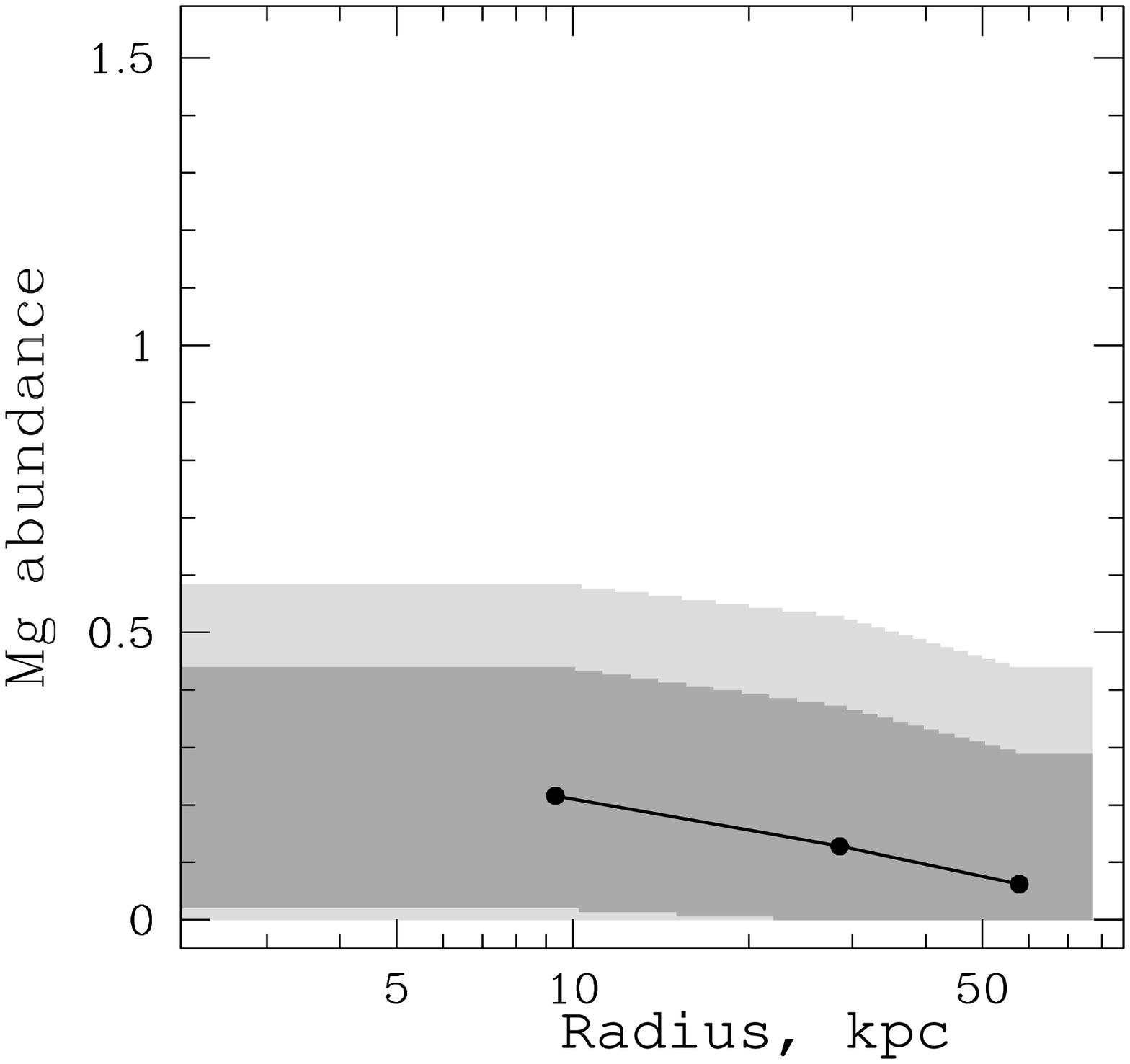} \hfill \includegraphics[width=2.2in]{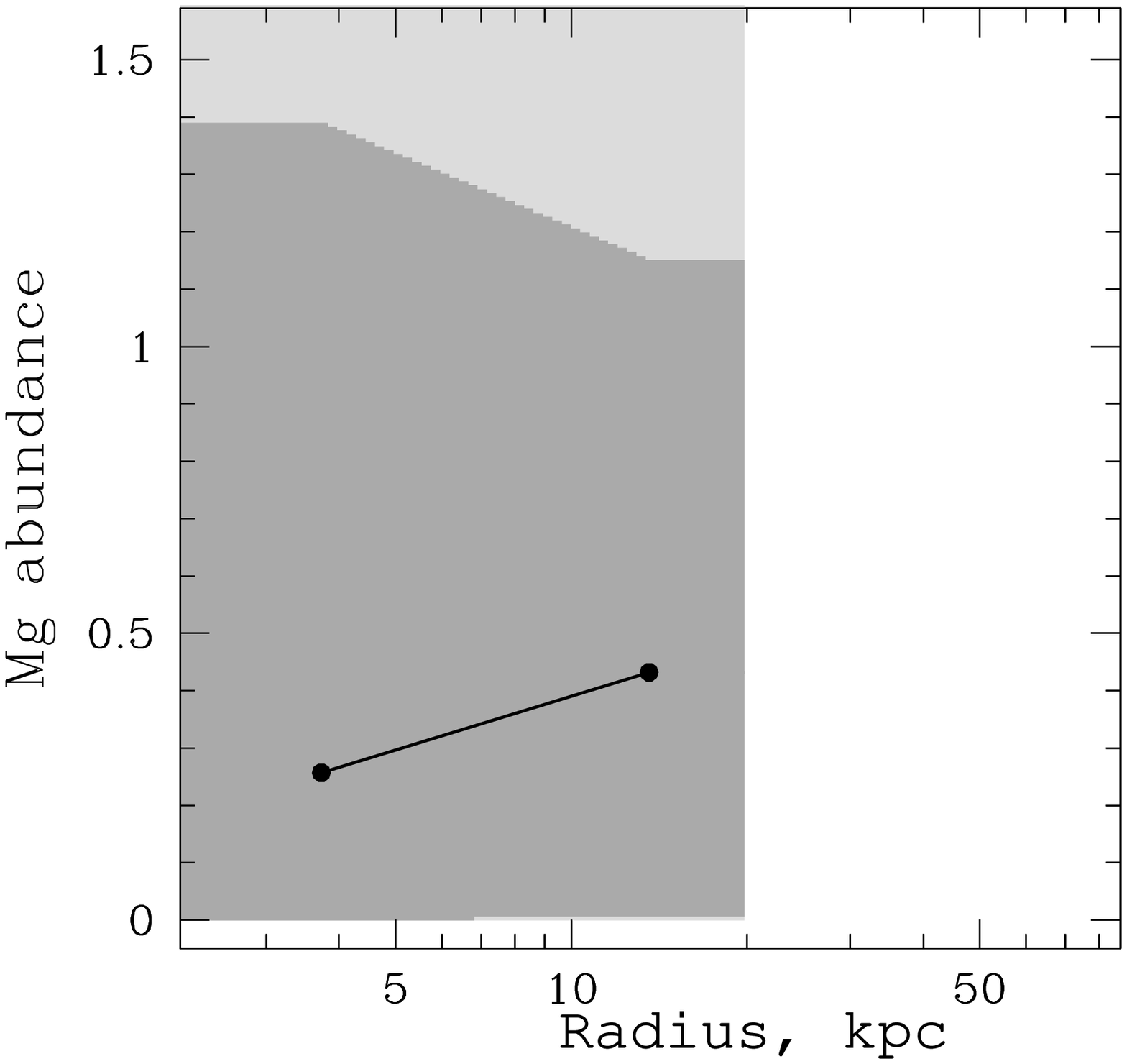} \hfill \includegraphics[width=2.2in]{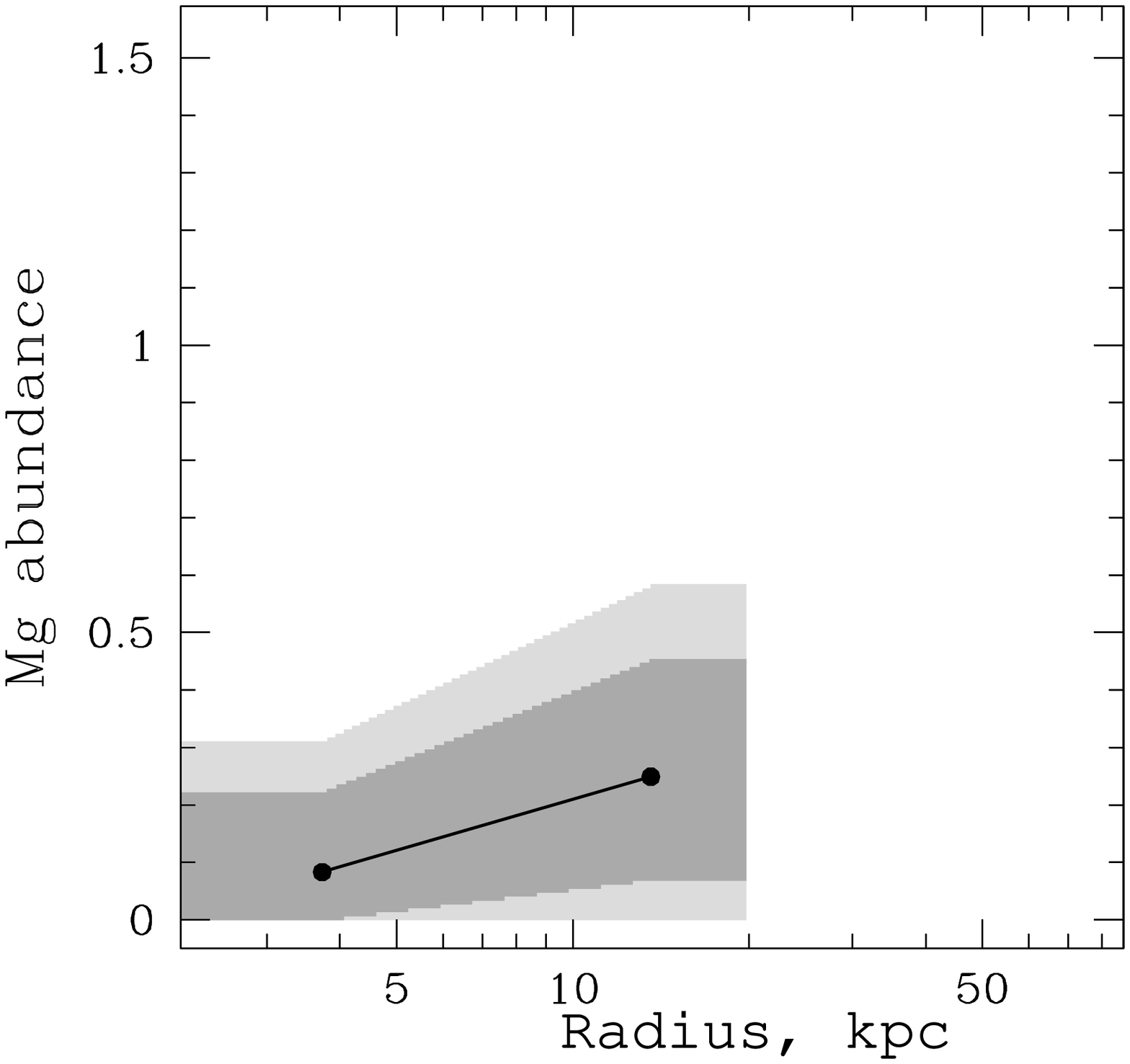}

\includegraphics[width=2.2in]{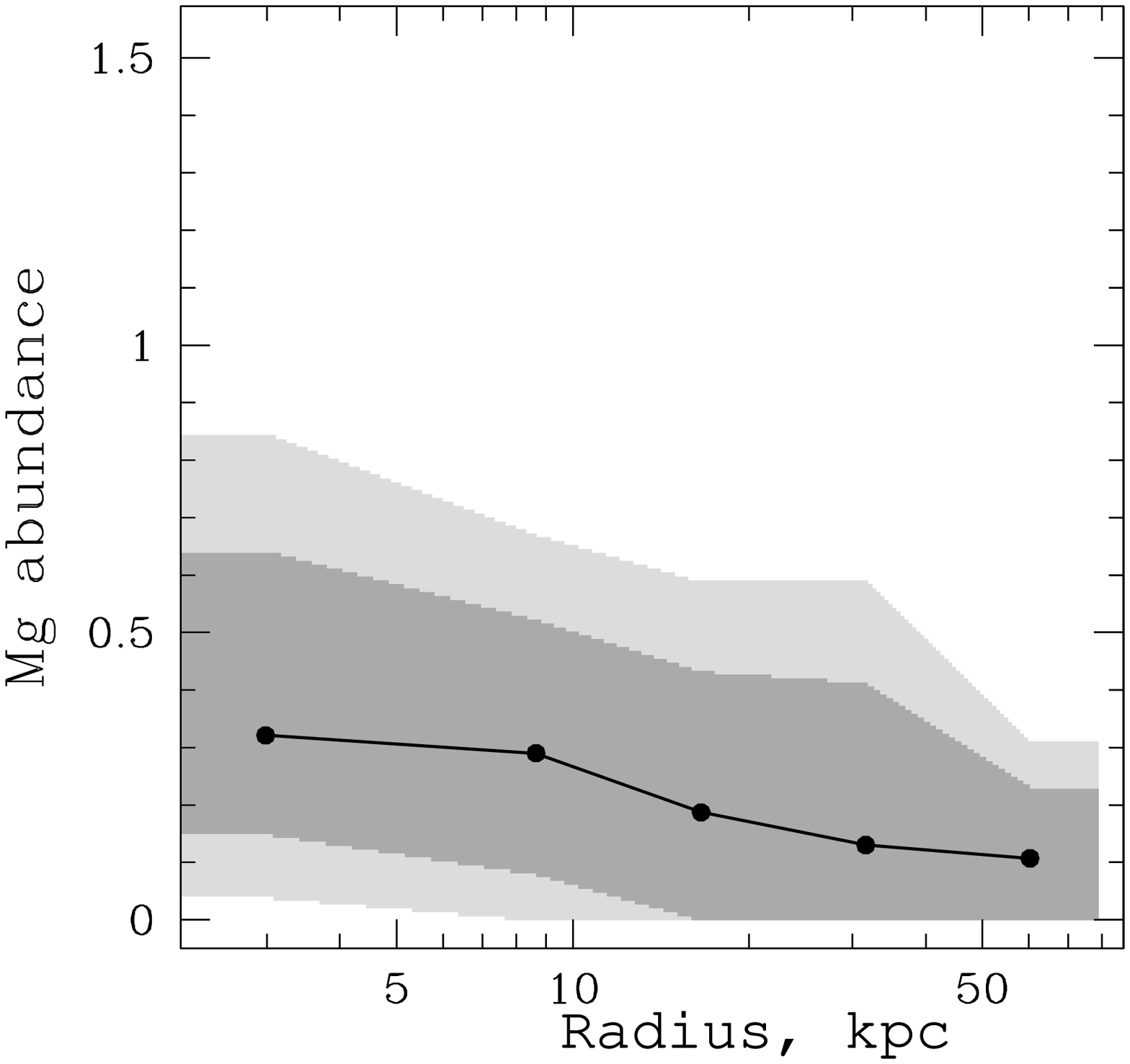} \hfill \includegraphics[width=2.2in]{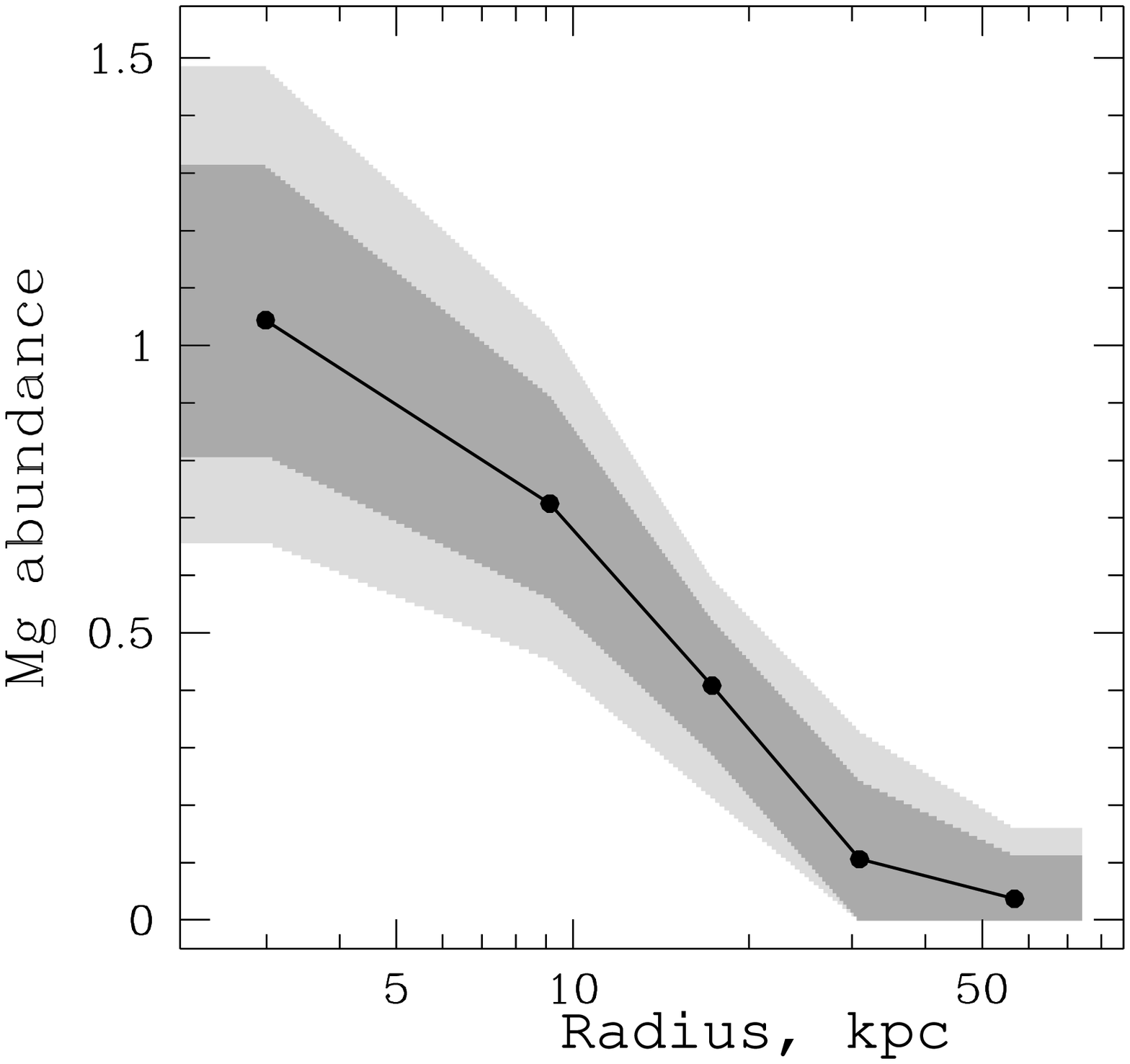} \hfill \includegraphics[width=2.2in]{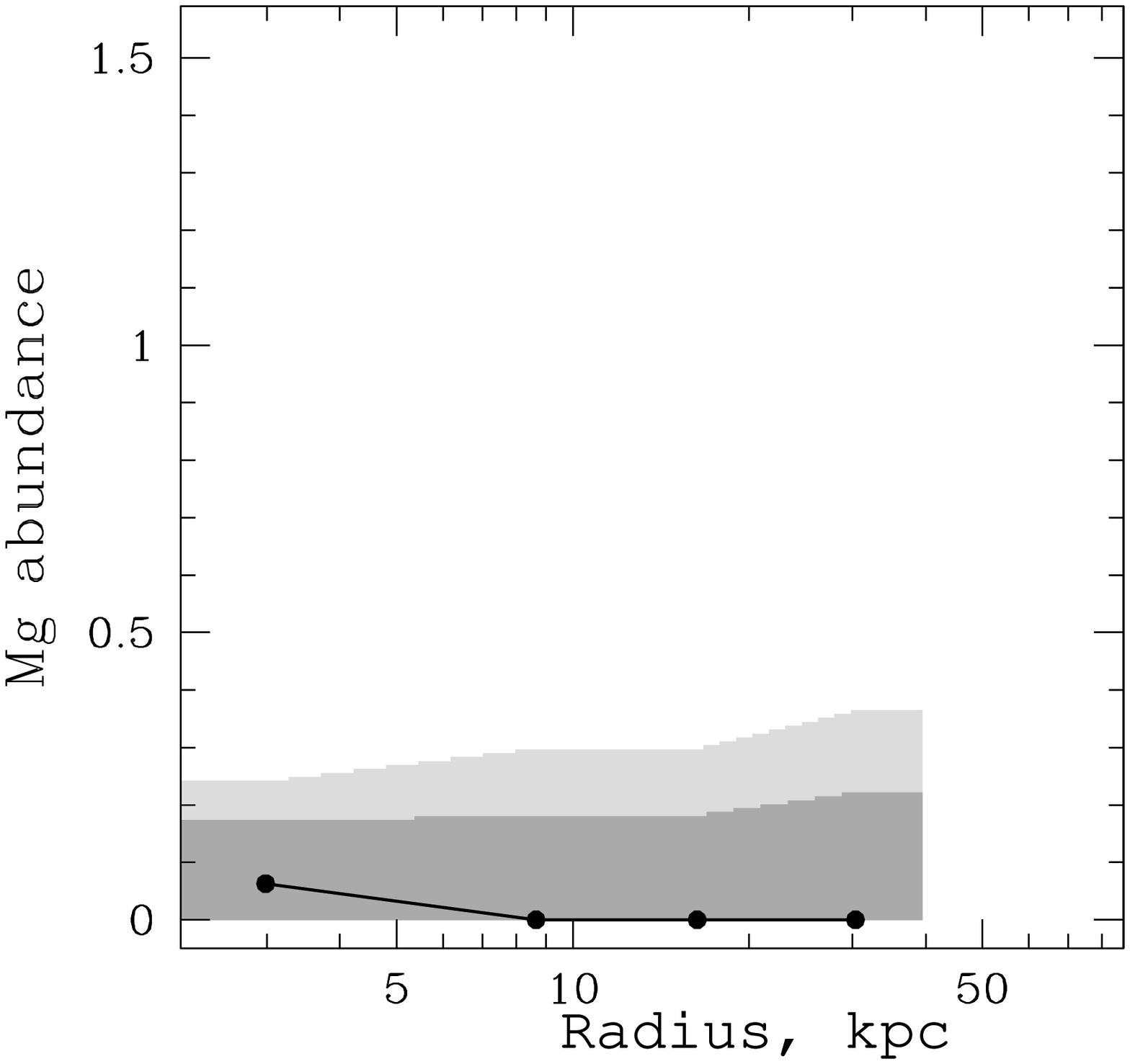}

\figcaption{Mg abundance profiles for Virgo early-type galaxies. Solid lines
  represent the best-fit curves describing ASCA results, with filled circles
  indicating the spatial binning used. Dark and light shaded zones around
  the best fit curve denote the 68 and 90 per cent confidence regions,
  respectively.
\label{virgo-mg}}

\vspace*{-10.9cm}

{\it \hfill NGC4261\hspace*{1.1cm} \hfill NGC4365\hspace*{1.1cm} \hfill NGC4374\hspace*{0.4cm}}

\vspace*{5cm}

{\it \hfill NGC4472\hspace*{1.1cm} \hfill NGC4636\hspace*{1.1cm} \hfill NGC4649\hspace*{0.4cm}}

%\vspace*{2cm}

\end{figure*}

In Table \ref{tab-xray} we summarize the basic X-ray and optical quantities
for the galaxy sample. Col.(1) identifies the galaxy; (2) the adopted
distance for \h0; (3) blue luminosity; (4) stellar mass to blue light ratio
(values from Lauer 1995, scaled to the adopted distance). Columns (5--7)
summarize our results from fitting the ROSAT surface brightness profiles for
galaxies with the beta-model.  Whenever a single component does not describe
the profile, we introduce a second, which is found to have the same slope,
but a different core radius ($r_{c2}$).  When we do not resolve the core, we
assume a value of 0.01\amin\ for $r_{c1}$.  Values for M86, given in columns
(5--7) are derived from emission to the south-west of the galaxy. Col.(8)
gives our estimation of the total mass within central 50 kpc.

%\begin{table}
{
\begin{center}
\footnotesize
\tabcaption{\centerline{\footnotesize
X-ray and optical quantities of the sample}
\label{tab-xray}}

\begin{tabular}{lclccccc}
\hline
\hline
NCG & D & $L_B$ &  $M/L_B$ & $\beta$ & $r_{c1}$ & $r_{c2}$ & $M_{tot}$\\
\#  & Mpc & $10^{10}$\lsun &  \msun/\lsun &   & (\amin) & (\amin) & $10^{12}$\msun\\
\hline
4261 & 53 & 12.9 &   8  &   0.40 & 0.01 & --- & 3.2 \\
4365 & 17 &  2.8 &  16  &   0.61 & 0.47 & --- & 3.9 \\
4374 & 17 &  3.5 &  11  &   0.38 & 0.01 & --- & 2.0 \\
4406 & 17 &  4.3 &  12  &   0.36 & 0.44 & --- & 2.1 \\
4472 & 17 &  8.5 &  13  &   0.51 & 0.03 & 0.8 & 3.8 \\
4486 & 17 &  6.6 &  14  &   0.49 & 0.38 & 3.7 & 5.7 \\
4552 & 17 &  2.2 &  11  &   0.62 & 0.15 & --- & 1.8 \\
4636 & 17 &  4.8 &  16  &   0.46 & 0.26 & --- & 2.0 \\
4649 & 17 &  5.4 &  18  &   0.55 & 0.21 & --- & 3.2 \\
\hline
\end{tabular}
\end{center}
}
%\end{table}

In spatially binning the data for our spectral analysis, we used a central bin of 1\amin\
radius, that corresponds to the cusp of the ASCA PSF. The outer radius is
determined by the extent of detectable emission. Logarithmic steps in radius were chosen
to provide similar statistical quality in the parameters derived for
different radii.

In Fig.\ref{virgo-te} we present the results on spatially resolved
temperature estimations in our sample. A general behavior of the radial
temperature profile, found in this and other studies, is the increase of the
temperature from a central cool region, with sometimes a small decline in
the temperature at large radii.  This general picture describes NGC4472,
NGC4636 and NGC4649.

ROSAT PSPC temperature determinations, presented in Fig.\ref{virgo-te} are
taken from NGC4365: Jones \etal (1997); NGC4472: Forman \etal (1993);
NGC4636: de-projected results inside 8\amin\  (also allowing for a hard
component in the center) plus a determination at 8--18\amin\  radii for 20\%
cosmic abundance from Trinchieri \etal (1994); NGC4649: results, allowing
for a hard component in the center, from Trinchieri \etal (1997). We provide
ROSAT measurements for the center of NGC4261 (three inner points), and
for NGC4374 and M86. All ROSAT errors correspond to 90\% confidence limits.

For the ROSAT spectral analysis, we extract spectra with IRAF {\it qpspec},
which was upgraded for particle background estimation, as in Plucinsky \etal
(1993).  Since this approach precludes use of the {\ssd} band in further
spectral analysis, we analyze data only from the {\ad} energy band.

While ASCA and ROSAT temperature measurements are in remarkable agreement
for all the galaxies, a comment should be made for NGC4261. It is one of the
low-power radio galaxies from the sample of Worrall and Birkinshaw (1994),
where a hard component is assumed, due to the AGN. In our ASCA analysis we
included this component, yet our temperature seems to disagree with the
ROSAT measurement at larger radii by Davis \etal (1995). To reconcile both
measurements, the gas temperature for NGC4261 after an initial rise, should
decline, as was measured for NGC5846 (F99).

%\begin{table}[H]
{
\footnotesize
\centering
\tabcaption{\footnotesize
\centerline{ASCA SIS temperature measurements$^{\dag}$}
\label{table-te}}

\begin{tabular}{lclc}
\hline
\hline
Annulus (\amin) & $kT_e$ & Annulus (\amin) & $kT_e$ \\
\hline
 & & & \\
\multicolumn{2}{c}{NGC4365 \hspace*{0.8cm}}   &   \multicolumn{2}{c}{NGC4374 \hspace*{0.8cm}}    \\                
0.0---1.5 &  0.533 (0.33:0.78) &  0.0---1.5 & 0.715 (0.67:0.76)  \\ 
1.5---4.0 &  1.516 (1.22:2.35) &  1.5---4.0 & 1.043 (0.88:1.23)  \\ 
 & & & \\                                            
\multicolumn{2}{c}{NGC4472 \hspace*{0.8cm}}          &  \multicolumn{2}{c}{NGC4636 \hspace*{0.8cm}}    \\                                          
0.0---1.2 &  0.761 (0.73:0.77) &  0.0---1.2 & 0.510 (0.46:0.54) \\
1.2---2.3 &  1.251 (1.09:1.51) &  1.2---2.5 & 0.769 (0.73:0.81) \\
2.3---4.4 &  1.191 (1.10:1.28) &  2.5---4.5 & 0.824 (0.77:0.88) \\
4.4---8.4 &  1.285 (1.02:1.54) &  4.5---8.0 & 1.026 (0.95:1.10) \\
8.4---16. &  1.397 (1.35:1.44) &  8.0---15. & 0.791 (0.76:0.82)  \\
 & & & \\                                            
\multicolumn{2}{c}{NGC4261 \hspace*{0.8cm}}         & \multicolumn{2}{c}{NGC4649 \hspace*{0.8cm}}    \\                  
0.0---1.2 &  0.710 (0.67:0.74)  & 0.0---1.2 & 0.776 (0.74:0.81)  \\
1.2---2.5 &  1.115 (0.79:1.33)  & 1.2---2.3 & 1.190 (1.09:1.37)  \\
2.5---5.0 &  1.621 (1.54:1.69)  & 2.3---4.3 & 1.254 (1.05:1.46)  \\
 &                              & 4.3---8.0 & 1.098 (0.83:1.35)  \\
\multicolumn{2}{c}{NGC4552 \hspace*{0.8cm}}  & & \\                                     
0.0---4.0 &   0.778 (0.73:0.82)     \\                                            
\hline                                               
\end{tabular}                                        

\begin{enumerate}
\item[{$^{\dag}$}]{\footnotesize ~ Errors are given at 68\% confidence level
    for one parameter of interest. MEKAL plasma code is used for spectral fitting.}
\end{enumerate}
}                                                    
%\end{table}                                         

Uncertainty in our measurements of the central temperature and
abundance for NGC4365 does not allow us to distinguish between the
different ROSAT results obtained by Jones \etal (1997) and Fabbiano
\etal (1994). The former finds a central temperature of 0.75
(0.51--1.37) keV, when the abundance is a free parameter, while in the
latter, the temperature is determined to be 0.2 (0.15--0.26) keV, when
the abundance is fixed to Solar and a hard component is added.

We were able only to provide a single temperature determination for NGC4552,
which is presented, along with measurements for other galaxies in Table
\ref{table-te}.

The introduction of a ``multi-phase'' medium has been used to model
the integrated ASCA spectra (Buote \& Fabian 1998; Buote 1999). As is seen for
NGC4636, NGC4472 and NGC4649 in Fig.\ref{virgo-te}, NGC4406 (in
Fig.\ref{m86-temap}), NGC5846 (F99) and NGC5044 (FP99), this need arises
from temperature gradients, previously resolved only by ROSAT and now also
by ASCA.

%\begin{table}[H]
{
\footnotesize
\centering
\tabcaption{\footnotesize
\centerline{ASCA SIS heavy element abundance measurements$^{\dag}$}
\label{table-ab}}

\begin{tabular}{llll}
\hline
\hline
Annulus (\amin) & $Mg/Mg_{\odot}$ & $Si/Si_{\odot}$ & $Fe/Fe_{\odot}$ \\
\hline
 & & & \\
\multicolumn{2}{c}{NGC4261} & & \\
0.0---1.2 &  0.216 (0.02:0.44) &  0.337 (0.11:0.59) &  0.308 (0.26:0.38)  \\ 
1.2---2.5 &  0.128 (0.00:0.37) &  0.190 (0.00:0.43) &  0.028 (0.00:0.17)  \\ 
2.5---5.0 &  0.062 (0.00:0.29) &  0.091 (0.00:0.23) &  0.196 (0.15:0.24)  \\ 
 & & & \\
\multicolumn{2}{c}{NGC4365} & & \\
0.0---1.5 &  0.257 (0.00:1.39) &  0.063 (0.00:1.41) &  0.688 (0.12:1.01)   \\  
1.5---4.0 &  0.432 (0.01:1.15) &  0.127 (0.00:0.46) &  0.041 (0.00:0.15)   \\  
 & & & \\
\multicolumn{2}{c}{NGC4374} & & \\                                                 
0.0---1.5 &  0.083 (0.00:0.22) &  0.307 (0.21:0.42) &  0.096 (0.07:0.13)  \\ 
1.5---4.0 &  0.249 (0.07:0.45) &  0.229 (0.05:0.42) &  0.000 (0.00:0.08)  \\ 
 & & & \\
\multicolumn{2}{c}{NGC4472} & & \\                                                 
0.0---1.2 &  0.321 (0.15:0.64) &  0.508 (0.33:0.81) &  0.296 (0.20:0.49)  \\  
1.2---2.3 &  0.290 (0.08:0.52) &  0.610 (0.43:0.81) &  0.481 (0.35:0.68)  \\  
2.3---4.4 &  0.187 (0.00:0.43) &  0.652 (0.48:0.89) &  0.533 (0.41:0.69)  \\  
4.4---8.4 &  0.130 (0.00:0.41) &  0.581 (0.30:0.86) &  0.405 (0.14:0.66)  \\  
8.4---16. &  0.107 (0.00:0.23) &  0.497 (0.41:0.57) &  0.265 (0.22:0.31)  \\
 & & & \\
\multicolumn{2}{c}{NGC4552} & & \\                                                 
0.0---4.0 &  0.120 (0.03:0.22) &  0.085 (0.01:0.17) &  0.049 (0.03:0.07)  \\
 & & & \\
\multicolumn{2}{c}{NGC4636} & & \\                                                 
0.0---1.2 &  1.044 (0.81:1.31) &  1.053 (0.84:1.29) &  0.456 (0.23:0.59)  \\ 
1.2---2.5 &  0.725 (0.56:0.91) &  0.888 (0.75:1.03) &  0.564 (0.48:0.65)  \\ 
2.5---4.5 &  0.408 (0.29:0.52) &  0.551 (0.44:0.65) &  0.383 (0.31:0.47)  \\ 
4.5---8.0 &  0.106 (0.00:0.24) &  0.325 (0.24:0.42) &  0.151 (0.11:0.20)  \\ 
8.0---15. &  0.037 (0.00:0.11) &  0.098 (0.06:0.15) &  0.088 (0.07:0.11)  \\ 
 & & & \\
\multicolumn{2}{c}{NGC4649} & & \\                                                 
0.0---1.2 &  0.063 (0.00:0.17) &  0.091 (0.02:0.18) &  0.117 (0.08:0.17)  \\
1.2---2.3 &  0.000 (0.00:0.18) &  0.139 (0.00:0.32) &  0.293 (0.19:0.48)  \\
2.3---4.3 &  0.000 (0.00:0.18) &  0.355 (0.15:0.53) &  0.164 (0.07:0.30)  \\
4.3---8.0 &  0.000 (0.00:0.22) &  0.644 (0.42:0.99) &  0.000 (0.00:0.15)  \\
\hline  
\end{tabular}

\begin{enumerate}
\item[{$^{\dag}$}]{\footnotesize ~ Definition of the solar units is 3.8, 3.55 and
  4.68 $\times10^{-5}$ for the number abundance of Mg, Si and Fe relative to
  H.  Errors are given at 68\% confidence level for one parameter of
  interest. MEKAL plasma code is used for spectral fitting.}
\end{enumerate}
}
%\end{table}

The problem posed by the Buote papers is whether the gas at the centers of
galaxies is truly multiphase, or does it only appear multiphase as a result
of projection and the coarse spatial resolution of ASCA.  Spatially resolved
measurements, that account for the gas projection effects, result in central
temperatures for NGC4636, NGC4472, NGC4649, NGC5846, NGC5044 of 0.51
(0.42:0.56), 0.76 (0.71:0.78), 0.78 (0.71:0.83), 0.59 (0.56:0.62), and
0.52 (0.49:0.57) keV, respectively. By comparison, for these galaxies, the low
temperatures in multiphase models are 0.52 (0.46:0.57), 0.72
(0.69:0.75), 0.67 (0.59:0.74), 0.63 (0.59:0.66), 0.70 (0.67:0.72) keV. For M86
the projected results give 0.70 (0.59:0.75) keV, while a multiphase analysis
gives 0.60 (0.37:0.70). All errors are quoted on the 90 \% confidence level.
The temperature determinations are very similar, with the exceptions of
NGC4649, where the ``multiphase'' value is slightly lower, and NGC5044
located at the center of an X-ray luminous group, where the ``multiphase''
temperature is higher. The latter can be explained by the pronounced,
spatially extended cooling flow in NGC5044, where the majority of emission
at 0.7 keV is identified in the spatially resolved analysis (cf FP99).

Since our measurements of central temperatures agree with ROSAT for the
galaxies presented in this paper, and also agree with low temperatures from
the two-component model of Buote (1999), there is presently no observational
need for a multiphase component.  However, the abundance determination in
the central spatial bin is likely to be controversial. Precise abundance
values for the central galaxy regions will remain a subject for discussion
until XMM RGS results are available. However, because of these
uncertainties, we exclude from further analysis our central abundance
measurements for NGC4472, NGC4649 and NGC4406, that disagree with the Buote
measurements.

For NGC4636 our systematic errors of 10 per cent are smaller than the 20 per
cent systematic errors used by Matsushita \etal (1997) in their spectral
analysis of the same observation. Nevertheless, their simplified treatment
of the emission does not allow Matsushita \etal (1997) to resolve the
central temperature component properly. Acceptable $\chi^2$ values are
obtained in our spectral fitting. As shown in Fig.\ref{virgo-te} our
temperature results agree well with the previous ROSAT determinations. The
Matsushita \etal (1997) abundance results agree with ours, but disagree with
those by Mushotzky \etal (1996). However, since our abundances are higher
than those by Mushotzky \etal (1996), but lower than those by Matsushita
\etal (1997), our 90 percent confidence area overlaps both measurements.

The complexity of the surface brightness and temperature distribution for
NGC4406 complicates the spectral analysis. We devote a special section
(\ref{sec:m86}) to its analysis.

\begin{table*}
{
\begin{center}
\footnotesize
\tabcaption{\centerline{\footnotesize
ASCA SIS measurements of M87$^{\dag}$}
\label{tab-m87}}

\begin{tabular}{llllll}
\hline
\hline
Annulus (\amin) & $kT_e$, keV & $Ne/Ne_{\odot}$ & $Si/Si_{\odot}$ & $S/S_{\odot}$ & $Fe/Fe_{\odot}$ \\
\hline
0.0---2.0  &  1.320 (1.25:1.40) &                   & 0.833 (0.78:0.88) &                    &  0.541 (0.52:0.56) \\  
2.0---5.0  &  1.520 (1.45:1.60) &                   & 0.692 (0.65:0.72) &                    &  0.447 (0.43:0.46) \\  
2.0---5.0  &  2.459 (2.43:2.49) &  0.491 (0.45:0.53)& 0.658 (0.62:0.70) &  0.498 (0.46:0.54) &  0.477 (0.45:0.51) \\  
5.0---10.  &  2.474 (2.44:2.51) &  0.434 (0.39:0.46)& 0.599 (0.56:0.63) &  0.375 (0.34:0.42) &  0.396 (0.37:0.43) \\ 
10.---15.  &  2.495 (2.48:2.51) &  0.392 (0.34:0.43)& 0.546 (0.51:0.58) &  0.257 (0.22:0.30) &  0.291 (0.28:0.30) \\  
40.---50.  &  2.711 (2.58:2.85) &  0.795 (0.39:1.26)& 0.440 (0.23:0.66) &  0.338 (0.08:0.59) &  0.225 (0.16:0.29) \\ 
\hline                                         
\end{tabular}                                  
\end{center}

\begin{enumerate}
\item[{$^{\dag}$}]{\footnotesize ~ Definition of the solar units is 12.3,
    3.55, 1.62 and 4.68 $\times10^{-5}$ for the number abundance of Ne, Si,
    S and Fe relative to H.  Errors are given at 68\% confidence level for
    one parameter of interest. MEKAL plasma code is used for spectral
    fitting.}
\end{enumerate}
}
\end{table*}

Abundance profiles of Mg, Si and Fe for NGC4472, NGC4636, NGC4649,
NGC4261, NGC4374 and NGC4552 are presented in Figs.\ref{virgo-fe},
\ref{virgo-si} and \ref{virgo-mg} and in Table \ref{table-ab}.  The
central Fe abundance is less than 0.7 solar, when meaningfully
constrained (this excludes NGC4365; see our abundance table
definition). In most galaxies, the central iron abundance is higher
than in the outskirts of the galaxy, where the abundance is always
below 0.3 solar and in half the galaxies in this sample, below 0.1 of
solar.  Except for NGC4636 whose abundances decrease with radius, the
radial abundance profiles of Mg and Si are consistent with being flat
with averaged values of 0.5 solar for NGC4472 and lower for the
rest of the galaxies. The ratio of $\alpha$-elements to iron is nearly
constant with radius for NGC4261, NGC4374, NGC4552 and NGC4636, while
this ratio increases with decreasing Fe abundance in NGC4472 and NGC4649.

Fe abundance gradients, although clearly established in NGC4636,
NGC4472, and NGC4649, may not be a common feature. Therefore a question
for future observations will be to resolve the gradients (or show
their absence) in galaxies with low abundance, like M84. One
difference, already seen, is that the inward increase in iron
abundance for relatively high metallicity systems, always corresponds
to an underabundance of alpha elements, while for galaxies with low iron
abundance (\eg\  M84), alpha-elements become
overabundant.

Before addressing the overall implications of the abundance results,
the following subsections discuss M87, M84 and M86 in detail.

\subsubsection{M87}

In our study of heavy element enrichment, M87 has a special place, being a
cD galaxy at the X-ray center of the Virgo cluster.  cD galaxies dominate
the light within 200 kpc from the cluster center. Their stellar mass loss
and gas mass can explain the observed heavy element abundance within the
central 50 kpc region. Schindler, Binggeli and Boehringer (1999) studied
two-dimensional galaxy distribution in Virgo and find that it centers aside
from M87. Therefore, strong element abundance gradients for M87, presented
in Fig.\ref{m87-ab-fig}, are characteristic of dead stellar population in
M87 with extra metals added by SNe. Results from our analysis of M87 are
summarized in Table \ref{tab-m87}. In addition to the ASCA observation
centered on M87, we include a 40\amin\ offset observation.

\medskip 

\centerline{\includegraphics[width=3.35in]{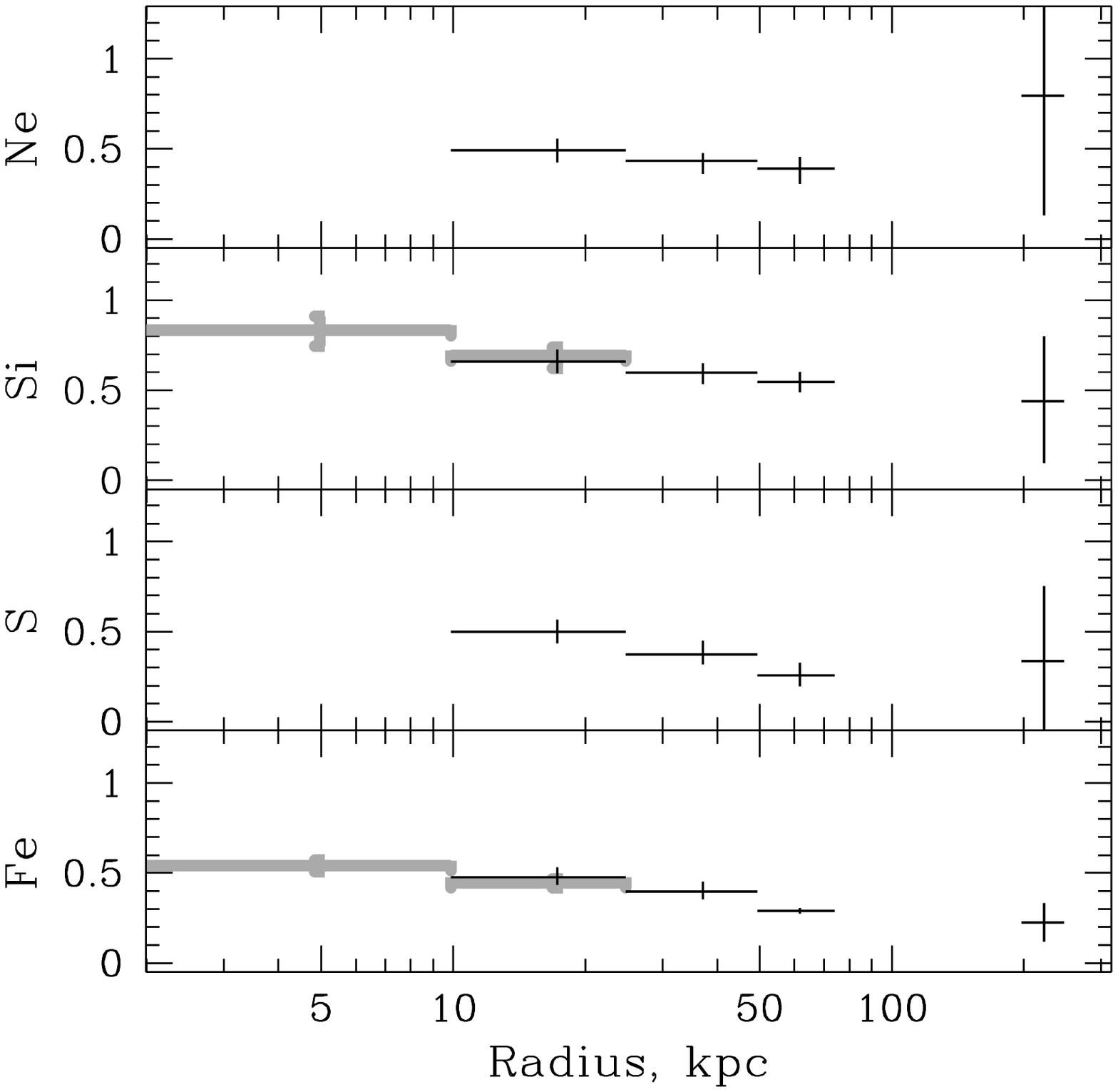}}

\figcaption{Abundance profiles for M87. Grey bars represent quantities
derived from a cooler component. Error bars are shown at 90\% confidence
levels.
\label{m87-ab-fig}}
\medskip

Although, a low temperature for the cool component could reflect the
galaxy's potential, the similarity in the abundance pattern between cool and
hot components, observed in Si and Fe, probably indicates a cooling flow.
However a detailed study of the link between dying stellar population of cD
and both phases of X-ray emitting gas is needed to fully understand that. A
study of the X-ray emission from M87 also was done in the PhD thesis of
Matsumoto (1998), who identified two temperature components and suggested
these were due to a multi-phase medium. Our results from 3-dimensional
modeling show that the hot component does not exist in a detectable amount
in the central spatial bin (inside 10 kpc), while the cool component is
localized within 25 kpc, which restricts any multi-phase region to a 10--25
kpc shell.

We searched the ROSAT image for counterparts to a possible multi-phase
shell. We found significant departures from spherical symmetry in the East,
North-West and South-West directions from M87 center, within the radius of
this shell (see Fig.\ref{m87-imh}). For the brighter regions, spatially
resolved spectroscopy with ROSAT/PSPC yields an average temperature of
$1.60\;(1.50:1.73)$ keV, using MEKAL code, while the remaining emission
within $2'-4'$ has a temperature of $2.11\;(1.95:2.33)$ keV (68\% confidence
intervals are cited). The temperatures differ at a confidence level better
then 99\%. The ROSAT determination for the hotter component is slightly
lower then ASCA value, but still within the 90\% error. Similar features,
previously identified for \eg\ NGC5044, were suggested to be cooling wakes,
produced as the central galaxy, which is the focus of the cooling flow,
moves in the cluster core. (David \etal 1994).  The observations also are
consistent with magnetic field confinement of the plasma (cf Makishima
1997). One could attempt to use the strong elemental abundance gradients
observed M87 to distinguish between these two scenaria when high spatial
resolution observations with at least CCD energy resolution are obtained by
Chandra or XMM.

\medskip 

\centerline{\includegraphics[width=3.35in]{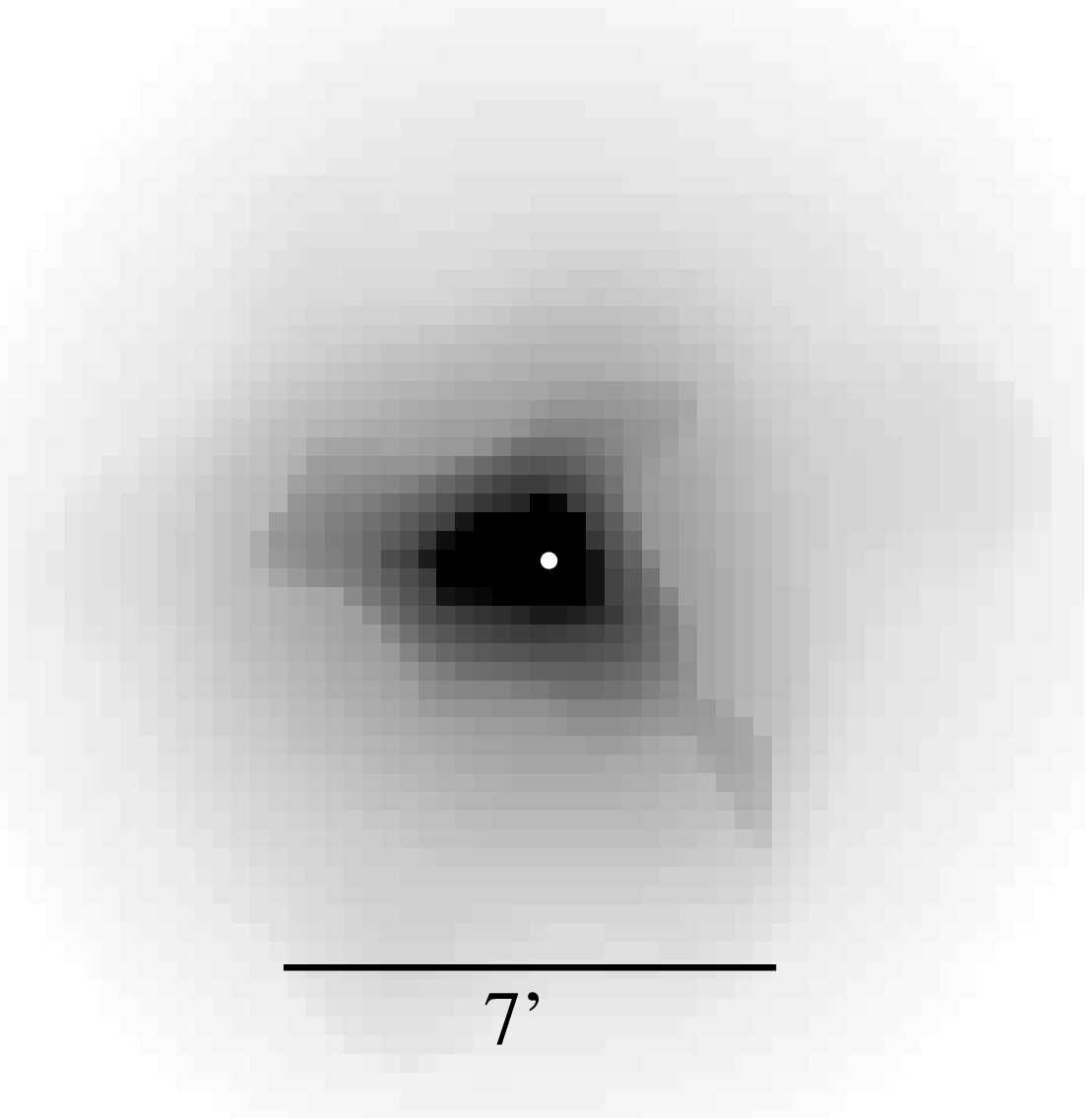}}

\figcaption{Wavelet-decomposition of the surface brightness map of M87 in
the 0.5--2.0 keV energy band. Asymmetric features are seen between radii of
$2 '-5'$ from the center at a confidence level exceeding 99\%. A white dot
denotes the position of the X-ray peak at 187.704 +12.393 (RA Dec., J2000).
\label{m87-imh}}
\medskip

\subsubsection{NGC4374 (M84)}

NGC4374 is an E1 galaxy with a heliocentric velocity of 1026~km sec$^{-1}$
and therefore probably lies within the core of the Virgo cluster.  Previous
X-ray analysis by Forman, Jones \& Tucker (1985) showed the presence of the
quasar 1222+131 2.4\amin\  south-east of the center of M84. We used the
position ($12^h25^m11.95^s$ +12\deg51\amin53.3\asec, J2000, taken from Bowen
\etal 1994) of the quasar to check possible boresight offsets within the
ROSAT image. The measured X-ray position is $12^h25^m11.65^s$
+12\deg52\amin00.60\asec\ (J2000), so $\Delta x=3$\asec\  and $\Delta
y=7.3$\asec\  with a statistical error of 4\asec. After applying a correction
of $\Delta x=2.34$\asec\  and $\Delta y=4.54$\asec, taken from ROSAT
XRT/detector boresighting results (Briel \etal 1993), the positions agree
within the uncertainties. The quasar spectrum, derived from ROSAT data is
characterized by a power law with photon index 2.13 (1.98--2.27; 90\%
confidence) and F$_{\rm 1 keV}=8.04\pm1.1\times10^{-5}$\phscmkev. In
modeling ASCA data, we subtract the quasar spectrum.

Our central abundance determination for M84 has a smaller error bar, compared
to the analysis of Matsushita (1998).  This results from different treatments of the hard
component (she fits in an additional bremsstrahlung model, while we use
ROSAT results for the quasar spectrum) and the larger systematic error adopted there.

Radio observations of M84 at 1.4 and 4.9 GHz show a large scale reversal in
the magnetic field of the medium situated in front of the radio source
(Laing \& Bridle 1987). Although, this medium can contribute to the outer
parts of the M84 diffuse X-ray emission, to estimate its emission we need
spatially resolved spectroscopy on a 10\asec\  scale, which will be provided
with Chandra observations.

\subsubsection{Temperature and Abundance Structure in M86.}\label{sec:m86}

\begin{figure*}

\begin{minipage}{9cm}

%\bigskip
\includegraphics[width=3.25in]{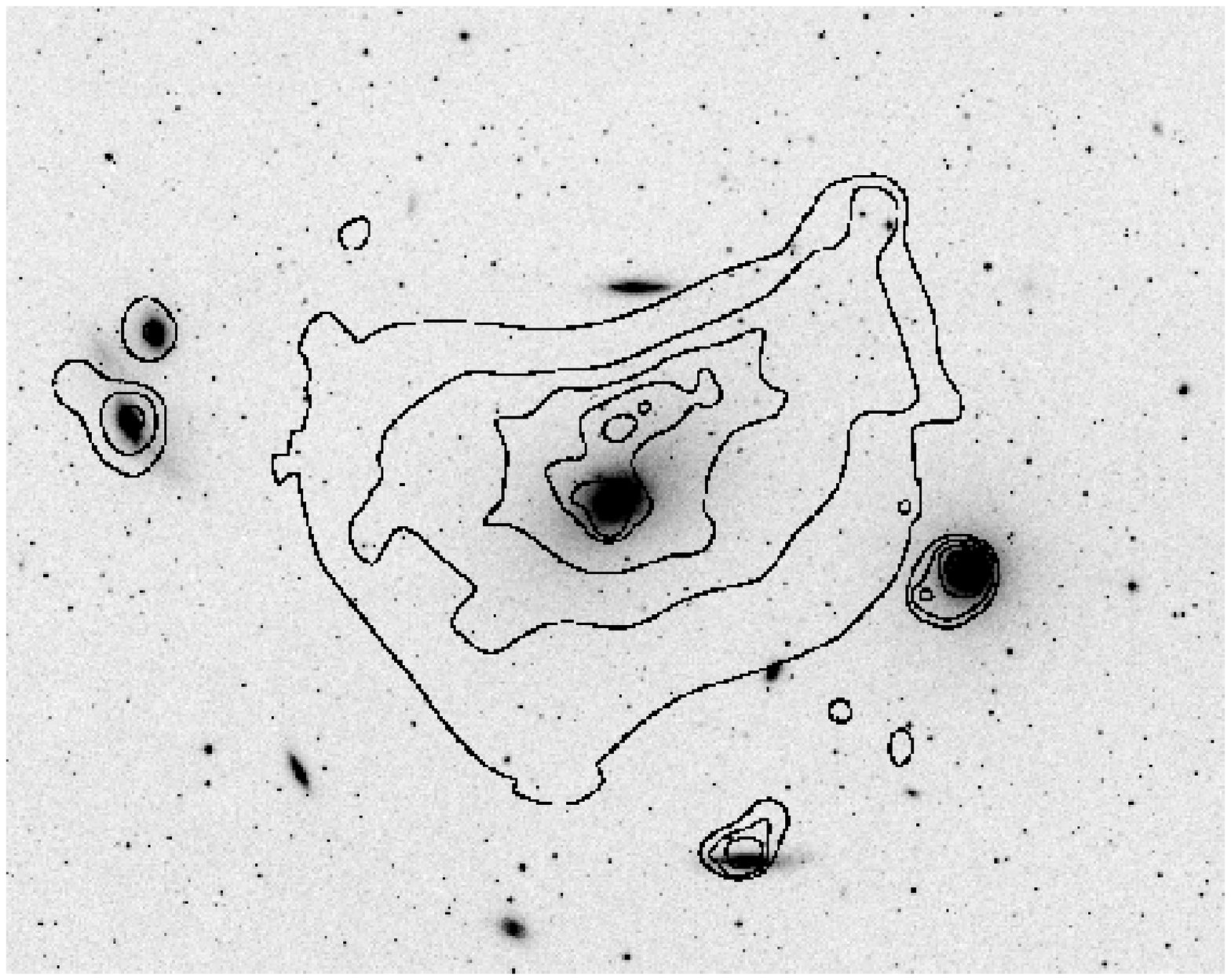}

\figcaption{Optical image of M86 with wavelet-decomposed X-ray
  contours from a ROSAT PSPC observation.
  The image is 64 by 50 arcminutes.  In addition to M86, emission is clearly
  seen from the galaxies M84, NGC4438/9 pair and NGC4388. North is
  towards the top, east is to the left.
\label{m86-imh}}
%  this is for single space version

\doublespace

\vspace*{-5.2cm}

\hspace*{0.cm} {\it NGC4438/9}

\hspace*{3.5cm} {\it M86}  \hspace*{2cm} {\it M84}

\vspace*{1cm}

\hspace*{4cm} {\it NGC4388}

\vspace*{2.7cm}
%  this is for double space version
% \vspace*{-5cm}
% 
% \hspace*{0.cm} {\it NGC4438/9}
% 
% \hspace*{3.5cm} {\it M86}  \hspace*{2cm} {\it M84}
% 
% \vspace*{1cm}
% 
% \hspace*{4cm} {\it NGC4388}
% 
% \vspace*{2.cm}

\end{minipage} \hfill \begin{minipage}{9cm}

{
% Labeling of the figure
\includegraphics[width=3.25in]{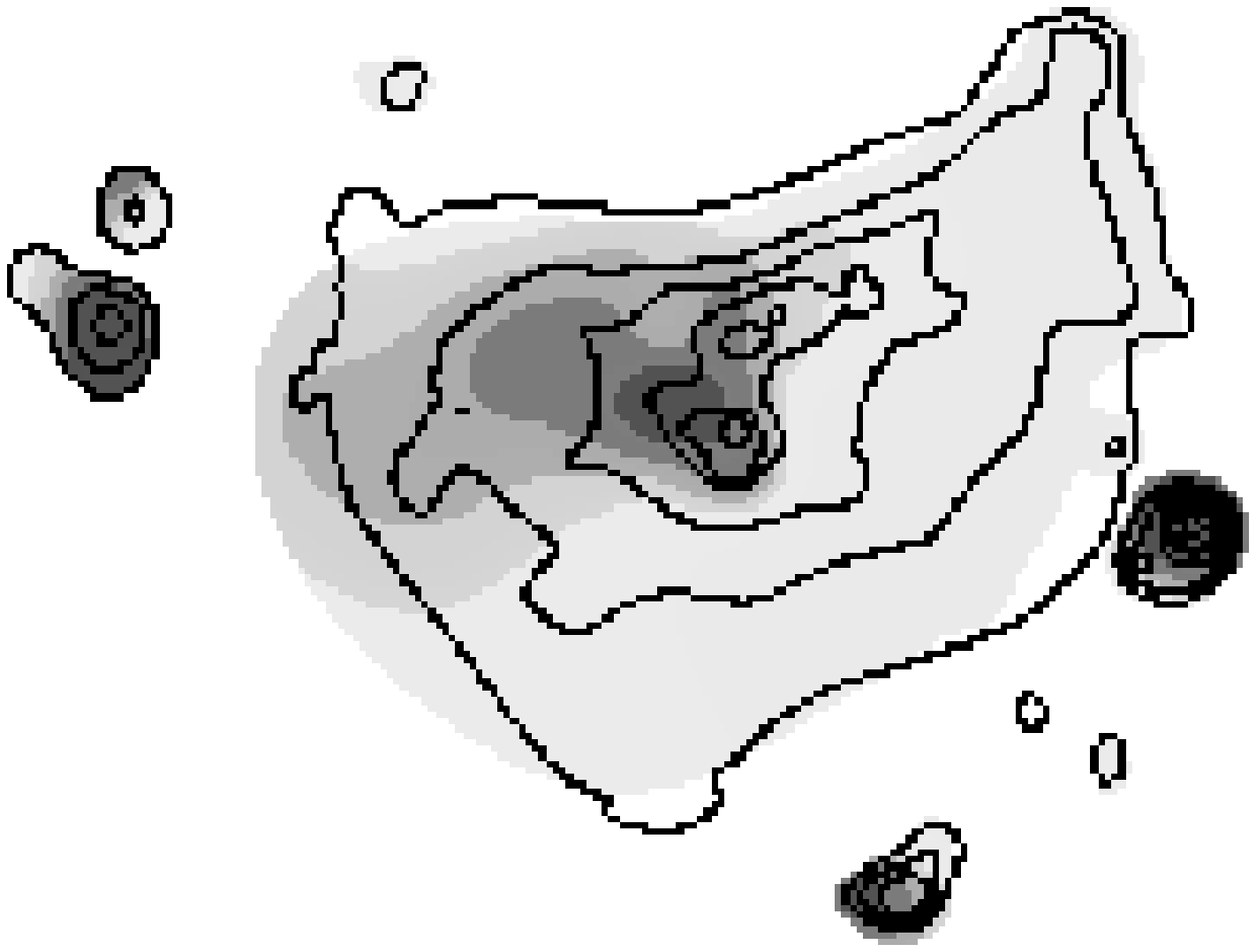}
\figcaption{Hardness ratio [R67-R45]/[R67+R45], derived from the ROSAT PSPC
  image. Adaptive smoothing was used with a minimal Gaussian width of
  1\arcmin\ ($\sigma$). For the range of {\kte}, Fe abundance, {\nhl},
  typical for M86, this map represent the temperature distribution of X-ray
  emitting gas around the galaxy. The cooler regions include the galaxy
  central region and extend to the north-east and east
  with slightly increasing temperature. The hottest
  regions are to the west and south-west of M86. The picture also indicates
  the ``temperatures'' of the nearby galaxies, which are similar to the
  center of M86. Numbers indicate the regions chosen for spectral analysis.
  Temperature determinations are given in Table \ref{m86-rosat-tab}.
  Repeated numbers mean that spectral information was averaged for these
  points.
\label{m86-temap}}
\pspicture(0.,9.)(0.,9.)
%\rput(0.,11.7){\rput(5.1,3.6){1}
%\rput(0.,11.2){\rput(5.1,3.6){1}
\rput(0.09,12.65){\rput(5.1,3.6){1}
\psline(4.6,3.7)(5.0,3.6)
\rput(3.5,4.2){2}
\rput(3.9,3.9){2}
\rput(2.5,4.0){3}
\rput(2.5,4.8){4}
\rput(3.2,3.0){4}
\rput(3.5,2.2){4}
\rput(4.2,4.5){5}
\rput(5.5,4.8){5}
\rput(5.1,4.0){6}
\rput(5.8,4.0){6}
\rput(4.8,2.5){6}
\rput(6.3,5.1){7}
\rput(6.6,4.0){7}
\rput(6.0,2.9){7}
\rput(3.4,5.0){8}
\rput(3.4,3.5){8}
\psline(4.5,4.45)(4.5,5.3)
\rput(4.5,5.5){9}}
\endpspicture
% Labeling of the figure
\medskip
}

\end{minipage}
\end{figure*}

An X-ray study of M86 with {\einstein} (Forman \etal 1979) revealed a peak
of emission centered on M86 and a ``plume'' extending north-west of the
galaxy.  Ram-pressure stripping of the hot gas from M86 as the galaxy
traverses the Virgo cluster was suggested to explain the observed structure
(Forman \etal 1979; Fabian, Schwartz \& Forman 1980; Takeda, Nulsen \&
Fabian 1984). The {\einstein} spectrum of M86 was adequately represented by
radiation from an optically thin hot plasma with kT=0.8 keV, although no
estimate of the Fe abundance could be made (Forman, Jones \& Tucker 1985).

Spatial analysis of the X-ray brightness of M86 from the ROSAT PSPC and
HRI data was presented in Rangarajan \etal (1995; hereafter RWEF). In
addition to the plume, they identified a southern extension (from
the center to about 3\amin\  to the south), a Void (to the north of the
galaxy) and a north-eastern arm (extending from the center more than 5\amin\
to the north-east).  While the exact regions selected for spectral analysis
by RWEF differ from ours, the core and plume are both analyzed in common and
the results are in good agreement.  In this section we concentrate on the
temperature mapping of the diffuse emission around M86 and verify the
results of our ROSAT analysis through detailed modeling of
the ASCA SIS data.

In Figure \ref{m86-imh} we present the optical image of the M86 field with
contours from the wavelet-decomposition of the X-ray emission overlaid. At the
center of M86, the cluster emission is less than 5\% of the peak galaxy
emission.  The lowest contour shown around M86 is twice the cluster
emission, at the eastern part of M86, closest to the cluster center.  An
X-ray plume extends 19\amin\  north-west of the galaxy. Two outer contours of
X-ray emission resemble optical light seen in Malin's deep print (Malin
1981). Analysis of the optical image showed optical ``blobs'' in excess
of the elliptical shape associated with the X-ray plume, that could be
modeled as star formation regions within the cooling hot gas (Nulsen \&
Carter 1987). An alternate explanation by White \etal (1991) suggested that
optical starlight scattered from stripped dust could explain the elliptical
shape of the blue light from the plume and also the infrared emission.

In Figure \ref{m86-temap} we present the (R67--R45)/(R67+R45) hardness ratio
derived from the ROSAT PSPC image, overlaid on the X-ray surface brightness
contours (the broad band R47 image), where Rnm means summing of images from
Rn to Rm, as defined by Snowden \etal (1994). As was shown by Finoguenov
(1997), for a temperature in the range 0.5--1.5 keV, and metal abundances
exceeding 0.1 solar this ratio uniquely determines the gas temperature in a
single-phase plasma. The four gray scale levels correspond to different
temperatures with dark gray representing 0.7 keV and light gray the 1.0--1.1
keV interval.  Cool regions include the galaxy center and its immediate
environs (particularly to the north-east).  The cool region extends eastward
with slightly increasing temperature. The coolest region in the north-east
extension is associated with an enhancement in the X-ray emission, referred
to as the North-Eastern Arm in RWEF.  The plume temperature structure is not
uniform, but is instead cooler in the center toward M86 (immediately north
of the galaxy).  The hottest regions are those south-west of M86. The Figure
also indicates the ``temperatures'' of the nearby galaxies.

With the spatial resolution of the ASCA SIS data, we identify only five
regions in Figure \ref{m86-temap}, that correspond to ROSAT regions. We use
the ROSAT image as input for the ASCA modeling. Results of the spectral
analysis of the temperature structure in M86 from both ROSAT and ASCA are
presented in Table \ref{m86-rosat-tab}. {\nh} was fixed at the Galactic
value of $2.75\times10^{20}$ cm$^{-2}$ (Stark \etal 1992).

\medskip

%\begin{table}[H]
{
\footnotesize
\centering
\tabcaption{
\centerline{\footnotesize Temperature structure in M86$^{\dag}$}
\label{m86-rosat-tab}}

\begin{tabular}{rllrl}
\hline
\hline
{\# } & ROSAT $kT_e$, keV & ASCA $kT_e$, keV & L$^{\ddag}$ & Comments \\
\hline
 1& 0.752 (0.73:0.78) & 0.600 (0.37:0.70) & 249 & center\\
 2& 0.743 (0.71:0.78) & 0.691 (0.61:0.77) & 128 & the N-E Arm\\
 3& 0.834 (0.79:0.87) & & 131 & \\  
 4& 0.937 (0.85:1.05) & & 177 & East edge \\
 5& 0.867 (0.85:0.88) & 0.857 (0.80:0.89) & 223 & the Plume\\
 6& 1.076 (1.05:1.10) & 1.034 (0.91:1.14) & 286 & S-W to N4406\\
 7& 1.075 (1.04:1.10) & 1.116 (0.98:1.17) & 200 & West edge\\
 8& 0.871 (0.84:0.90) &  &     81 & \\
 9& 0.760 (0.72:0.80) & included in \# 5 &   90 & plume center\\
\hline
\end{tabular}
%\vspace{0.5pc}

\begin{enumerate}
\item[{$^{\dag}$}]{\footnotesize ~ Errors are given at 90\% confidence level for
  one parameter of interest. Raymond-Smith plasma code was used for ROSAT
  data, while MEKAL code was used to describe the ASCA data. Col.(1) points
  to the area of spectrum extraction denoted in Fig.\ref{m86-temap} with
  more comments in col.(5).}

\item[{$^{\ddag}$}]{\footnotesize ~ ROSAT/PSPC flux in $10^{-14}$ \ergscms\ in the
  0.5--2.0 keV band} 
\end{enumerate}
}
%\end{table}

\medskip

Spectral variations in the plume, seen in the ROSAT hardness ratio map
(Fig.\ref{m86-temap}) are quantified by the difference of $0.10\pm0.02$~keV,
found between spectral regions 9 and 5. In Table
\ref{m86-asca-tab} we give results for the heavy element abundance
measured in our ASCA analysis.  Given the complex temperature structure of
the M86 hot gaseous halo, these abundance determinations should be considered
with caution. In summary, within our uncertainties, the abundances of Fe,
Mg, and Si are each fairly uniform throughout M86, as well as the plume
region.

% 
% \vspace*{1cm}
% 
% \vspace*{1cm}
% 

%\begin{table}[H]
{
\footnotesize
\centering
\tabcaption{\footnotesize
ASCA SIS measurements of heavy element abundances in M86$^{\dag}$ 
\label{m86-asca-tab}}

\begin{tabular}{lllc}
\hline
\hline
$Mg/Mg_{\odot}$ & $Si/Si_{\odot}$ & $Fe/Fe_{\odot}$ & \# \\
\hline
 0.255 (0.14:0.46) & 0.489 (0.35:0.66) & 0.146 (0.08:0.23) & 1\\     
 0.266 (0.16:0.38) & 0.303 (0.14:0.42) & 0.098 (0.07:0.15) & 2\\     
 0.276 (0.18:0.37) & 0.440 (0.36:0.52) & 0.220 (0.19:0.25) & 5+9\\   
 0.334 (0.11:0.57) & 0.435 (0.25:0.66) & 0.148 (0.07:0.26) & 6\\     
 0.307 (0.09:0.53) & 0.246 (0.12:0.37) & 0.241 (0.20:0.29) & 7\\     
\hline
\end{tabular}
%\vspace{0.5pc}

\begin{enumerate}
\item[{$^{\dag}$}]{\footnotesize ~ Definition of the solar units is 3.8,
    3.55 and 4.68 $\times10^{-5}$ for the number abundance of Mg, Si and Fe
    relative to H.  Errors are given at 68\% confidence level for one
    parameter of interest. Col.(4) points to the area of spectrum extraction
    denoted in Fig.\ref{m86-temap}. MEKAL code was used for spectral
    modelling. }
\end{enumerate}

}
%\end{table}

%\medskip

%\clearpage

\section{ Discussion}\label{sec:disc-n}

Before proceeding to the theoretical modeling, we should first adopt an
appropriate abundance scale. Since abundance measurements in stars, as well
as theoretical modeling, are usually given using an abundance table
different from that used for X-ray analysis, we converted our results to
units of (3.95, 3.55, 3.31)$\times10^{-5}$ for the Mg, Si and Fe number
abundance relative to H for use throughout this section.

\subsection{ Comparison of the elemental abundance in Virgo 
  ellipticals with the stellar population in the Solar vicinity.}
\label{sec:stars}

\begin{figure*}

\includegraphics[width=3.25in]{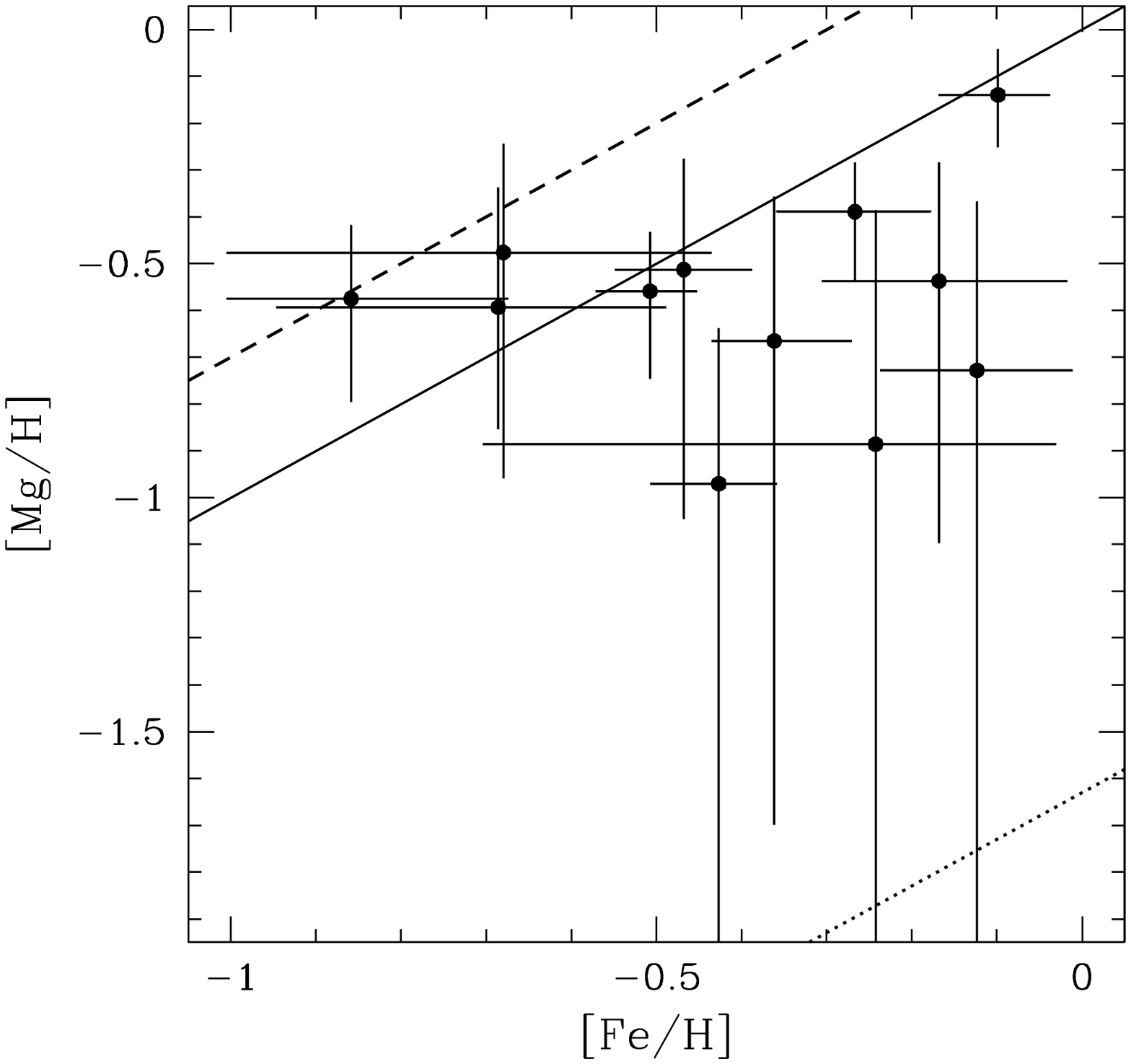} \hfill \includegraphics[width=3.25in]{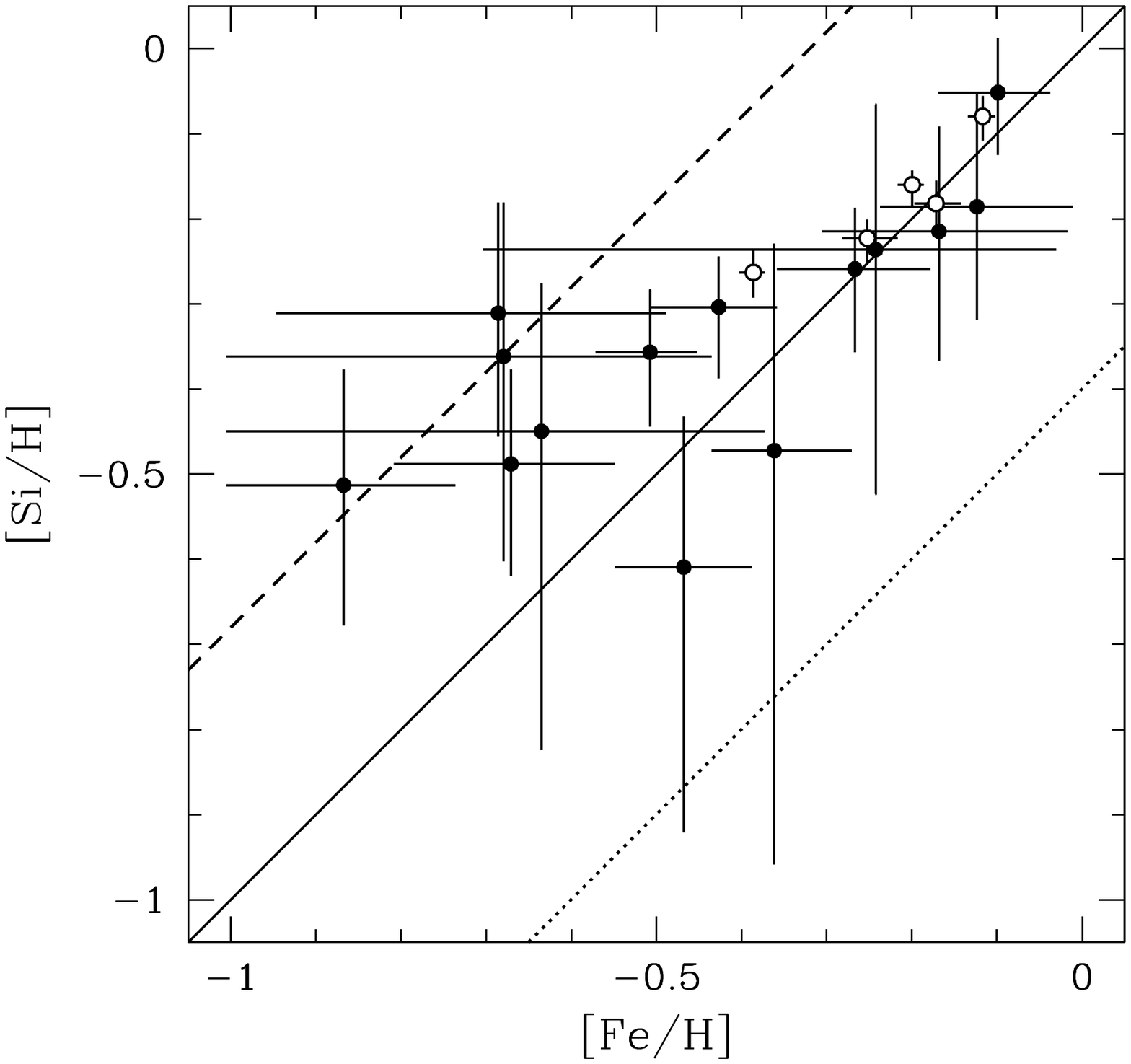}

\figcaption{[Mg/H] {\it (left panel)} and [Si/H] {\it (right panel)} vs
  [Fe/H]. Open circles indicate the best-fit values for M87 (measured for
both cool and hot gas components) and filled circles for the rest of the
galaxies in our sample. Error bars are shown at the 68 percent confidence
level. Meteoritic units of Fe abundance are used (3.31$\times10^{-5}$ number
abundance relative to H) and [] indicates logarithmic values.  The solid
line is the solar ratio, dotted line the SN~Ia ratio (Thielemann \etal
1993), and dashed line the averaged SN II yields from Tsujimoto \etal
(1995).
\label{si_vs_fe}}

\bigskip
\includegraphics[width=3.5in]{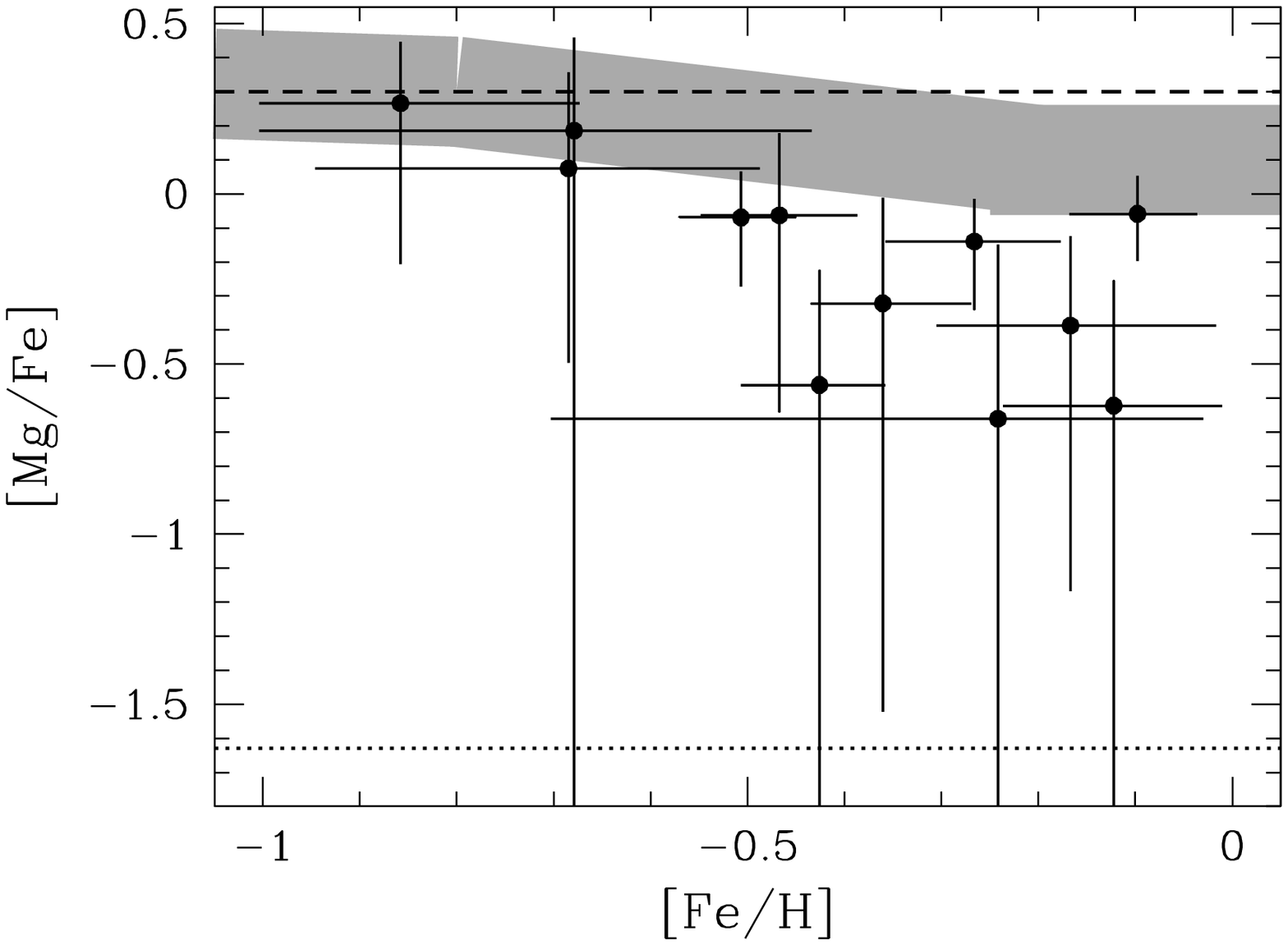} \hfill \includegraphics[width=3.5in]{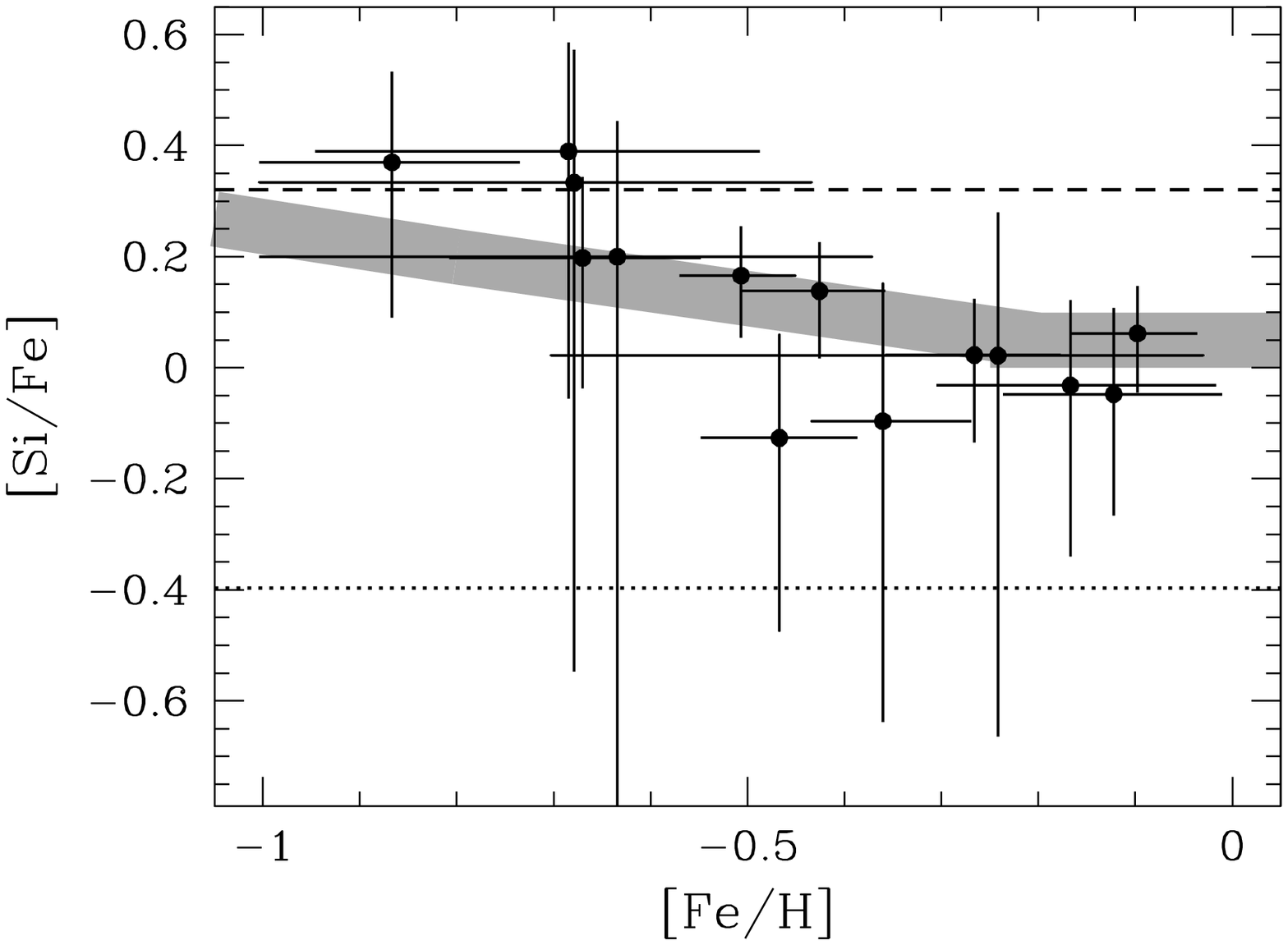}

\figcaption{ [Mg/Fe] vs [Fe/H] {\it (left panel)} and [Si/Fe] vs [Fe/H] {\it
    (right panel)} for Virgo galaxies. Error bars are shown at the 68 
  percent confidence level. Stellar measurements in the solar vicinity are
  represented by a gray line. The thickness of this line indicates the
  intrinsic spread in the measured values. Dotted line indicates SN~Ia
  element ratio (Thielemann \etal 1993), and dashed line is the average SN II yields
  from Tsujimoto \etal (1995).
\label{virgo-si2fe}}
\medskip

\end{figure*}

The abundance results presented here for a set of Virgo galaxies appear far
from uniform. Some galaxies have uniformly low abundance, while values for
others depend on the radius chosen for the measurement and are higher at the
centers. A hint toward the unification of the presented data arises from the
correlation of a low Fe abundance, either throughout the galaxy or in the
outskirts, with an overabundance of alpha-process elements, such as Mg and
Si studied here. In Fig.\ref{si_vs_fe} we plot the Mg and Si abundance vs
Fe, omitting points with very large uncertainties and restricting values of
Mg, Si and Fe to be above 0.1 solar. The observed trends of Mg vs Fe and the
plot of Si vs Fe at low Fe are not parallel to lines of constant element
ratios. As these plots show, for low Fe abundance, Mg and Si become
overabundant. However, there is a difference between Mg and Si in that the
Mg to Fe ratio is below the solar ratio at high values for iron. These plots
suggest an increasing importance of SN~Ia enrichment at solar metallicity,
compared to 1/10 solar, with SN~II enrichment dominant at low
metallicities. Evidence for SN~II prevalence at low metallicities arises
primarily from the abundance data for NGC4374 and the outskirts of NGC4472
and NGC4649. The innermost M87 points for Si and Fe are plotted for
comparison and show similar behavior.

Comparison of the X-ray determined abundance pattern for the ISM in
elliptical galaxies with that of the stellar population in the Solar
vicinity was suggested by the work by Wyse (1997) to search for any
differences in the IMF and to constrain the role of SN~Ia in element
enrichment. The choice of stellar determinations for our Galaxy is limited
by the absence of optical data on Si/Fe in early-type galaxies. Only a hint
toward an overabundance of Mg/Fe is found from the optical line-index
gradients by Kobayashi \& Arimoto (1999). In Fig.\ref{virgo-si2fe} we plot
[Mg/Fe] vs [Fe/H] and [Si/Fe] vs [Fe/H] to compare the elemental abundances
of Virgo ellipticals with optical measurements for stellar abundances in our
Galaxy, as compiled by Timmes, Woosley, and Weaver (1997, see references
therein).

Fig.\ref{virgo-si2fe} shows that measurements from different galaxies have
little intrinsic scatter, despite the diversity in radial abundance
profiles. Also, compared to the stellar data, at high Fe abundance, the
Virgo X-ray sample demonstrates equal or even a higher SN Ia contribution,
compared to the stellar data for our Galaxy (a spiral).  However, in many
theoretical studies, the abundance pattern of the stellar population in
early-type galaxies is assumed to follow SN~II yields. Thus, these
observations show the importance of SN~Ia's in the chemical enrichment of
the hot gas in ellipticals. To produce these results, we briefly discuss gas
accretion (cf Brighenti \& Mathews 1997), a prolonged duration of star
formation (cf Martinelli \etal 1998; F99) and post-formation SN~Ia activity.

Since the gas that is highly enriched by SNe~Ia lies in the central regions
of the galaxies, the likelihood of gas accretion to produce this effect is
low. As for the feasibility of prolonged star formation in the central
region, Martinelli \etal (1998) showed that, if the galaxy is not considered
as one zone, the central galaxy region could have much longer star
formation, accompanied by accretion of metal-poor gas toward the center.
This scenario, however, would need to address the problem of matching the
SN~Ia rates adopted in the theoretical modeling with the current X-ray
data. At present, most X-ray limits on abundances imply almost no
post-formation SN~Ia activity.

The merger hypothesis for the origin of the X-ray abundance gradients (Bekki
1998) may be more relevant for the galaxies with low iron abundance. Indeed,
the best candidates for galaxy mergers from our sample, NGC4374 and NGC4552
(Schweitzer \& Seizer 1992) have low Fe abundance and are good examples of
the overabundance of alpha-process elements. While a much more detailed
abundance study of galaxies from the Schweitzer \& Seizer (1992) list is
required to draw any firm conclusion, we mention that the observed tendency
is in excellent agreement with differences expected between star formation vs
merging scenario for galaxy formation.

\subsection{A note on the comparison of X-ray abundance
  determinations with optical data on elliptical galaxies.}

It is generally agreed that early type galaxies are currently passively
evolving stellar systems, where significant star formation was cut off at
early epochs by a galactic wind (\eg\ Ciotti 1991, David \etal 1991). The
large amount of hot gas presently found in E's is attributed to stellar mass
loss, and as such, should be characterized by stellar metallicity and
stellar velocity dispersion, with supernovae supplying additional elements
and energy into the interstellar medium. Previous works, where this scheme
was adopted, met a mismatch between stellar abundances, derived via modeling
of the optical measurements, and the tremendously low metallicities found in
the X-ray gas, which imply both a low metal content in stars and a SN~Ia
rate that is lower than measured in optical searches (\eg\ Arimoto \etal
1997).

Perhaps the first step towards resolving this apparent discrepancy was the
discovery of strong abundance gradients in the stellar content of
ellipticals (\eg\  Carollo, Danziger and Buson 1993).  Combining these
results with the presence of gas inflow, Loewenstein \& Mathews (1991) {\it
were able to explain} the low X-ray abundances, although the problem of low
SN~Ia rate persisted. A similar conclusion was derived in our analysis of
NGC5846 (F99).

The ``apparent inconsistency'' between optical and X-ray determinations,
claimed by Arimoto \etal (1997), can be attributed to the neglect of
significant differences in the spatial resolution of X-ray and optical
observations.  The ``central'' abundance value from X-ray measurements often
corresponds to a 3\amin\ radius region, compared to a region of only
$\sim2''$ in the optical. In terms of overdensities it corresponds to a
change by a factor of $\sim1000$, which \eg\ exceeds a difference between
clusters of galaxies and $Ly{\alpha}$ clouds (see Cen \& Ostriker 1999 for a
discussion of crucial importance of equal overdensity in metallicity
comparisons).  Moreover, the complex temperature structure of the gas at the
galaxy center often precludes precise abundance determinations at the center
(Buote 1999, FP99). Even in cases when we are fairly confident in our
abundance determinations, as e.g. NGC4374, the central region corresponds to
1\arcmin. An average optical abundance means a {\it luminosity averaged}
value typically within a few arcseconds, while the metallicity of X-ray gas
reflects {\it mass-averaged} metallicities from a dead stellar population,
that has been heavily smoothed by the inflow motion of the gas in a cooling
flow (cf Loewenstein \& Mathews 1991).

Moreover, as Bazan \& Mathews (1990) showed, the evolutionary tracks from
Van den Bergh \& Bell (1985) exhibit shorter main sequence lifetimes for low
metallicity stars due to their higher luminosities, thus these low
metallicity stars drop out faster from the ``luminosity averaged''
abundances while contributing more to the mass-loss averaged abundances.
Modeling of the SN~Ia rate should be adjusted to allow for this phenomena
(F99).

An alternative explanation proposed by Arimoto \etal (1997) criticized the
iron abundance determinations based on the Fe L-shell emission line
complex. A major effort to understand this issue is presented in Phillips
\etal (1999). As it follows from their work, for CCD-resolution spectroscopy
in the temperature range around 1 keV, uncertainty of the MEKAL code does
not exceed much the 10\% level at 0.5--10 keV energy band. However,
uncertainties increase strongly at softer energies and lower emission
temperatures. Therefore, our adopted level of systematic uncertainties
related to the usage of MEKAL code are justified.  In addition, we note that
the X-ray iron abundance determinations for middle-age SN Remnants in LMC,
that also depend on the modeling of L-shell emission complex, are in very
good agreement with optically determined values (Hughes, Hayashi \& Koyama
1998). Despite a more complex ionization balance in SNR emission, this class
of SNR exhibit a spectrum very similar to X-ray spectra of early-type
galaxies, that can be understood in view of the sufficient ionization ages
of these SNRs. 

Another strong argument comes from the similarly low abundances of Si and
Mg, as derived in the X-ray analysis of early-type galaxies. These are based
on abundance determinations using K-shell lines, so the criticism of Arimoto
\etal (1997) is not applicable. Once again, the negative abundance gradients
in the stellar population and inflow motions of the X-ray emitting gas
require that the X-ray abundance determination be lower in order to agree
with the optical results.

\subsection{Post-formation SN~Ia rate in Virgo ellipticals}

The low Fe abundance in NGC4374 (Loewenstein \etal 1994), implies strict
limits on the SN~Ia rate. This result, along with our abundance measurements
that favor SN~II production, suggests that at the observed distance from the
galaxy center (which is always in excess of $R_e$ for this sample), the
stellar population also has a low abundance and that no SN~Ia enrichment is
allowed. On the other hand, when a higher Fe abundance is observed (\eg\
central parts of NGC4636, NGC4472), SN~Ia enrichment becomes a necessity.

While the origin of the differences in enrichment pattern is probably due to
the presence of a surrounding group potential for NGC4636 and NGC4472, the
correlation of an overabundance of alpha-elements with low metallicity
suggests {\it metallicity dependent SN~Ia rates}. An example of related
theoretical work is that by Worthey and Weaver (1995) on the metal yields from
SN~II as a function of metallicity. The only theoretical work discussing the
SN~Ia behavior, as a function of metallicity, is by Kobayashi \etal (1998).
They considered a model for SN Ia progenitor system by Hachisu, Kato \&
Nomoto (1996) and concluded that no SN~Ia occur at metallicities below 0.1
solar. This result provides a good explanation (however not a unique one)
for the delay in SN~Ia enrichment for Galactic stars. To examine this model,
we perform a standard modeling of the X-ray data (as described \eg in
Loewenstein \& Mathews 1991).

\begin{figure*}
  
  \vspace*{-2cm}

\rput(3.2,2.25){\includegraphics[width=2.5in]{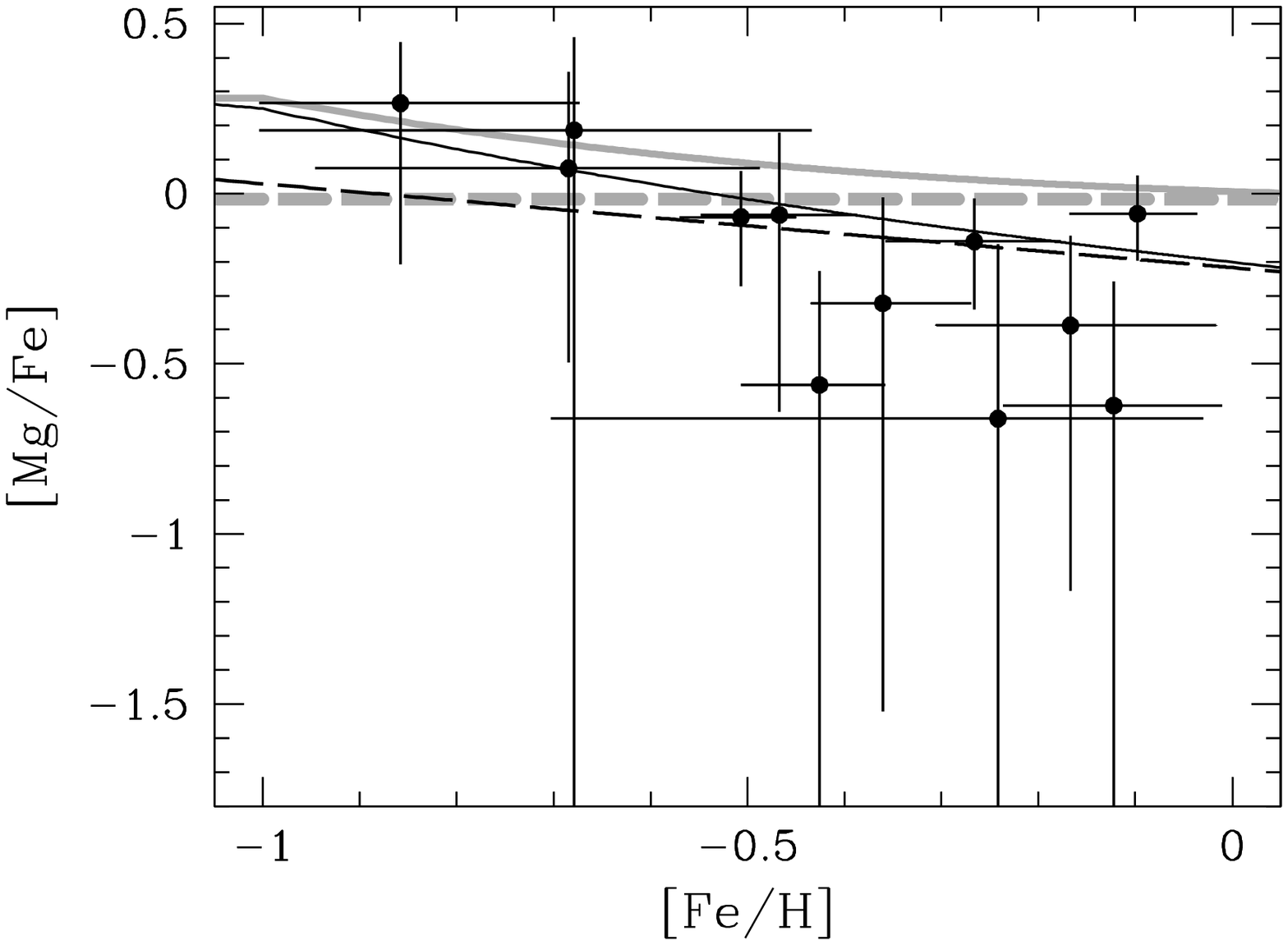}} \hfill \rput(3.2,2.25){\includegraphics[width=2.5in]{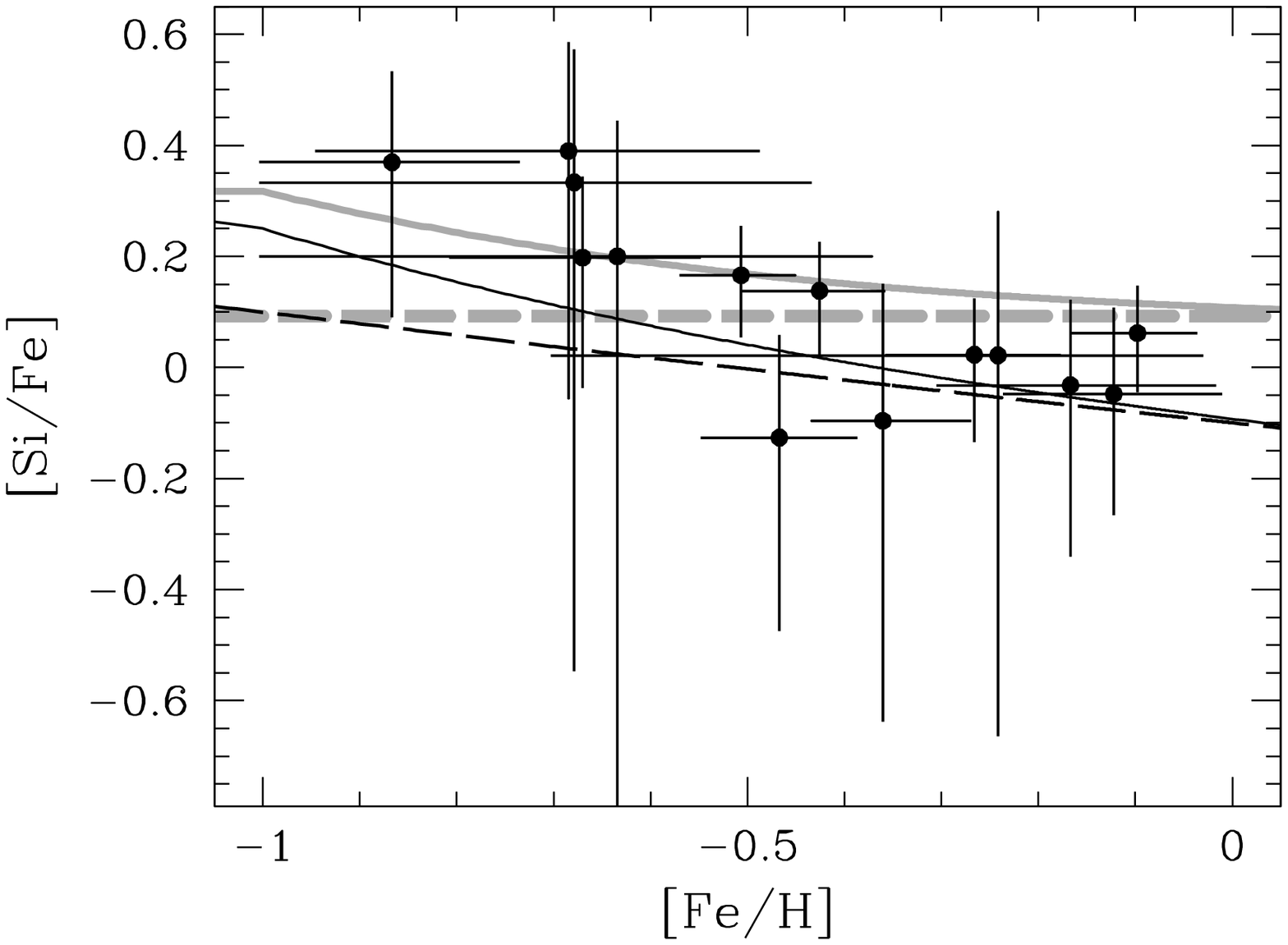}} \hfill \includegraphics[width=1.88in]{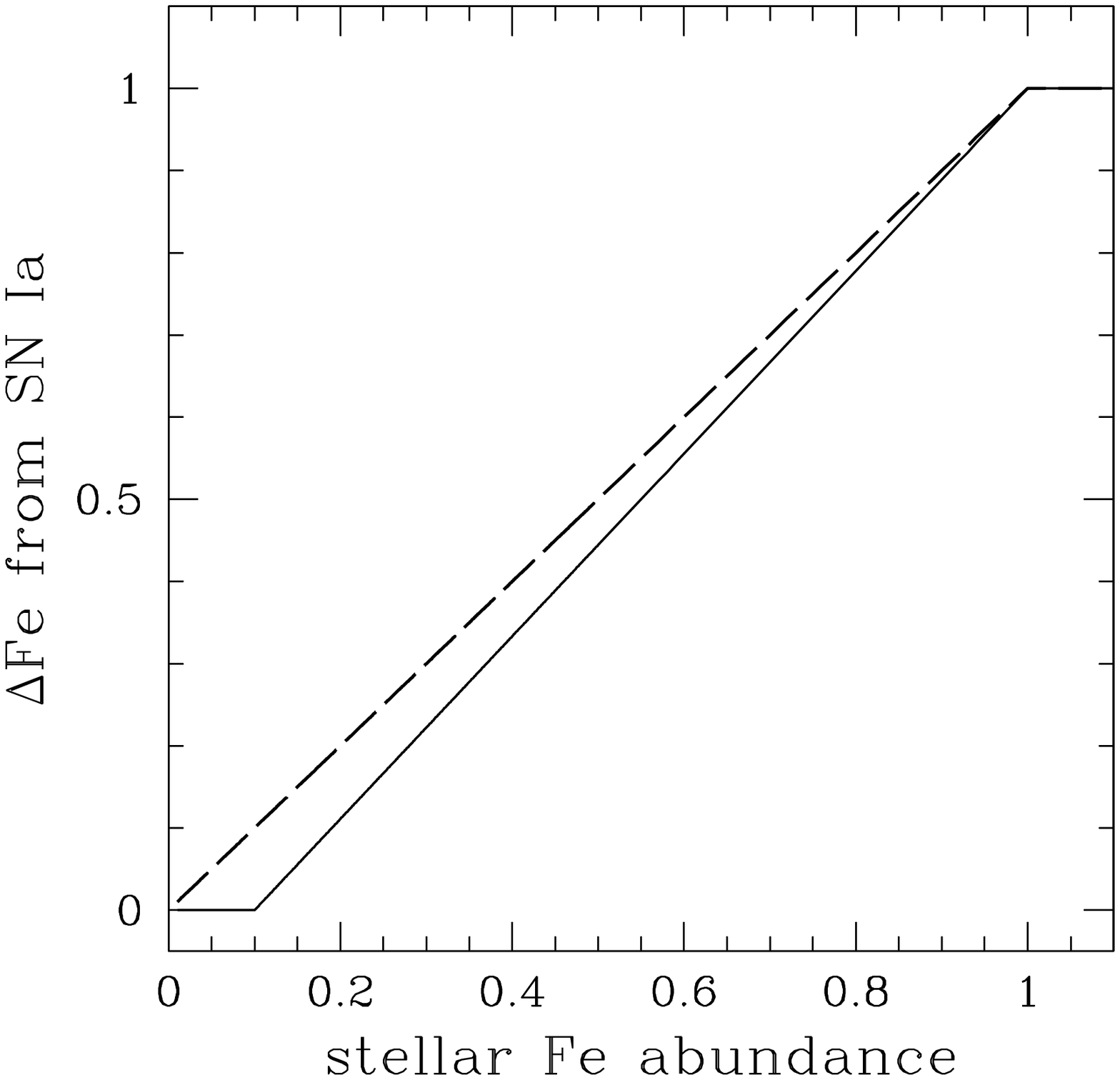}

\vspace*{0.5cm}

\rput(3.2,2.25){\includegraphics[width=2.5in]{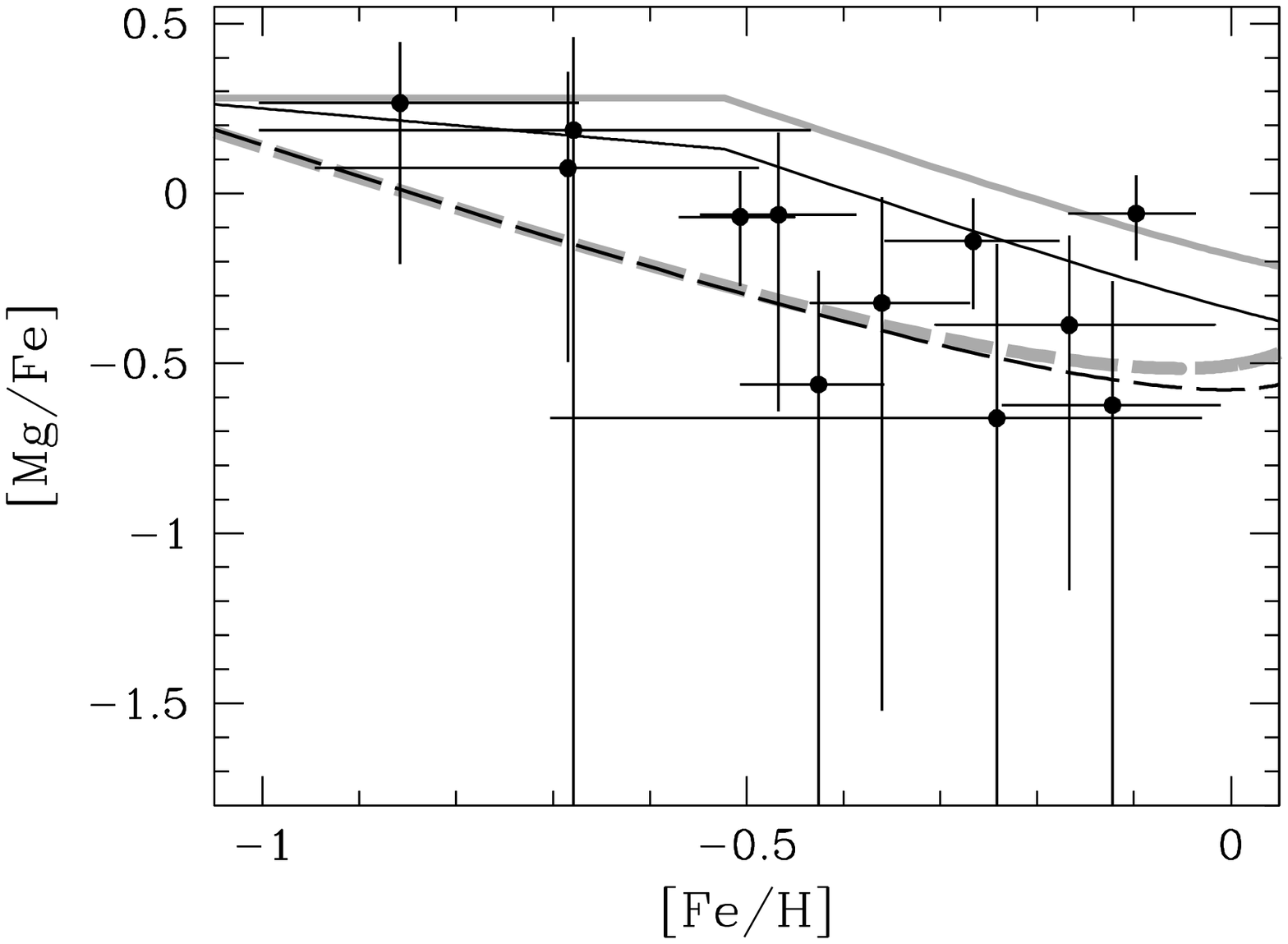}} \hfill \rput(3.2,2.25){\includegraphics[width=2.5in]{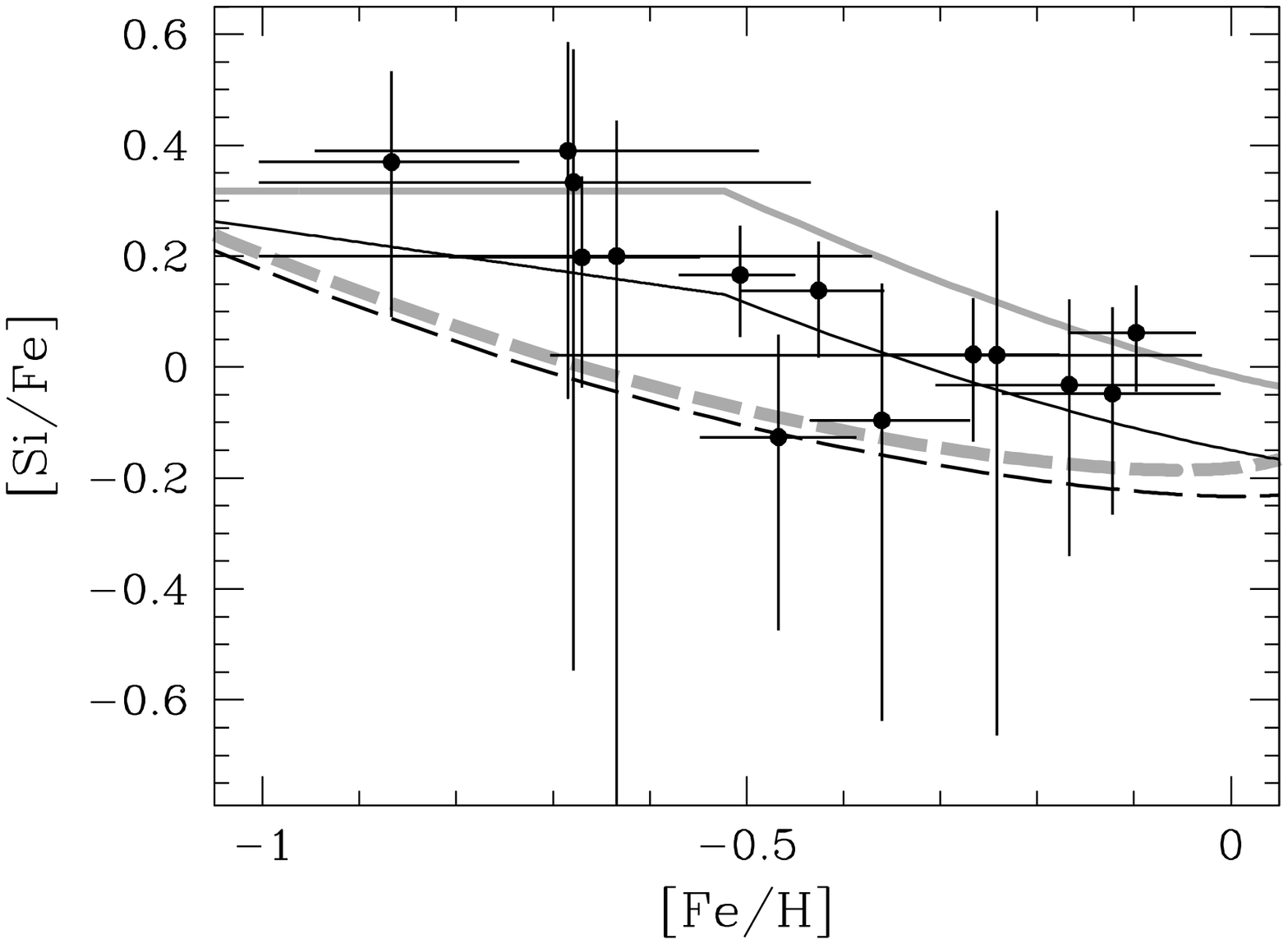}} \hfill \includegraphics[width=1.88in]{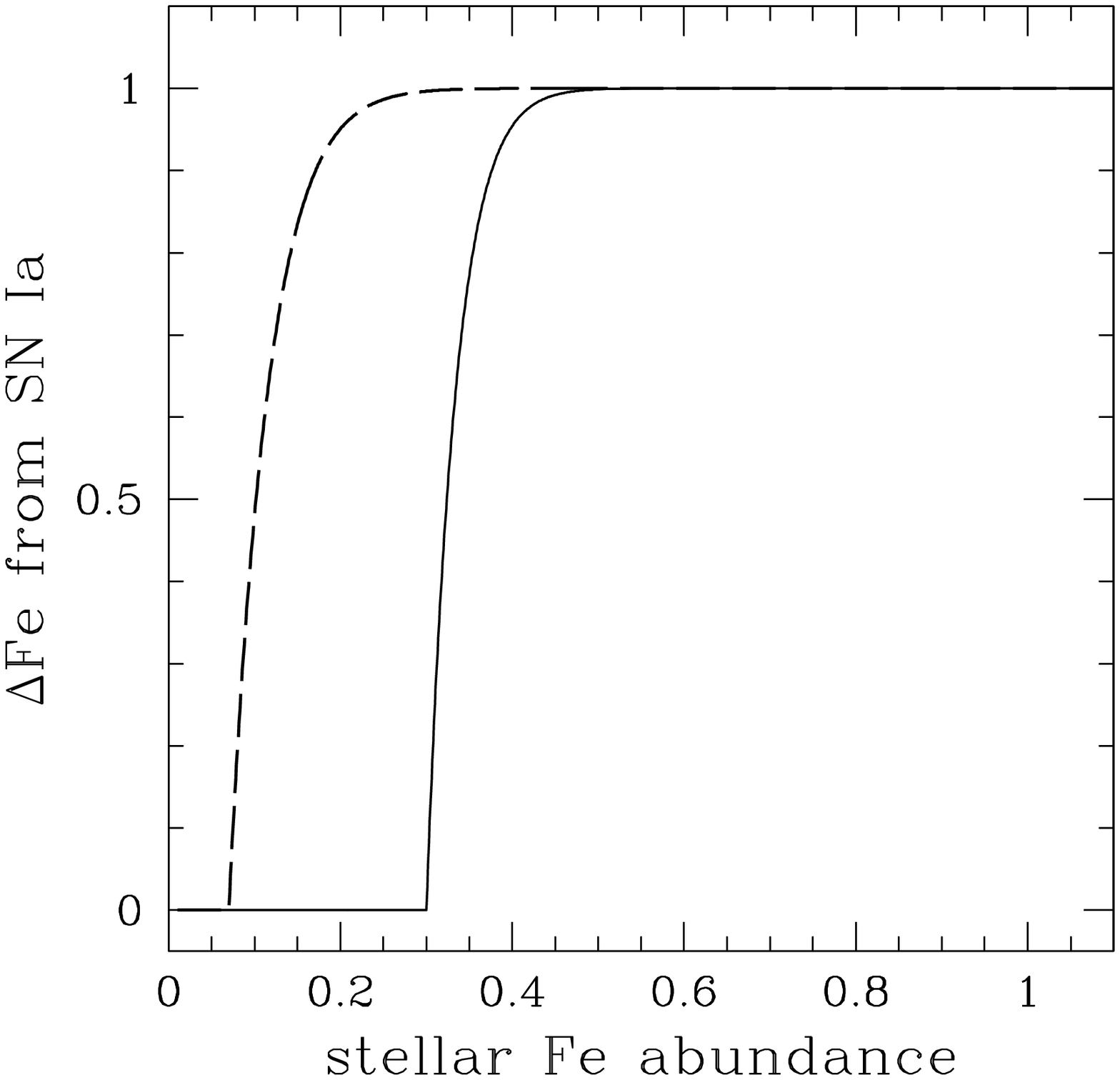}

\vspace*{0.5cm}

\rput(3.2,2.25){\includegraphics[width=2.5in]{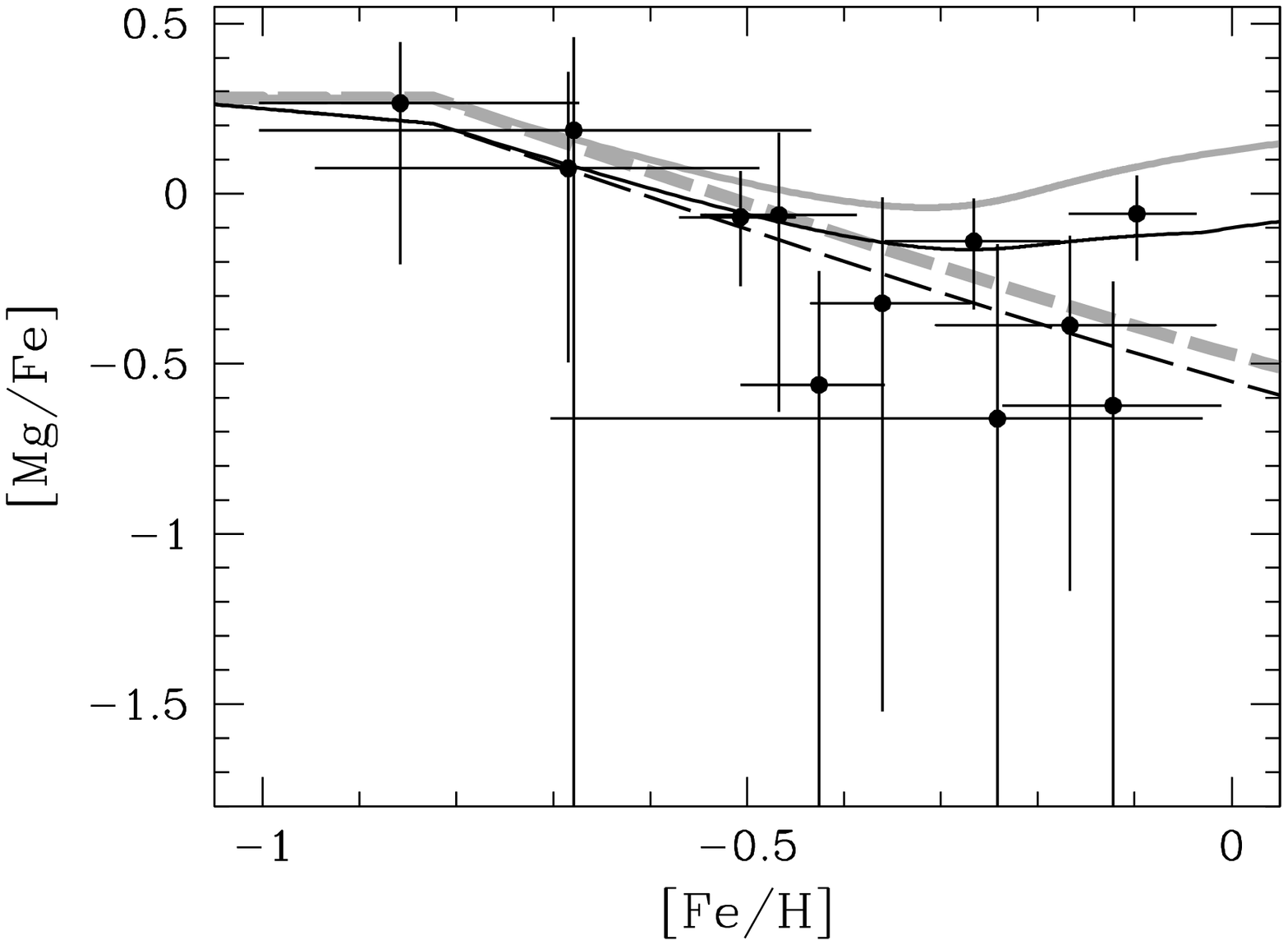}} \hfill \rput(3.2,2.25){\includegraphics[width=2.5in]{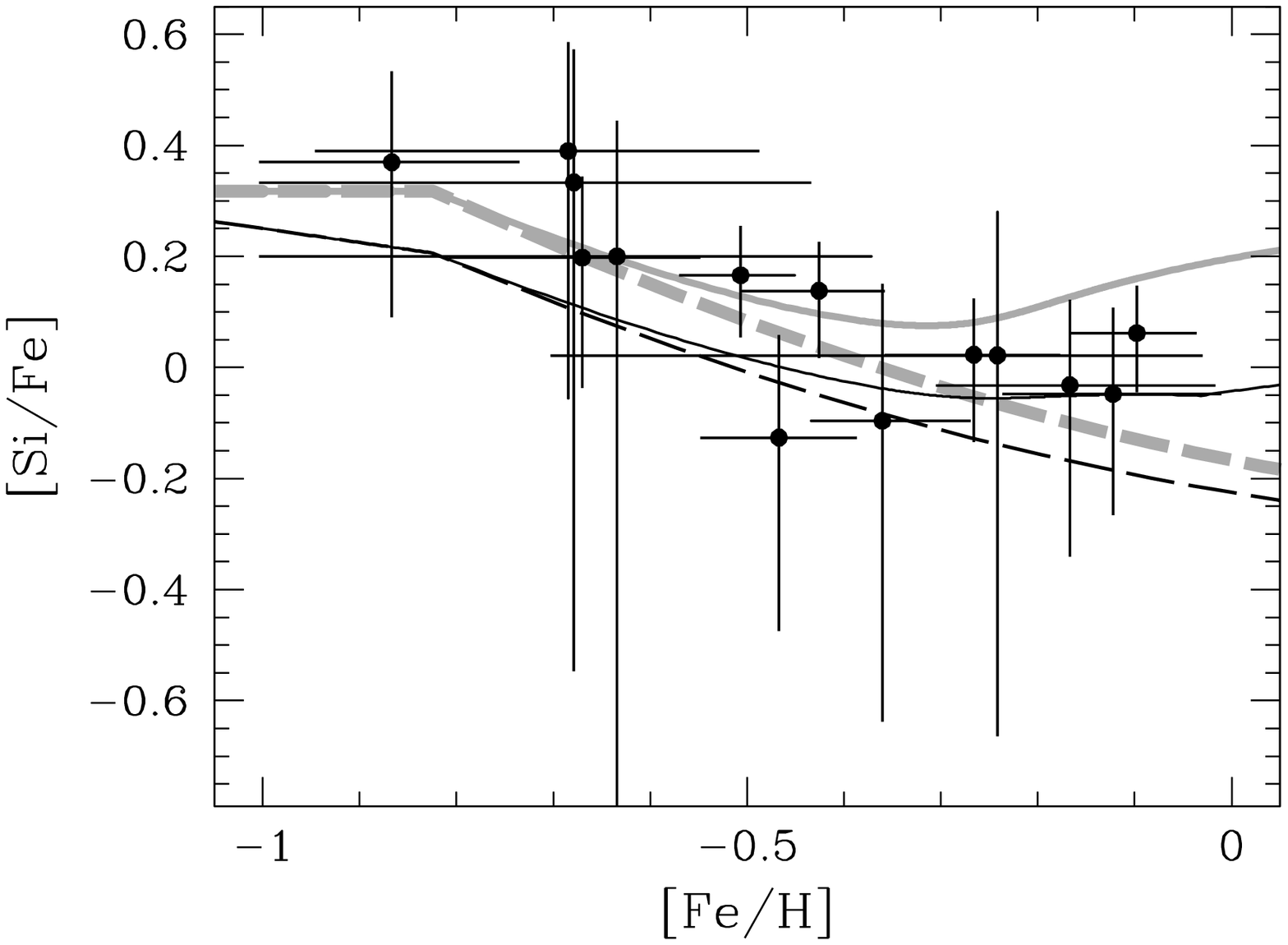}} \hfill \includegraphics[width=1.88in]{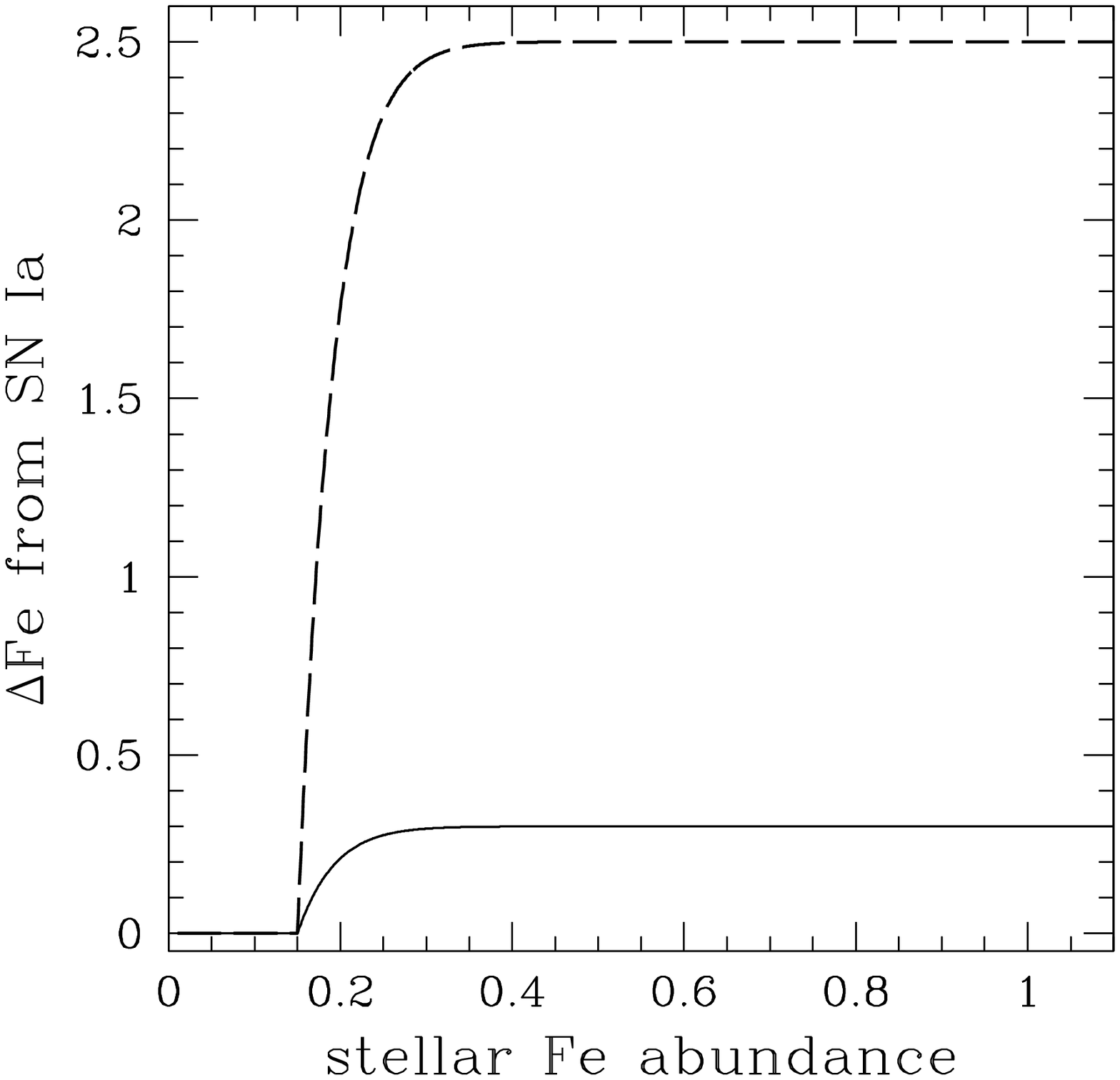}

\figcaption{ Modeling of the abundance data by varying the SN~Ia rate with
the progenitor metallicity. Error bars are shown for the 68 percent
confidence level. Left panels compare the model predictions with Mg vs Fe
data, central panels -- Si vs Fe, while panels to the right indicate the
model used.  Line coding is used to indicate the model applied (the latter
is plotted in the right panels). Black lines in abundance plots indicate the
use of Galactic stellar metallicities, while gray lines are the results
based on metallicities fixed at the integrated SN~II values.  The top panels
demonstrate a need for a steep change in the $\Delta Fe$ from SN~Ia as a
function of progenitor metallicity. The second row of panels exploits the
sensitivity of the data to the assumed value for the cut-off in metallicity
of the SN~Ia progenitor. The bottom panels demonstrate the sensitivity
toward the low SN~Ia rates. $\Delta Fe$ has the meaning of the product of
SN~Ia Fe yield and SN~Ia rate in the sense of Eq.\ref{eq:z}.
\label{fig:snia:mod}}
\end{figure*}

The X-ray derived abundances can be expressed as

\begin{equation} \label{eq:z}
[Z/H]=[Z/H]_* + {f_{\small SN} \; SNU \; M_{{\small SN},_{Z}} \over
{M_*\over L_B} \; {\dot{M}_* \over M_*}}\;,
\end{equation} 

where $f_{\small SN}$ is the SN~Ia rate in units of 1 event (100 yr)$^{-1}$
${ L_B}$/10$^{10}$ (SNU), $M_{{\small SN},_{Fe}}$ is the mass of iron
released in each SN~Ia event, and ${\dot{M}_* \over M_*}$ is the stellar
mass loss, that we adopt as 3 (for the Salpeter IMF) or 5 (for the Arimoto
\& Yoshii IMF) $\times10^{-20}$ sec$^{-1}$ (following calculations by
Mathews 1989). In this approximation, we implicitly assume that the SN~Ia
rate changes with time, similar to the stellar mass loss.  Z denotes the
metal considered. As was estimated in Arimoto \etal (1997), SN~Ia rates
found in optical measurements correspond to $\Delta Fe = 2.5$ for the second
term on the right of Eq.\ref{eq:z}, which we adopt.

Eq.\ref{eq:z} is simply the ratio of metals released by SNe~Ia and by
stellar mass loss to the amount of hydrogen released by stellar mass loss
(cf Loewenstein \& Mathews 1991). As Davis \& White (1996) discussed, the
X-ray iron abundance is correlated with the strength of the gravitational
potential, with SN~Ia ejecta escaping from low-mass systems.  Although the
iron abundance increases with galaxy mass and, in fact even higher iron M/L
is found in groups and even more in clusters (\eg\ Renzini \etal 1993,
FP99), to change the iron abundance (contrary to iron mass), the SN~Ia
ejecta should escape (or be stripped) {\it preferentially}.  Otherwise the
difference among the different systems is only the time for the gas to
accumulate, which cannot explain the low iron abundance in currently adopted
SN Ia models (Greggio \& Renzini 1983), where the SN~Ia rate evolves similar
to stellar mass-loss rate (\eg Loewenstein \& Mathews 1991).

In our modeling we consider two different dependences of Si/Fe on Fe. In one
we assume the dependence measured for the Galactic (spiral) stellar
population. In the other, we assume a constant ratio, corresponding to the
integrated yield from SNe~II (using the Tsujimoto \etal 1995 yields from
Gibson \etal 1997, where the adopted IMF slope was 1.35). The first
assumption implies prolonged, yet ineffective star formation, while the
second requires a very short period of star formation, so only the
contribution from Type II SNe appears in the stellar pattern.  Modeling the
Mg data is not very sensitive to the choice of the stellar abundance
pattern, while the Si data are already well described by the first
assumption. Fig.\ref{fig:snia:mod} demonstrates that a sharp change in the
input of SNe~Ia to the ISM, as a function of the progenitor metallicity, is
required by the measurements of Mg and Si vs Fe abundance.

As seen from Fig.\ref{fig:snia:mod}, within the assumed model our data are
sensitive to low rates of SN~Ia, although they cannot place upper limits
on the rate, contrary to the usual reference to X-ray abundance
determinations (cf Loewenstein \& Mathews 1991). Thus, in the current
study, X-ray observations of early-type galaxies suggest a sharp decline
in the SN~Ia rate as a function of the progenitor star metallicity, and
provide a lower limit on the SN~Ia rate. Our results imply a cut-off
metallicity in the range 0.07--0.3 solar. Lower values are ruled out by the
low abundance in the X-ray gas, while the upper limit is determined by the
change in the abundance pattern above 0.2 solar metallicity. To explain the
difference in the Si/Fe ratios between M84 and the outskirts of NGC4472 and
NGC4649 on one hand and the centers of NGC4472, NGC4649 and NGC4636 on the
other, the model requires a lower limit of 0.3 solar on the Fe contribution
from SN~Ia.

\bigskip
\centerline{\includegraphics[width=3.25in]{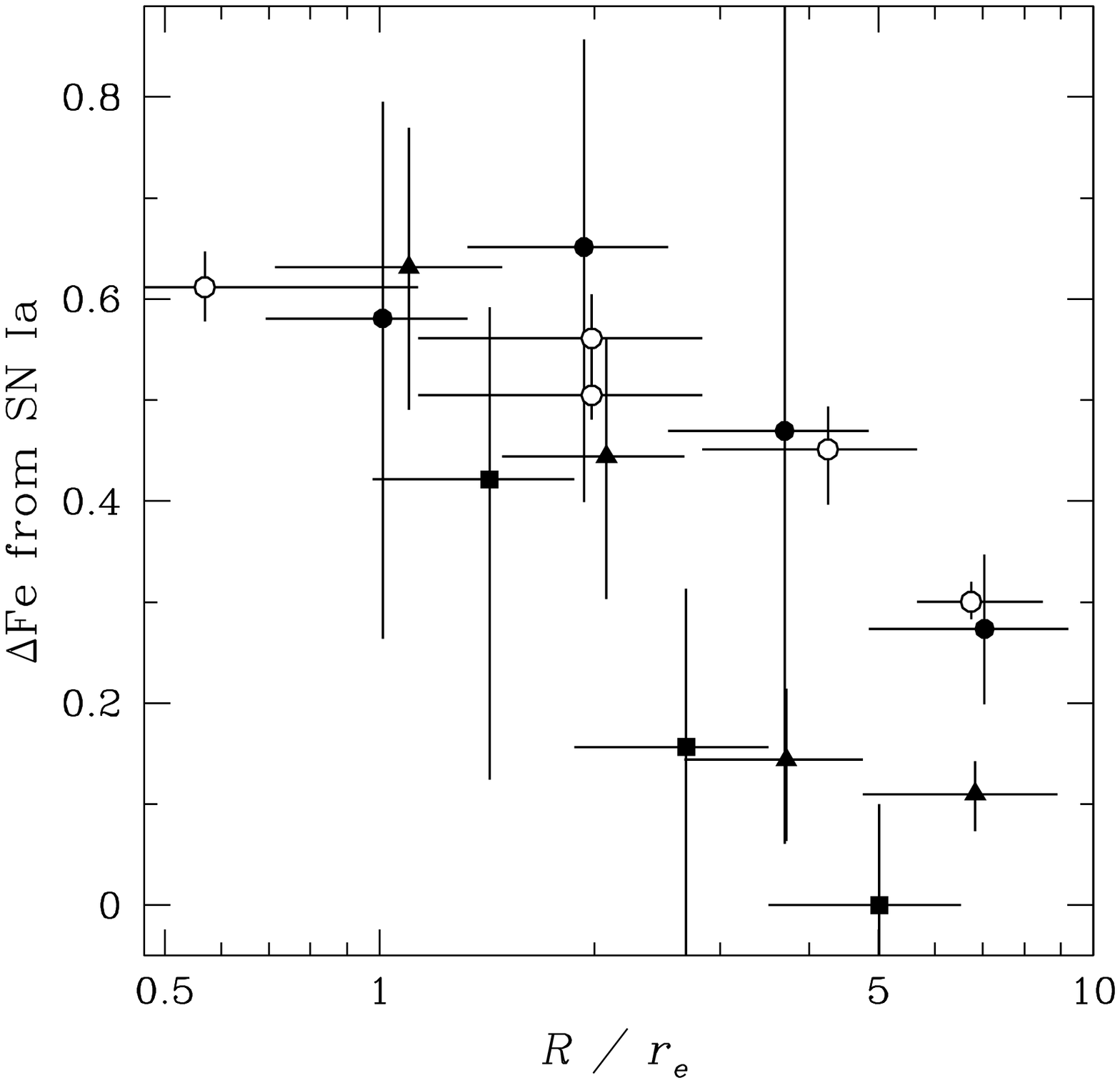}}

\figcaption{SN Ia contribution to the iron abundance in NGC4472 (filled
  circles), NGC4636 (filled triangles), NGC4649 (filled squares) and M87
(open circles, values are given for both cold and hot components) at radii
expressed in units of $r_e$. Error bars are shown for 68 percent confidence
level. The units of Fe abundance are $3.31\times10^{-5}$ relative to H in
number.
\label{sn1a_re}}
\medskip

Although any abrupt rise with radius in the metallicity of the X-ray gas
will be smeared inward by the gas inflow, studies of the SN~Ia rate as a
function of radius can provide additional constraints. In the following we
assume the stellar Si/Fe ratio corresponding to the SN~II yields from
Woosley \& Weaver (1995) with reduced Fe yields ($y_{Fe}=0.07$\msun\ Timmes
\etal 1995; Prantzos \& Silk 1998) and use our Si and Fe measurements to
separate the contributions from different types of SN. In Fig.\ref{sn1a_re}
we present our estimates for the SN Ia contributions to the iron abundance
in NGC4472, NGC4636, NGC4649 and M87 at radii expressed in units of $r_e$
(corresponding to 1.733\amin, 1.683\amin, 1.227\amin\ and 1.766\amin,
respectively). The average value among NGC4472, NGC4636, NGC4649
measurements within $1 r_e$ is $0.56\pm0.10$ solar (90 per cent confidence),
which is consistent with the estimate for M87 itself.  Beyond $6 r_e$, the
SN~Ia contribution to the iron abundance is less than 0.3 solar.  If we use
the Arimoto \& Yoshii IMF values, we find SN~Ia rates for individual
galaxies of $0.06\pm0.05$ SNU for NGC4472, $0.08\pm0.03$ SNU for NGC4636,
$0.05\pm0.05$ for NGC4649 and $0.11\pm0.01$ for M87 (uncertainties are at
the 90 percent confidence level). These can be compared with the SN~Ia rate
of $0.15\pm0.06$ SNU from SN~Ia searches (Cappellaro \etal 1997). We scale
our calculations to the $H_0=75$ km s$^{-1}$ Mpc$^{-1}$, used in their work
and assume 0.4\msun\ of Fe is released by each SN~Ia event (compared to the
usually assumed iron mass of 0.7\msun, Arimoto \etal 1997), which increases
our estimates for SN~Ia rates from the X-ray abundances (as well as usage of
reduced SN II Fe yields). Our choice for estimating stellar mass loss
corresponds to a flat IMF, that also favors higher SN~Ia rates.

Quite interestingly, not all SN~Ia found in optical searches are similar.
The estimated rate of 0.05 SNU applies to the ``faint'' SN~Ia (Cappellaro
\etal 1997). If we neglect these as a source of metals, then the effective
SN~Ia rate would be $0.10\pm0.06$ SNU. We note that to compare with the
optical rate, one should average the SN limits over all the galaxies from
the complete sample.  Thus, if low-luminosity galaxies have lower SN~Ia
rates compared to the bright cluster members, then SN~Ia optical
magnitude-limited searches will be biased toward detecting the higher rates.
From the current optical data on SN~Ia rates, we conclude that these are
consistent with X-ray measurements for bright early-type galaxies.

We conclude that models, in which the SN rate (or yields) depend on
metallicity, can solve the long standing problem of low X-ray limits on
SN~Ia rates compared to those derived from optical searches (cf Arimoto
\etal 1997 and references therein). In addition, the model of Kobayashi
\etal (1998) employed here, has predictions for high redshift SN~Ia searches 
that could be directly tested in the future.  The sample of
galaxies studied here consists of optically bright galaxies. An important
test for the SN Ia models could be made through X-ray studies of optically
faint galaxies, that are presently assumed to have both low metallicities
and an underabundance of alpha elements.

\section{ Conclusions}

We performed spatially resolved spectroscopy for a sample of nine early-type
galaxies in Virgo using X-ray observations from ROSAT and ASCA. Careful
treatment of the galaxy's X-ray emission reveals a temperature structure
consisting of a cool central region and an outward increase in temperature,
similar to earlier results. With the ASCA SIS we determine the radial
distribution of Mg, Si and Fe. Declining iron abundances with radius are
found for NGC4472, NGC4649 and confirmed for NGC4636. Profiles of Mg and Si
are on average flatter than for iron, leading to an underabundance of
alpha-process elements at the centers of these galaxies. The low iron
abundance of NGC4374, found in earlier ASCA studies (Loewenstein \etal 1994)
is confirmed by our analysis, with a detection of the corresponding
overabundance of alpha-process elements.

We compare the [Mg/Fe] and [Si/Fe] vs [Fe/H] dependences for our sample with
the stellar pattern of the Galaxy and conclude that SN Ia enrichment is
important for the hot gas in early-type galaxies.

We compare the abundance measurements for our sample with the implications
of the low-metallicity inhibition of the SNe~Ia, proposed by Kobayashi \etal
(1998). We conclude that within this model it is possible to solve the
long-standing problem of high SN~Ia rates implied in the optical searches
compared with the lower X-ray upper limits. In terms of this model, our data
characterize the SN~Ia rate by a cut-off in the progenitor star metallicity
in the range 0.07--0.3 solar and a {\it lower} limit on SN~Ia rate
corresponding to an Fe abundance of 0.3 solar (corresponding to 0.04
$h_{75}^{3}$ SNU for galaxies in our sample).

Adopting an average SN~II element pattern for the stellar population in our
sample, we illustrate the radial input of SN~Ia ejecta into the X-ray gas.
Our estimates for the SN~Ia rate at the centers of our brightest galaxies
give value of $0.08\pm0.03h_{75}^{3}$ SNU. The rates inferred from optical
searches should be corrected for the presence of ``faint'' SN Ia events,
since these release little metal and therefore are not ``detected'' in the
X-ray spectra.  With this correction the present-day SN~Ia rate in
early-type galaxies is $0.10\pm0.06$ $h_{75}^{2}$ SNU (Cappellaro \etal 1997)
and is therefore comparable with the X-ray estimates. 

This work was supported by the Smithsonian Institution and NASA grant
NAG5-2588. AF thanks Prof. Chugai, Prof. Makishima and Dr. Matsushita for
their discussion of the results presented in this paper. Authors thank the
anonymous referee for pointing out to a necessity of a discussion regarding
the optical/X-ray abundance determinations.

\end{document}